\documentstyle{mn}

\newcounter{parentequation}
\def\beglet{
  \addtocounter{equation}{1}%
  \setcounter{parentequation}{\value{equation}}%
  \setcounter{equation}{0}%
  \def\theequation{\arabic{parentequation}\alph{equation}}%
  \ignorespaces
}
\def\endlet{
  \setcounter{equation}{\value{parentequation}}%
  \def\theequation{\arabic{equation}}%
}
\def\begletA{
  \addtocounter{equation}{1}%
  \setcounter{parentequation}{\value{equation}}%
  \setcounter{equation}{0}%
  \def\theequation{A\arabic{parentequation}\alph{equation}}%
  \ignorespaces
}
\def\endletA{
  \setcounter{equation}{\value{parentequation}}%
  \def\theequation{\arabic{equation}}%
}
\def\begletB{
  \addtocounter{equation}{1}%
  \setcounter{parentequation}{\value{equation}}%
  \setcounter{equation}{0}%
  \def\theequation{B\arabic{parentequation}\alph{equation}}%
  \ignorespaces
}
\def\begletC{
  \addtocounter{equation}{1}%
  \setcounter{parentequation}{\value{equation}}%
  \setcounter{equation}{0}%
  \def\theequation{B\arabic{parentequation}}%
  \ignorespaces
}
\def\endletB{
  \setcounter{equation}{\value{parentequation}}%
  \def\theequation{\arabic{equation}}%
}
\def\ltsima{$\; \buildrel < \over \sim \;$}
\def\gtsima{$\; \buildrel > \over \sim \;$}
\def\simlt{\lower.5ex\hbox{\ltsima}}
\def\simgt{\lower.5ex\hbox{\gtsima}}
\def\kms{{\rm kms}^{-1}}
\def\etal{{\it et al.}\rm}

\begin{document}

\title[Supernovae Feedback in Galaxy Formation]{A Model of Supernovae
Feedback in Galaxy Formation}

\author[G. Efstathiou]{G. Efstathiou \\
Institute of Astronomy, Madingley Road, Cambridge, CB3 OHA.}

\maketitle

\begin{abstract}
A model of supernovae feedback during disc galaxy formation is
developed. The model incorporates infall of cooling gas from a halo
and outflow of hot gas from a multiphase interstellar medium
(ISM). The star formation rate is determined by balancing the energy
dissipated in collisions between cold gas clouds with that supplied by
supernovae in a disc marginally unstable to axisymmetric
instabilities.  Hot gas is created by thermal evaporation of cold gas
clouds in supernovae remnants, and criteria are derived to estimate
the characteristic temperature and density of the hot component and
hence the net mass outflow rate. A number of refinements of the model
are investigated, including a simple model of a galactic fountain, the
response of the cold component to the pressure of the hot gas,
pressure induced star formation and chemical evolution. The main
conclusion of this paper is that low rates of star formation can expel
a large fraction of the gas from a dwarf galaxy.  For example, a
galaxy with circular speed $\sim 50\;\kms$ can expel $\sim 60$--$80\%$
of its gas over a time-scale of $\sim 1$ Gyr, with a star formation
rate that never exceeds $\sim 0.1 M_\odot$/year. Effective feedback
can therefore take place in a quiescent mode and does not require
strong bursts of star formation. Even a large galaxy, such as the
Milky Way, might have lost as much as $20\%$ of its mass in a
supernovae driven wind.  The models developed here suggest that dwarf
galaxies at high redshifts will have low average star formation rates
and may contain extended gaseous discs of largely unprocessed
gas. Such extended gaseous discs might explain the numbers,
metallicities and metallicity dispersions of damped Lyman alpha
systems.
\vskip 0.2 truein
\end{abstract}


\section{Introduction}

Since the pioneering paper of White and Rees (1978), it has been clear
that some type of feedback mechanism is required to explain the shape
of the galaxy luminosity function in hierarchical clustering theories.
The reason for this is easy to understand; if the power spectrum of
mass fluctuations is approximated as a power law $P(k) \propto k^n$,
the Press-Schechter (1974) theory for the distribution of virialized haloes
predicts a power law dependence at low masses
\begin{equation}
{d N(m) \over dm} \propto m^{-(9-n)/6}. \label{Int1}
\end{equation}
For any reasonable value of the index $n$ ($n \approx -2$ on the
scales
relevant to galaxy formation in cold dark
matter (CDM) models), equation~(\ref{Int1}) predicts a much steeper
mass spectrum than the observed faint end slope of the galaxy
luminosity function, $dN(L)/dL
\propto L^\alpha$, with $\alpha \approx -1$ (Efstathiou, Ellis and
Peterson 1988, Loveday \etal$\;$ 1992, Zucca \etal$\;$ 1997).
Furthermore, the cooling times of collisionally ionised gas clouds
forming at high redshift are short compared to the Hubble time (Rees
and Ostriker 1977, White and Rees 1978). Thus, in the absence of
feedback, one would expect that a large fraction of the baryons would
have collapsed at high redshift into low mass dark matter haloes,  in
contradiction with observations.

In reality, there are a number of complex physical mechanisms that can
influence galaxy formation and these need to be understood if we are
to construct a realistic model of galaxy formation.  In the `standard'
cold dark matter model ({\it i.e.} nearly scale invariant adiabatic
perturbations), the first generation of collapsed objects will form in
haloes with low virial temperatures ($T \simlt 10^4$K, characteristic
circular speeds $v_c \simlt 20 \kms$). Molecular hydrogen is the
dominant coolant at such low temperatures and so an analysis of the
formation of the first stellar objects requires an understanding of
the molecular hydrogen abundance and how this is influenced by the
ambient ultraviolet radiation field (Haiman, Rees and Loeb 1997,
Haiman, Abel and Rees 1999).  As the background UV flux rises, the
temperature of the intergalactic medium will rise to $\sim 10^4$K
({\it e.g.} Gnedin and Ostriker 1997) and the UV background will
reduce the effectiveness of cooling in low density highly ionized gas
(Efstathiou 1992). A UV background can therefore suppress the collapse
of gas in regions of low overdensity. It is this
low density photoionized gas that we believe
accounts for the Ly$\alpha$ absorption lines (Cen \etal$\;$ 1994,
Hernquist \etal$\;$ 1996, Theuns \etal$\;$ 1998, Bryan \etal$\;$
1999). Photoionization can also suppress the collapse of gas in haloes
with circular speeds of up to $v_c \sim 20$--$30\kms$. However,
numerical simulations have shown that a UV background cannot prevent
the collapse of gas in haloes with higher circular speeds, though it
can reduce significantly the efficiency with which low density gas is
accreted onto massive galaxies (Quinn, Katz and Efstathiou 1996, Navarro and Steinmetz
1997).

To explain the galaxy luminosity function, feedback is required in
galaxies with circular speeds $v_c \simgt 50 \;\kms$ with
characteristic virial temperatures of $\simgt 10^5\;$K. Energy
injection from supernovae is probably the most plausible feedback
mechanism for systems with such high virial temperatures. Winds from
quasars might also disrupt galaxy formation (Silk and Rees 1998) or,
more plausibly, limit the growth of the central black hole (Fabian
1999). Here we will be concerned exclusively with supernova driven
feedback and will not consider feedback from an active nucleus.
Simple parametric models of supernovae feedback were developed by
White and Rees (1978) and White and Frenk (1991) and form a key
ingredient of semi-analytic models of galaxy formation ({\it e.g.}
Kauffmann, White and Guiderdoni 1993, Lacey
\etal$\;$ 1993, Cole \etal$\;$ 1994, Baugh \etal$\;$ 1996, 1998,
Somerville and Primack 1999).  In this paper, we develop a more
detailed model of the feedback process itself.  Previous papers on
supernovae feedback include those of Larson (1974), Dekel and Silk
(1986) and Babul and Rees (1992). These authors compute the energy
injected by supernovae into a uniform interstellar medium (ISM) and
apply a simple binding energy criterion to assess whether the ISM
will be driven out of the galaxy. The feedback process in these models
is explosive, operating on the characteristic timescale of $\sim 10^6$
--$10^7$ yrs for supernova remnants to overlap.  This is much shorter
than the typical infall timescale of hot gas in the halo, begging the
question of how a resevoir of cold gas accumulated in the first place.
The present paper differs in that we model the ISM as a two-phase
medium consisting of cold clouds and a hot pressure confining medium,
{\it i.e.} as a simplified version of the three-phase model of the ISM
developed by McKee and Ostriker (1977, hereafter
MO77). The cold component contains most of the gaseous mass of the
disc and is converted into a hot phase by thermal evaporation in
expanding supernovae remnants. In this type of model, the cold phase
can be lost gradually in a galactic wind as it is slowly converted into
a hot phase.

The main result of this paper is that low rates of star formation can
expel a large fraction of the baryonic mass in dwarf galaxies over a
relatively long timescale of $\sim 1$ Gyr. We therefore propose that
effective feedback can operate in an steady, unspectacular mode;
strong bursts of star formation and superwind-like phenomena ({\it
e.g.} Heckman, Armus and Miley 1990) are not required, although
galaxies may experience additional feedback of this sort. In fact,
hydrodynamic simulations suggest that nuclear starbursts are
ineffective in removing the ISM from galaxies with gas masses $\simgt
10^6 M_\odot$ (Mac Low and Ferrara 1999, Strickland and Stevens 1999)
because hot gas generated in the nuclear regions is a expelled in a
bipolar outflow without coupling to the cool gas in the rest of the
disc.  This result provides additional motivation for investigating a
``quiescent'' mode of feedback.  Silk (1997) describes a model which
is similar, in some respects, to the model described here. However,
the  model described here is  more detailed and allows a crude
investigation of the radial properties of a disc galaxy during
formation.

The layout of this paper is as follows. A simple model of star
formation regulated by disc instabilities is described in Section
2. This is applied to `closed box' ({\it i.e.}  no infall or outflow
of gas) models of disc galaxies neglecting feedback. Section 3
describes a model of the interaction of expanding supernovae shells in
a two-phase ISM. This section is based on the model of MO77, but
instead of focussing on equilibrium solutions that might apply to our
own Galaxy, we compute the net rate of conversion of cold gas to hot
gas incorporating the model for self-regulating star formation. This
yields the temperature and density of the hot phase as a function of
time and radius within the disc.  Section 4 revisits the model of
Section 2, but includes simultaneous infall and outflow of gas.  This
model is extended in Section 5 to include a galactic fountain, the
pressure response of the cold ISM to the hot phase, and a model of
chemical evolution. Section 6 describes some results from this model
and discusses the effects of varying some of the input parameters.  In
addition, the efficiency of feedback is computed as a function of the
circular speed of the surrounding dark matter halo. Our conclusions
are summarized in Section 7. Although we focus on disc galaxies in
this paper, a similar formalism could be applied to the formation of bulges
if the assumption that gas conserves its angular momentum during
collapse is relaxed.

\section{Star Formation Regulated by Disc Instabilities}

\subsection{Rotation curve for the Disc and Halo}

The dark halo is assumed to be described by the Navarro, Frenk and White
(1996, hereafter NFW) profile
\begin{equation}
 \rho_H(r) = {\delta_c \rho_c \over (cx) (1 + cx)^2}, \qquad x\equiv r/r_v,
 \label{Rot1}
\end{equation}
where $\rho_c$ is the critical density, $r_v$ is the virial radius
at which the halo has a mean overdensity of $200$ with respect to the
background and $c$ is a concentration parameter (approximately $10$ for
CDM models). The circular speed corresponding to this profile is
\begin{equation}
 v^2_H(r) = v^2_{v} {1 \over x} { \left [ {\rm ln} (1 + cx) - cx/(1 + cx)
\right ] \over  \left [ {\rm ln} (1 + c) - c/(1 + c)
\right ]}, \quad v^2_v \equiv {G M_v \over r_v}, \label{Rot2}
\end{equation}
where $M_v$ is the mass of the halo within the virial radius.

We assume that the disc surface mass density distribution
is described by an exponential,
\begin{equation}
 \mu_D(r) = \mu_0 \; {\rm exp}(- r/r_D), \qquad M_D \equiv 2\pi \mu_0 r_D^2,
\label{Rot3}
\end{equation}
where $M_D$ is the total disc mass. The rotation curve
of a cold exponential disc is given by (Freeman 1970)
\begin{eqnarray}
 v^2_D(r) =2 v^2_c y^2
\left [I_0(y)K_0(y)- I_1(y)K_1(y)\right], \\
\quad y \equiv {1 \over 2}{r \over r_D}, \quad v^2_c = {G M_D \over r_D}. \nonumber
\label{Rot4}
\end{eqnarray}

To relate the disc scale length, $r_D$, to the virial radius of the
 halo $r_v$, we assume that the angular momentum of the disc material
 acquired by tidal torques is approximately conserved during the
 collapse of the disc (see Fall and Efstathiou 1980). This fixes the
 collapse factor
\begin{equation}
 f_{coll} = {r_V \over r_D} 
\label{Rot5}
\end{equation}
in terms of the dimensionless spin parameter $\lambda_H \equiv J \vert
E \vert^{1/2} G^{-1}M^{-5/2}$ of the halo component. The spin
parameter is found to have a median value of $\approx 0.05$ from
N-body simulations (Barnes and Efstathiou 1987), and for the models
described here, this value is reproduced for collapse factors of
around $50$.  A more detailed calculation of the collapse factor of
the disc is given in Section 4.

\subsection{Vertical scale height of the disc}

The velocity dispersion of the cold gas clouds in the vertical
direction is assumed to  be constant and equal to $\sigma^2_g$. The
equations of stellar hydrodynamics then give the following solution
\begin{equation}
 \rho(z) = {\mu_g \over 2 H_g} {\rm sech}^2 \left ( {z \over H_g} \right), \label{Vert1}
\end{equation}
where $\mu_g$ is the surface mass density of the gas and the scale height
is given by
\begin{equation}
 H_g = {\sigma^2_g \over \pi G \mu_g}. \label{Vert2}
\end{equation}
Equation (\ref{Vert2}) must be modified to take into account the
stellar disc. We do this approximately by assuming `disc pressure
equilibrium' (Talbot and Arnett 1975)
\begin{equation}
 H_g = {\sigma^2_g \over \pi G \mu_g} { 1 \over (1 + \beta/\alpha)}, \label{Vert3}
\end{equation}
where the quantities $\alpha$ and $\beta$ relate the vertical velocity
dispersion $\sigma^2_*$ and surface mass density $\mu_*$ of the stars
to those of the gas clouds
\beglet
\begin{eqnarray}
\sigma_* = \alpha \sigma_g,   \\
\mu_* = \beta \mu_g. 
\end{eqnarray}
\endlet

\subsection{Stability of a two-component rotating disc}

The stability of rotating discs of gas and collisionless particles
to axisymmetric modes has been analysed in classic papers by Goldreich
and Lynden-Bell (1965) and by Toomre (1964). Here we use the results of Jog
and Solomon (1984) who analysed the stability of a rotating disc
 consisting of two isothermal fluids of sound speeds $c_1$, $c_2$
and surface mass densities $\mu_1$ and $\mu_2$. These authors find
that such a disc is stable to axisymmetric modes of wavenumber $k$ if
\begin{equation}
 x = {2 \pi G \mu_1 \over \kappa^2} { k \over ( 1 + k^2c_1^2/\kappa^2)}
+ {2 \pi G \mu_2 \over \kappa^2} { k \over ( 1 + k^2c_2^2/\kappa^2)} < 1,
\label{Stab1}
\end{equation}
where $\kappa$ is the epicyclic frequency
\begin{equation}
 \kappa = 2\omega \left ( 1 + {1 \over 2} {r \over \omega} {d\omega \over dr}
\right )^{1/2}. \label{Stab2}
\end{equation}
Equation (\ref{Stab1}) yields a cubic equation for the most unstable
mode $k_m$.  Solving this equation in terms of the parameters $\alpha$
and $\beta$ of equations (10), and ignoring the small differences
between a gaseous and collisionless disc, we can write the stability
criterion for a two-component system as
\begin{equation}
 \sigma_g =  {\pi G \mu_g \over \kappa} g (\alpha, \beta). \label{Stab3}
\end{equation} 
This is identical to the Goldreich-Lynden-Bell criterion except for
the factor $g(\alpha, \beta)$. This factor is plotted in Figure 1 for
various values of $\alpha$ and $\beta$.

\subsection{Star formation and supernovae energy input}

We assume a stellar initial mass function (IMF) of the standard
Salpeter (1955) form
\begin{eqnarray}
 {dN_* \over dm} = Am^{-(1+x)}, \quad m_l < m < m_u, \quad x  = 1.35, 
\label{SN1}\\
 m_l = 0.1 M_\odot, \quad m_u = 50M_\odot, \quad\quad\quad\quad\nonumber
\end{eqnarray} 
and that each star of mass greater than $8M_\odot$ releases
$10^{51}E_{51}$ ergs in kinetic energy in a supernova explosion. For
the IMF of equation~(\ref{SN1}), one supernova is formed for every
$125M_\odot$ of star formation.  The energy injection rate is
therefore related to the star formation rate by
\begin{equation}
 \dot E_{sn} =  2.5 \times 10^{41} E_{51}\dot M_* \; {\rm erg/sec}, \label{SN2}
\end{equation} 
where $\dot M_*$ is the star formation rate in $M_\odot$ per year.

\begin{figure}

\vskip 2.8 truein

\includegraphics{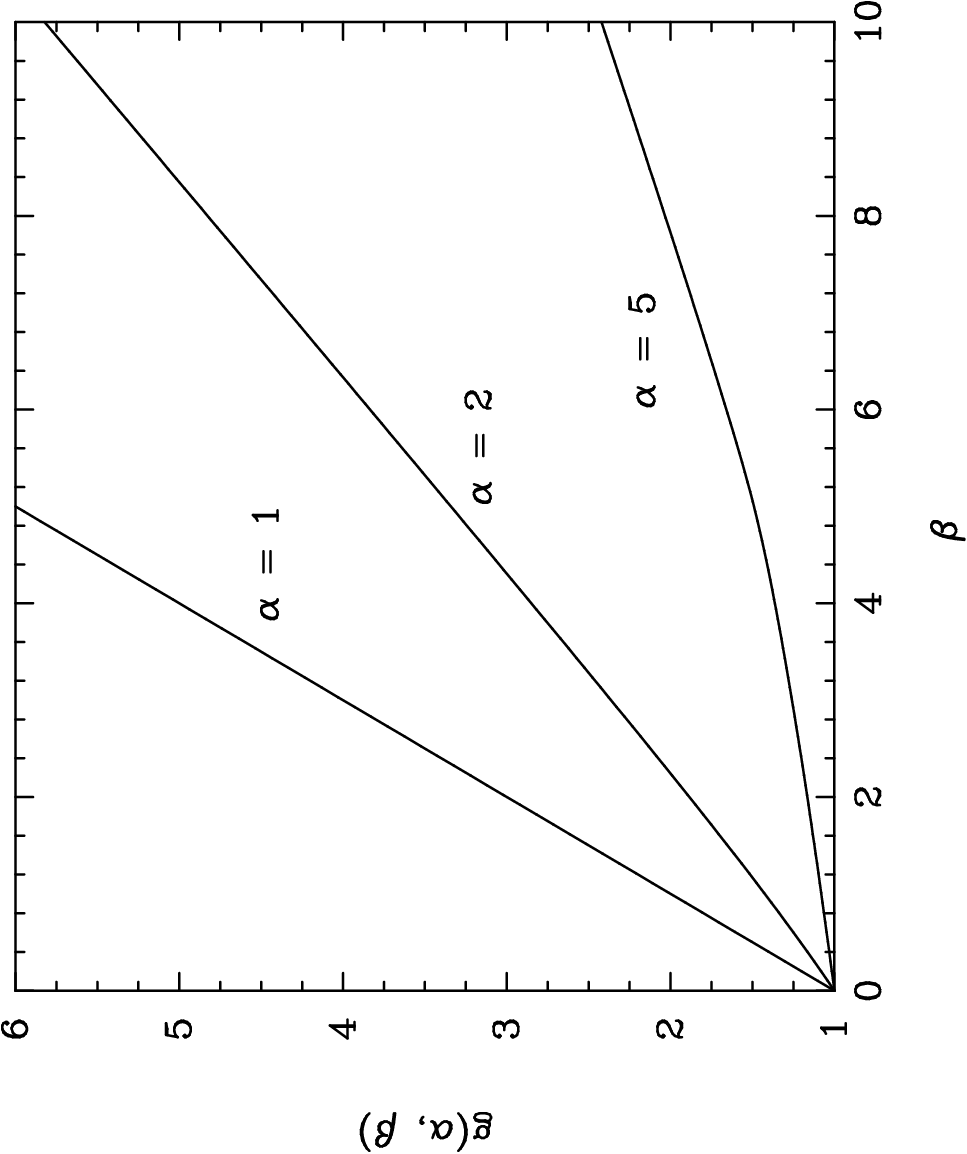}

\caption
{The factor $g(\alpha, \beta)$ appearing in the stability criterion of
equation (13) plotted against $\beta$ for three values of $\alpha$}
\label{figure1}
\end{figure}

\begin{table*}
\bigskip
\centerline{\bf Table 1: Parameters of Model Galaxies}
\begin{center}
\begin{tabular}{ccccccccc} \hline \hline
\noalign{\medskip}
   & $v_c$ (km/s) & $v_{\rm max}$ (km/s) & $v_v/v_c$ & $r_D$ (kpc) &
$M_D$ ($M_\odot$) &  $f_{coll}$     & $c$  & $\lambda_H$ \\
Model MW   &  $280$ & $212$ & $0.45$  & $3.0$       &  $5.5 \times 10^{10}$ &
$50$ & $10$ & $0.065$ \\
Model DW   &  $70$ &  $53$ & $0.45$  & $0.2$       &  $2.3 \times 10^{8}$ &
$50$ & $10$  & $0.065$\\ \hline
\noalign{\medskip}
\end{tabular}
\end{center}
\end{table*}

\subsection{Energy dissipated by cloud collisions}

We assume cold clouds of constant density 
$\overline \rho_c = 7 \times10^{-23}$ g/cm$^3$ with a distribution of cloud radii
\begin{eqnarray}
 {dN_{ca} \over da} = N_0a^{-4}, 
\quad a_l < a < a_u, \\
 \quad  a_l = 0.5 \;{\rm pc}, \quad a_u = 10\;
{\rm pc}
\nonumber,
\label{Diss1}
\end{eqnarray} 
(MO77). Following MO77, the clouds are assumed to have an isotropic
Gaussian velocity distribution with velocity dispersion independent of
cloud size and that they lose energy through inelastic collisions. The
rate of energy loss per unit volume is given by
\begin{eqnarray}
 {dE_{coll} \over dtdV} = 24 \pi^{3/2} \overline \rho_c N_{cl}^2
a_l^5 \sigma_g^3 I_a, \\
 I_a = {1 \over 2} \int_1^{a_u/a_l}
\int_1^{a_u/a_l} {(x+y)^2 \over (x^3 + y^3)} {dx \over x} {dy \over y}
\nonumber \label{Diss2},
\end{eqnarray} 
where $N_{cl}$ is the local cloud density $N_{cl} =
N_0/3a_l^3$. Integrating equation (17) over the vertical
direction and using equation (\ref{Vert3}) for the scale height, the
rate of energy loss per unit surface area $\dot E^\Omega_{coll}$ is
\begin{eqnarray}
 \dot E^\Omega_{coll}  = 5.0 \times 10^{29} \left ( 1 + {\beta \over \alpha}
\right )
\mu_{g5}^3 \sigma_{g5} \;\; {\rm erg}/{\rm sec}/{\rm pc}^2 \label{Diss3},
\end{eqnarray} 
where $\mu_{g5}$ is the surface mass density of the gas component in
units of $5 M_\odot/{\rm pc}^2$ and $\sigma_{g5}$ is the  cloud
velocity dispersion in units of $5$ km/sec. These values are close to
those observed in the local solar neighbourhood. To estimate the
efficiency with which supernovae accelerate the system of clouds, we
normalize to the observed net star formation rate of the Milky
Way. Assuming that the gas distribution has a flat surface mass
density profile to $R_{max} = 14$ kpc (Mihalas and Binney 1981),
$\beta \approx 10$,  $\alpha \approx 5$, and equating the integral
of (\ref{Diss3}) to $\epsilon_c \dot E_{sn}$ (equation \ref{SN2}),
we find
\begin{equation}
 \epsilon_c E_{51} \dot  M_* = 0.004. \label{Diss4}
\end{equation} 
An efficiency parameter of $\epsilon_c = 0.01$ produces a net star
formation rate of $0.4M_\odot$/yr which is reasonable for a Milky
Way-like galaxy. We will therefore adopt a constant value of
$\epsilon_c=0.01$ in the models of the next subsection. The value of
$\epsilon_c$ will, of course, depend on the properties of the clouds,
ISM and star formation rate. For example, in the model of MO77 the
clouds are accelerated by interactions with the cold shells
surrounding supernova remnants and they find efficiencies $\epsilon_c$
of typically a few percent. We investigate the effect of varying
$\epsilon_c$ in Section 6.

\subsection{Self-regulating models without inflow or outflow}

The equations derived above allow us to evolve an initially gaseous
disc and to compute the local star formation rate, cloud velocity
dispersion {\it etc}. The system of stars and gas is constrained to
satisfy the stability criterion of equation (\ref{Stab3}), which fixes
the cloud velocity dispersion $\sigma_g$.  There is some empirical
evidence that star formation in nearby galaxies is regulated by a
stability criterion of this sort ({\it e.g.} Kennicutt 1998).  The
energy lost in cloud collisions (equation \ref{Diss3}) is balanced
against the energy input from supernovae assuming a constant
efficiency factor $\epsilon_c = 0.01$. We assume further that $\alpha
= 5$ (equation 10a) {\it i.e.} that stars are instantaneously
accelerated to higher random velocities than the system of gas clouds,
and that the properties of the gas clouds (mass spectrum, internal
density, {\it etc}) are independent of time. These are clearly
restrictive assumptions, but they allow us to generate simple models
of self-regulating star formation with only one free parameter
$\epsilon_c$.

 We study the evolution of two model galaxies with parameters listed
in Table 1. Model MW has parameters roughly similar to those of the
Milky Way and model DW has parameters similar to those of a relatively
high surface brightness dwarf galaxy.

\begin{figure*}

\vskip 3.8 truein

\includegraphics{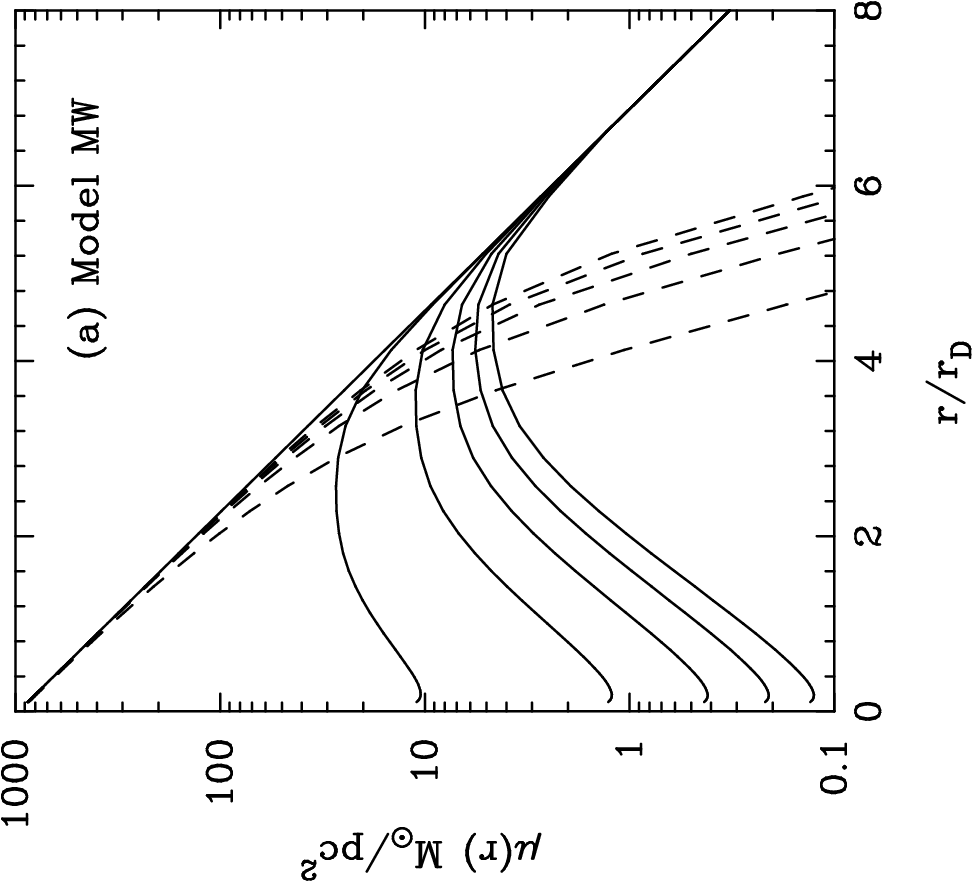}
\includegraphics{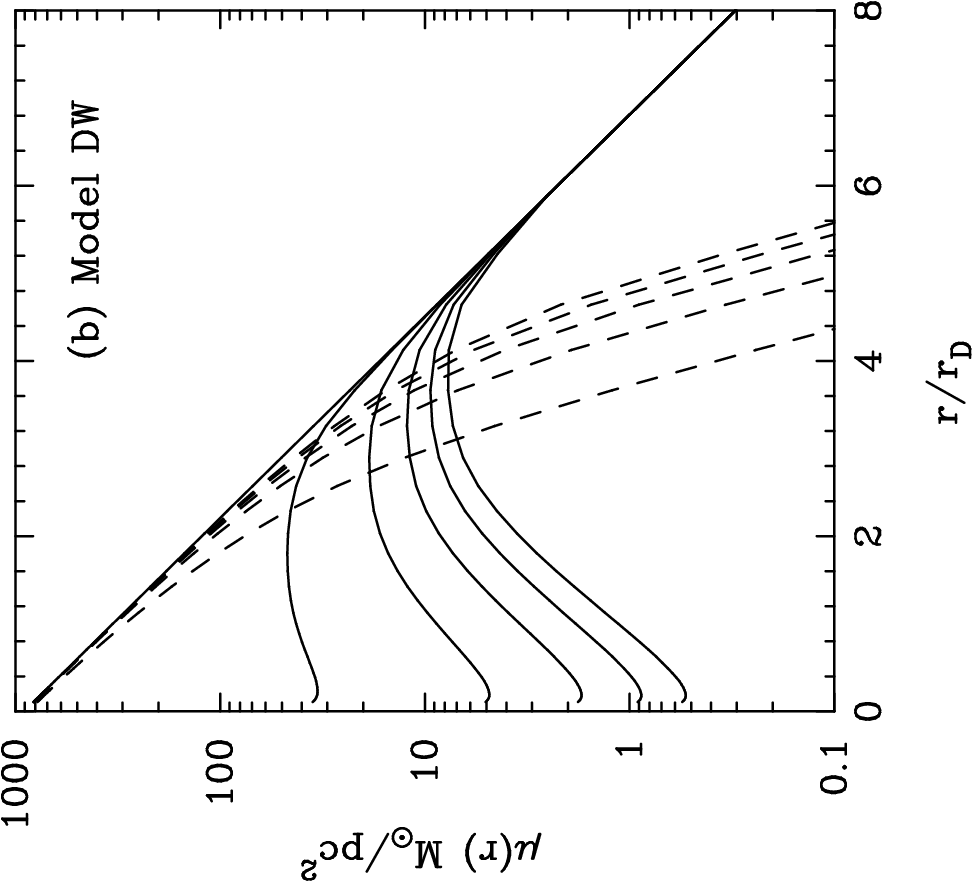}

\caption
{The evolution of the gas (solid lines) and stellar (dashed lines)
surface mass density distributions according to the simple
self-regulating model described in this section. The results are
shown for ages of $0$, $0.1$, $1$, $3$, $6$ and $10$ Gyr.}
\label{figure2}
\end{figure*}

Figure 2 shows the evolution of the gas and stellar surface mass
densities of the two models. The net star formation rates, gas
fractions and mean gas cloud velocity dispersion are plotted in Figure
3. In model MW, the star formation rates are initially high ($> 100
M_\odot$/yr) and hence the timescale for star formation is short; half
the disc mass is converted into stars in $10^7$ years. The
star formation rate declines rapidly to less than $1 M_\odot$/yr after
a few Gyr. As figure 2 shows, the star formation at early times is
concentrated to the inner parts of the disc which have  a high surface
density and hence the gas distribution develops a characteristic
surface density profile with an inner `hole', similar to what is seen
in the HI distributions in real galaxies (see Burton 1991).  The
stellar disc is truncated at about the Holmberg (1958) radius ($r/r_D
\approx 5$), in rough agreement with observations. The truncation
arises because the gas disc becomes thick at large radii (equation
\ref{Vert3}) and the rate of energy lost in cloud collisions can be
balanced by a very low star formation rate.

\begin{figure*}

\vskip 4.5 truein

\includegraphics{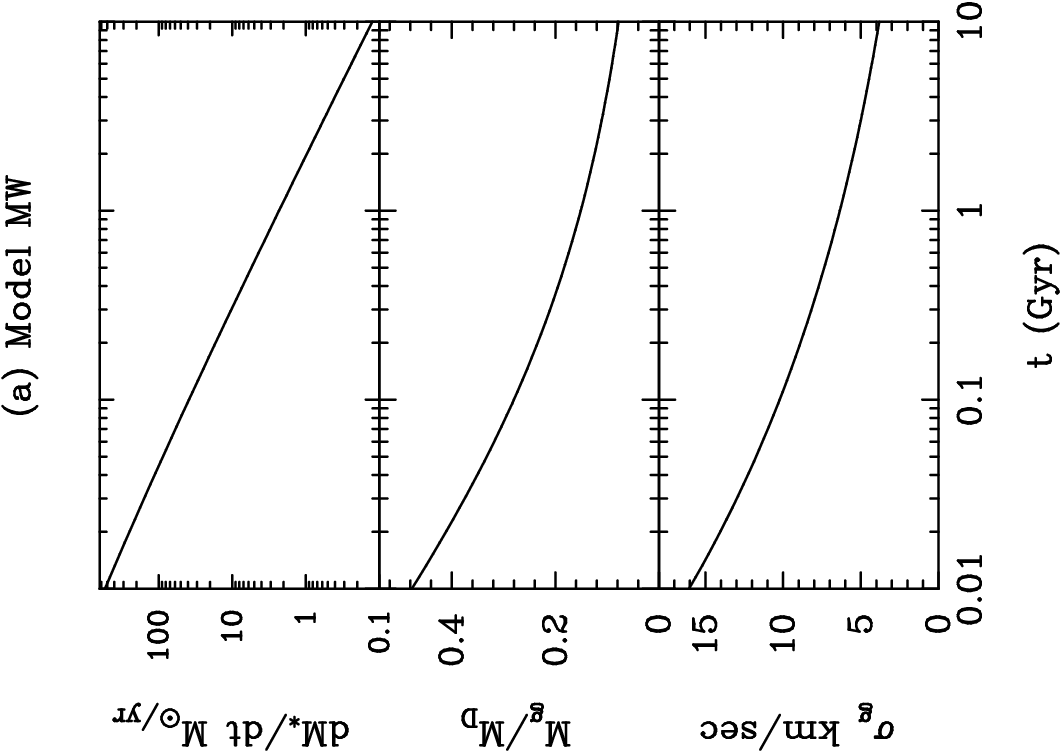}
\includegraphics{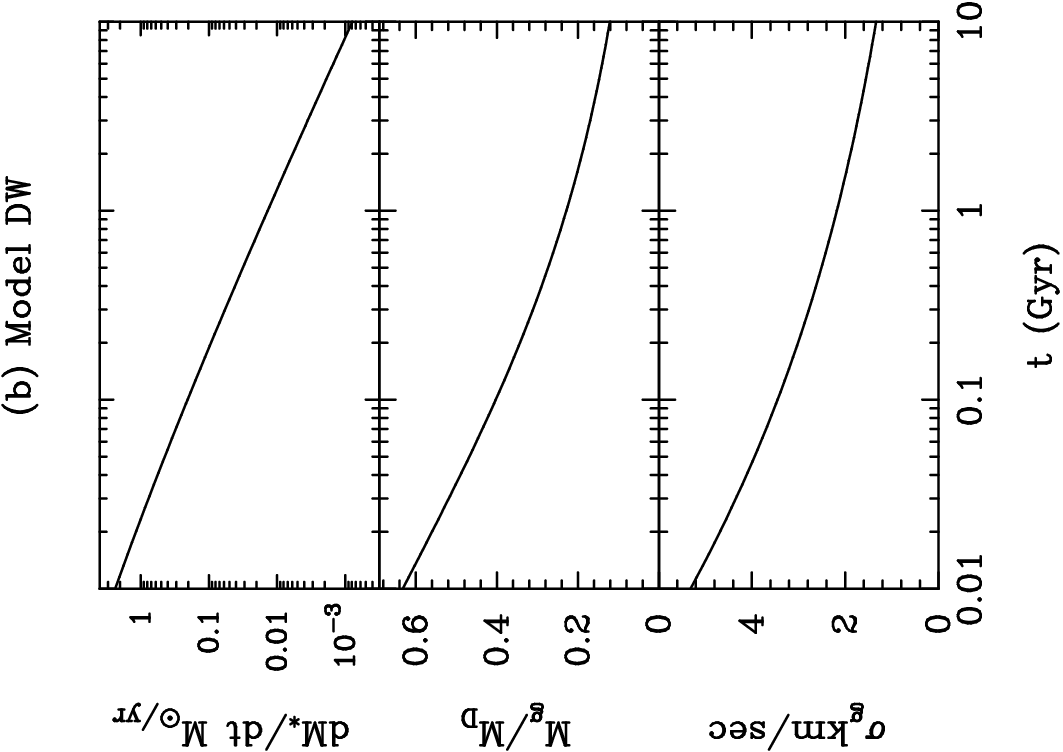}

\caption
{Evolution of the star formation rate, gas fraction and gas cloud
velocity dispersion in the self-regulating model.}
\label{figure3}
\end{figure*}	

The evolution of model DW is qualitatively similar, though the star
formation rate is scaled down roughly in proportion to the disc mass.
Half the gas is converted to stars by $3 \times 10^7$ yr, and the gas
fraction is $0.12$ after $10^{10}$ yr, similar to the final gas fraction
of $0.13$  in model MW.

Neither of these models is satisfactory. The star formation rate in
model MW is too high at early times to be compatible with deep number
counts (see {\it e.g.} Ellis 1997), which require more gentle star
formation rates in typical $L^*$ galaxies. Model SW converts most of
its gas into stars on a short timescale and so does not solve the
problem raised in the introduction of explaining the flat faint end
slope of the luminosity function in CDM-like models. As we will see in
later sections, infall of gas provides the solution to the former
problem, since this allows the disc to build up gradually on a cooling
or dynamical timescale. Outflow of hot gas heated by supernovae
provides a solution to the latter problem.

\section{Evolution of a two-phase ISM}

In this Section we consider the interaction of a multiphase
interstellar medium with expanding supernova remnants following the
model of MO77 and discuss the conditions under which a protogalaxy can
form a wind. The key ingredients of the model are as follows. Most of
the cold gaseous mass is assumed to be in cold clouds with properties
as given in Section 2.5. Supernovae explode and their remnants
propagate evaporating some of the cold clouds and forming a low
density hot phase of the ISM. The star formation rate therefore
determines the evaporation rate and hence the rate of conversion of
the cold phase to a hot phase. A wind from the galaxy can result if
the hot phase is: (i) sufficiently pervasive (filling factor of order
unity), (ii) low density (so that radiative cooling is unimportant)
and (iii) the temperature of the hot phase exceeds the virial
temperature of the galaxy. In this section we follow closely the
theory of the ISM developed by MO77 and we use their notation where
possible.

\subsection{Evaporation of cold clouds}

An expanding supernovae remnant will evaporate a mass of 
\begin{equation}
 M_{ev} \approx 540\; E_{51}^{6/5} \Sigma^{-3/5} n_h^{-4/5} \; M_\odot,
 \label{Eva1}
\end{equation}
where $n_h$ is the density interior to the supernovae remnant and
$\Sigma$ (in pc$^2$) is the evaporation parameter introduced by MO77
\begin{equation}
 \Sigma = { \gamma \over 4 \pi a_l N_{cl} \phi_\kappa} .
 \label{Eva2}
\end{equation}
Here the parameter $\gamma$ relates the blast wave velocity to the
isothermal sound speed ($v_b = \gamma c_h$, $\gamma \approx 2.5$) and
the parameter $\phi_\kappa$ quantifies the effectiveness of the
classical thermal conductivity of the clouds ($\kappa_{eff} = \kappa
\phi_\kappa$) and so is less than unity if the conductivity is reduced
by tangled magnetic fields, turbulence {\it etc}. Using equations
(\ref{Vert1}) and (\ref{Vert3}) to estimate the mean cloud density we
find
\begin{equation}
 \Sigma \approx 280 {\sigma_{g5}^2 \over \mu_{g5}^2} {1 \over (1+
\beta/\alpha)} {1 \over \phi_\kappa} \;{\rm pc}^2 =
f_\Sigma{\Sigma_\odot}, \quad \Sigma_\odot \approx 95\; {\rm pc}^2,
\label{Eva3}
\end{equation}
where $\Sigma_\odot$ is the evaporation parameter characteristic of
the local solar neighbourhood ($\beta/\alpha \approx 2)$. 

Evaluating equation (\ref{Eva1}), we find
\begin{equation}
 M_{ev} \approx 1390\; E_{51}^{6/5} f_\Sigma^{-3/5} \phi_\kappa^{3/5} n_{h-2}^{-4/5} \; M_\odot,
 \label{Eva4}
\end{equation}
where $n_{h-2}$ is $n_h$ in units of $10^{-2} {\rm cm}^{-2}$ (a
characteristic value for the hot component). Thus, provided thermal
conduction is not highly suppressed,  a single supernovae remnant can
evaporate a much larger mass than the $125\;M_\odot$ formed in
stars per supernovae for a standard Saltpeter IMF (Section 2.4). If a
significant fraction of this evaporated gas can escape in a wind, then
star formation will be efficiently suppressed.

\subsection{Temperature and density of the hot phase}

To compute the properties of the hot phase we assume that the
disc achieves a state in which the porosity parameter $Q$
is equal to unity. The disc is then permeated by a network of
overlapping supernovae remnants. Ignoring cooling interior to the
remnants (which we will see is a reasonable approximation for 
an ISM with low metallicity) the age, radius and temperature of
a SNR when $Q=1$ are given by
\beglet
\begin{eqnarray}
t_o = 5.5 \times 10^6 S_{13}^{-5/11} \gamma^{-6/11} E_{51}^{-3/11} n_h^{3/11}
\;\; {\rm yr},   \\
R_o = 100 S_{13}^{-2/11} \gamma^{2/11} E_{51}^{1/11} n_h^{-1/11}\;\;
 {\rm pc}, \\
T_o = 1.2 \times 10^{4} S_{13}^{6/11} \gamma^{-6/11} E_{51}^{8/11} 
n_h^{-8/11}\;\; {\rm K}.
\end{eqnarray}
\endlet
where $S_{13}$ is the supernova rate in units of $10^{-13} {\rm pc}^{-3}
{\rm yr}^{-1}$.
The density of a remnant at $t_o$ ($n_h^o \approx M_{ev}/(4/3\pi R_o^3)$),
gives an approximate estimate of the density of the ambient hot phase
\begin{equation}
 n_h^o \approx 4.3 \times 10^{-3} S_{13}^{0.36} \gamma^{-0.36} E_{51}^{0.61} 
f_\Sigma^{-0.393}\;\; {\rm cm}^{-3}.
 \label{Temp2}
\end{equation}
Inserting this estimate into equations (24) we find
\beglet
\begin{eqnarray}
t_o = 1.2 \times 10^6 S_{13}^{-0.36} \gamma^{-0.64} (E_{51}
f_\Sigma)^{-0.11}
\;\; {\rm yr},   \\
R_o = 164 (S_{13}/\gamma)^{-0.21}  E_{51}^{0.04} f_\Sigma^{0.035}\;\;
 {\rm pc}, \\
T_o = 6.6 \times 10^{5} (S_{13} E_{51} f_\Sigma/\gamma)^{0.29}\;\; {\rm K},
\end{eqnarray}
\endlet
and the rate at which clouds are evaporated is
\begin{equation}
 \dot M_{ev} = 
2.7 \times 10^{-10} S_{13}^{0.71} \gamma^{0.29} E_{51}^{0.71} 
f_\Sigma^{-0.29}\;\; {\rm M}_\odot {\rm pc}^{-3} {\rm yr}^{-1}. 
 \label{Temp3}
\end{equation}
Integrating equation (\ref{Temp3}) over the scale height of the disc gives
the evaporated mass per unit area,
\begin{eqnarray}
 \dot M_{ev}^\Omega \approx  
1 \times 10^{-7} \left ({\sigma_{g5}^2 \over
\mu_{g5} (1 + \beta/\alpha) } \right ) \times  \qquad \nonumber \\ 
\qquad  \qquad S_{13}^{0.71} \gamma^{0.29} E_{51}^{0.71} 
f_\Sigma^{-0.29}\;\; {\rm M}_\odot {\rm pc}^{-2} {\rm yr}^{-1}. 
 \label{Temp4}
\end{eqnarray}

Adopting a cooling rate of $\Lambda \approx 2.5 \times 10^{-22}
T_5^{-1.4}$ erg cm$^{3}$ s$^{-1}$
for $10^{5} \simlt T \simlt 10^{6}$ for a gas with primordial composition,
the ratio of $t_o$ to the cooling time $t_{\rm cool}$ is
\begin{equation}
 {t_o \over t_{\rm cool} } \approx 0.5 T_5^{-2.4} f_\Sigma^{-0.5},
 \label{Temp5}
\end{equation}
Thus if the temperature of the hot phase is higher than about $10^5$K,
the assumption that cooling can be neglected will be valid. A cooling
function for a gas with primordial composition will be used throughout
this paper. As the metallicity of the gas builds up, the cooling time
of the hot component will shorten and more of the supernovae energy
will be lost radiatively. This effect  will reduce the efficiency of 
feedback in galaxies with high metallicity but is not included in this
paper.

\begin{figure*}

\vskip 7.0 truein

\includegraphics{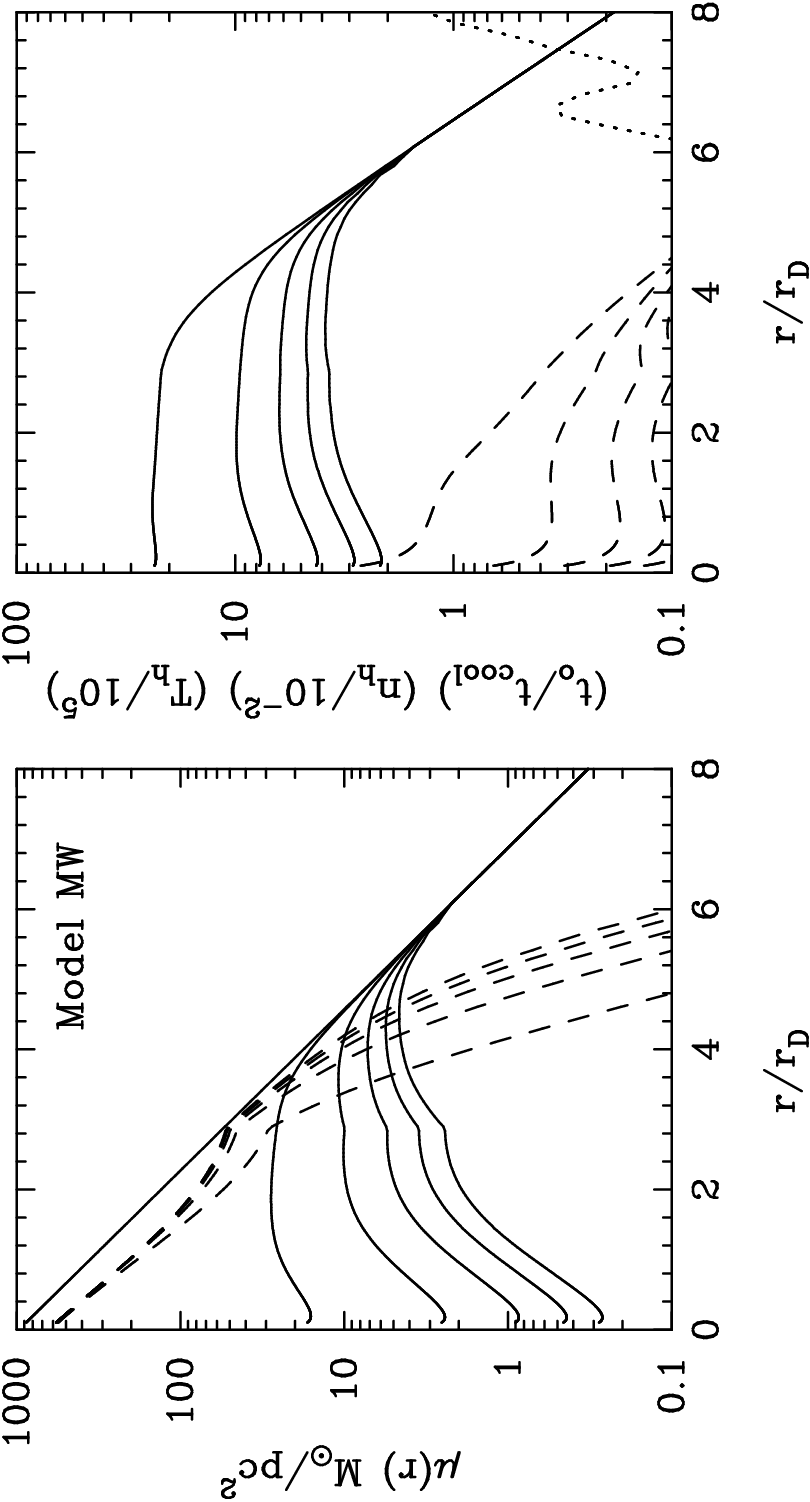}
\includegraphics{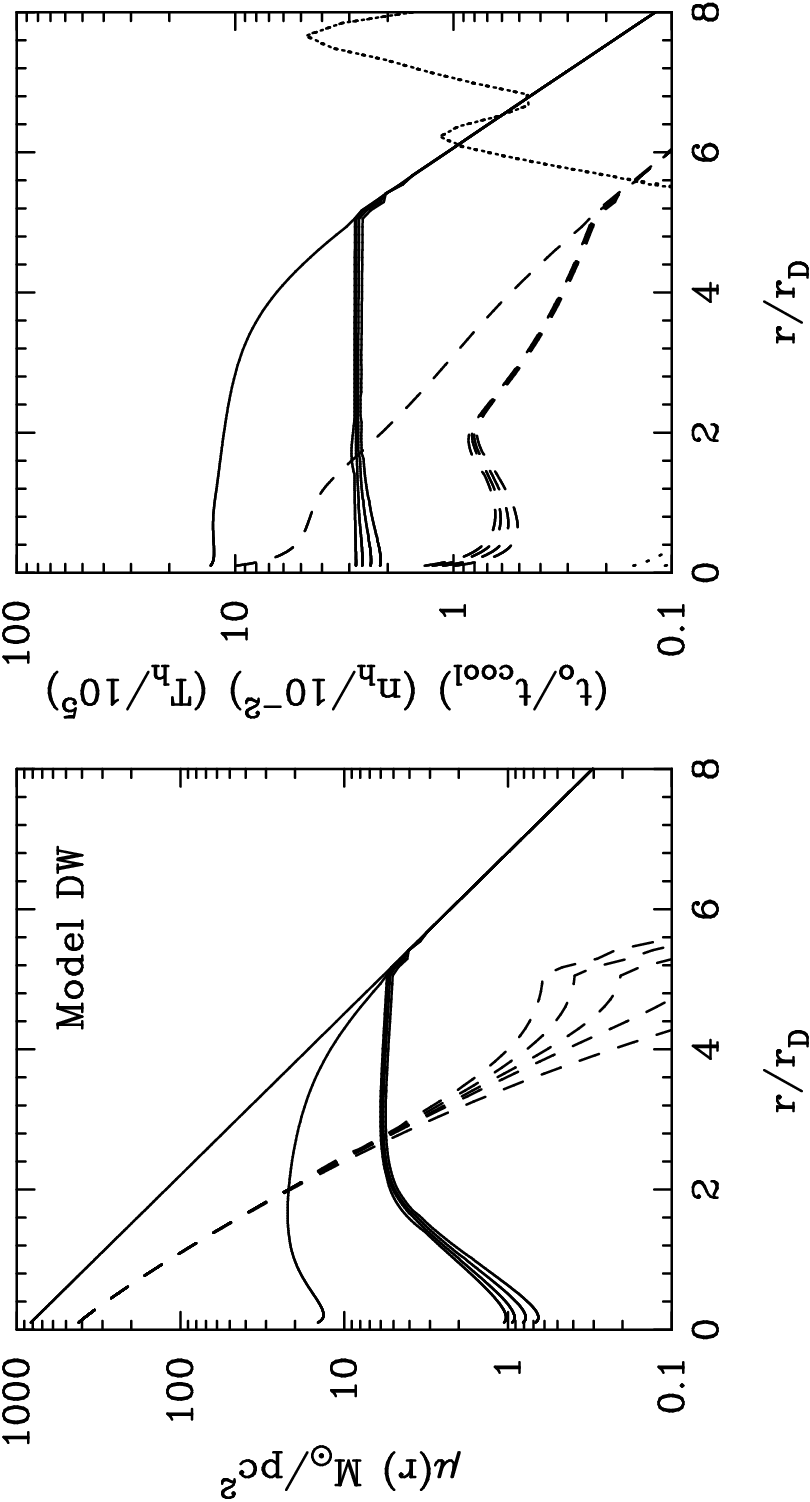}

\caption
{The left hand panels show the evolution of the gas (solid lines) and
stellar (dashed lines) surface mass density distributions for ages of
$0$, $0.1$, $1$, $3$, $6$ and $10$ Gyr as in figure 2. The panels to
the right show various properties of the hot gas component as a
function of the disc radius $r/r_D$. The solid lines show the
temperature, dashed lines show the density and the dotted lines show
the ratio of overlap to cooling timescales, $t_o/t_{\rm cool}$.}
\label{figure4}
\end{figure*}

\begin{figure*}

\vskip 4.8 truein

\includegraphics{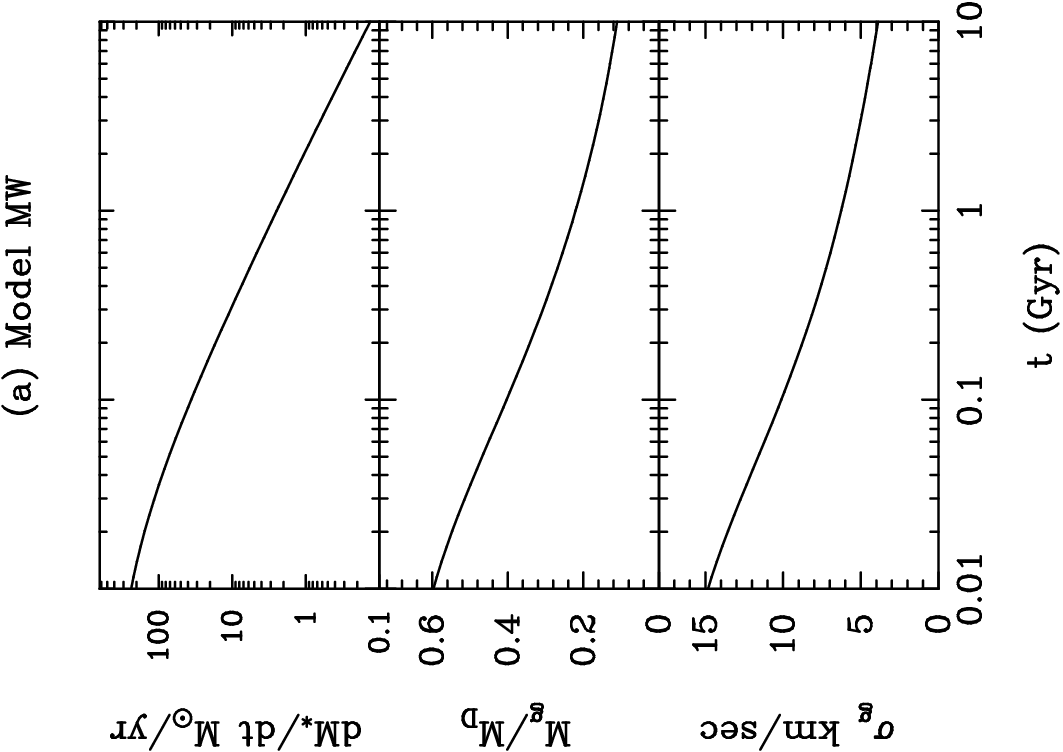}
\includegraphics{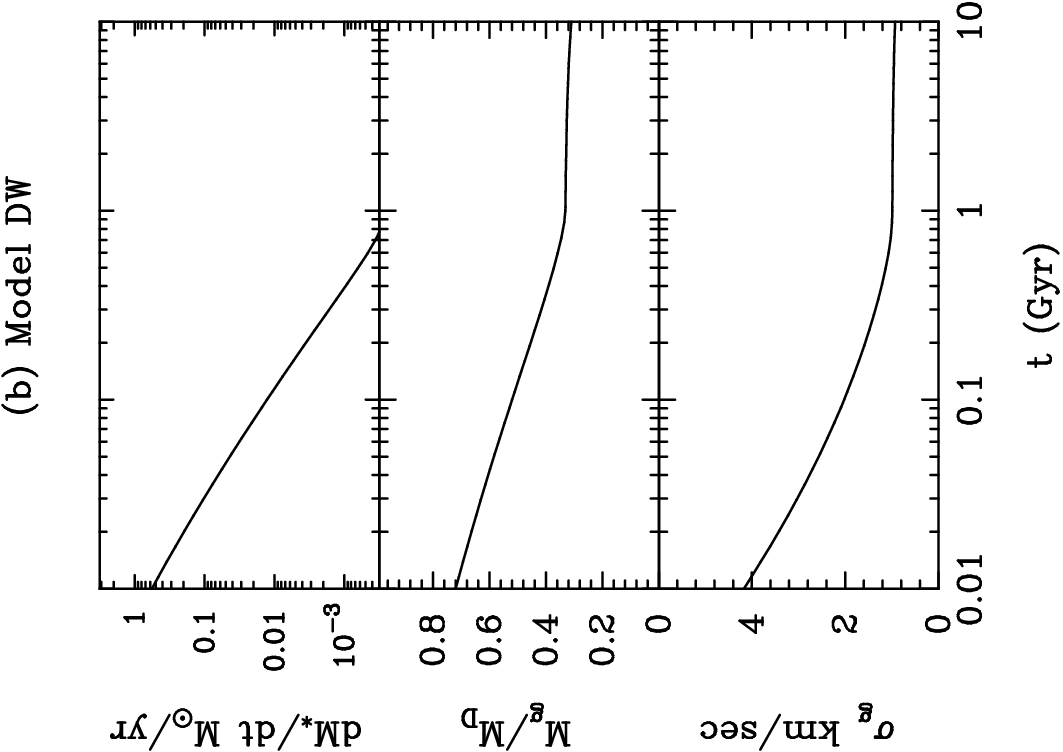}

\caption
{Evolution of the star formation rate, gas fraction and gas cloud
velocity dispersion for the models shown in figure 4.}
\label{figure5}
\end{figure*}	

\subsection{Simple self-regulating model with outflow}

In this section we apply the results of the previous paragraphs to
construct a simplified self-regulating model with outflow. The star
formation rate is governed by the self-regulation algorithm as in
Section 2.6 with the parameter $\epsilon_c = 0.01$. This provides an
estimate of the local supernova rate per unit volume which we insert
in equations (26) to compute the properties of the hot phase, adopting
a value $\phi_\kappa = 0.1$ in equation (\ref{Eva2}) for the
conduction efficiency parameter. The hot gas will be lost from
the system if its specific enthalpy
$$
{1\over 2} v^2 + {5 \over 2} {p \over \rho}   
$$
exceeds to within a factor of order unity
its gravitational binding energy per unit mass. If the gas has
an initial isothermal sound speed of $c_i = (kT/\mu_p) = 37 T_5^{1/2}
{\rm km} {\rm s}^{-1}$ (for a mean mass per particle of $\mu_p = 0.61
m_p$), conservation of specific enthalpy implies that the wind will
reach a bulk speed of $v_w \approx \sqrt{5}c_i$. Some of the thermal
energy will be lost radiatively, and in fact the spherical steady
wind solutions described in Appendix B suggest that a more accurate
criterion for the wind to escape from a galaxy is $v_w \approx
\sqrt{2.5} c_i > v_{esc}$, where $v_{esc}$ 
is the escape speed from the centre of the halo (neglecting
the potential of the disc). If $\sqrt{2.5} c_i < v_{esc}$, the
hot phase is returned instantaneously to the cold phase. This type
of binding energy criterion for outflow has been adopted in previous
studies ({\it e.g.} Larson 1974, Dekel and Silk 1986) and is clearly
oversimplified, as are the assumptions of instantaneous mass loss and
return of cold gas.  These points will be discussed further in Section
6, but for the moment these assumptions will be adopted to illustrate
the qualitative features of the model.  As gas is lost from the system,
the circular speed of the disc component (equation 5) is simply
rescaled by the square root of the mass of the disc that remains.

The evolution of the surface mass densities for the two disc models is
illustrated in Figures 4 and 5. In model MW, the evolution is similar
to that without outflow shown in Figure \ref{figure2}. With the simple
prescription for mass loss used here, no hot gas is lost unless the
temperature of the hot phase exceeds $T_{crit} \approx 5 \times
10^6$K. This does happen at early times when the star formation rate
is high, and about $25$\% of the galaxy mass is lost within $10^7$
yr. Thereafter, no more mass is lost and a nearly exponential disc is
built up with a gas distribution containing a central hole as in
Figure 2. The star formation rate in this model declines strongly with
time, exceeding $100M_\odot$/yr in the early phases of evolution.

The behaviour of model DW is qualitatively different. Here the
critical temperature for mass loss is much lower, $T_{crit} \approx 3
\times 10^5$ K, hence half the mass of the galaxy is expelled by $\sim
10^8$ yr and and $66\%$ by $1$ Gyr.  After $1$ Gyr, the temperature of
the hot phase drops below $T_{crit}$ and the galaxy settles into a
stable state with a low rate of star formation.

The wind prescription in these models, and particularly the assumption
that gas below the critical temperature necessary for escape is
returned instantaneously to the cold phase, is clearly oversimplified
and so the mass loss fractions should not be taken too seriously. A
more detailed model is developed in Section 6.  A more serious deficiency of
the model presented here is that the entire gas disc is assumed to
have formed instantaneously at $t=0$.  This is unrealistic and leads
to high rates of star formation and gas ejection at early times.  A
simple infall model, similar to those adopted in semi-analytic models
(White and Frenk 1991, Cole \etal$\;$ 1994) is included in the next
Section.

\section{Infall Model}

\subsection{Conservation of specific angular momentum}

\begin{figure*}

\vskip 3.6 truein

\includegraphics{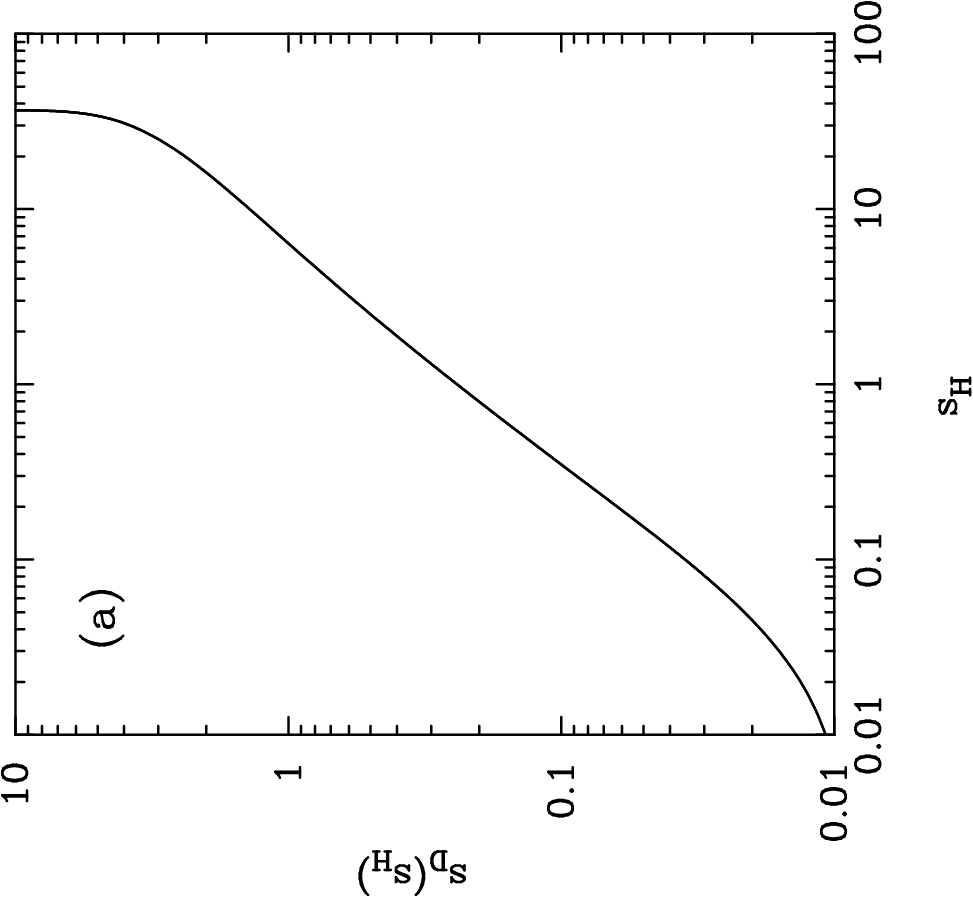}
\includegraphics{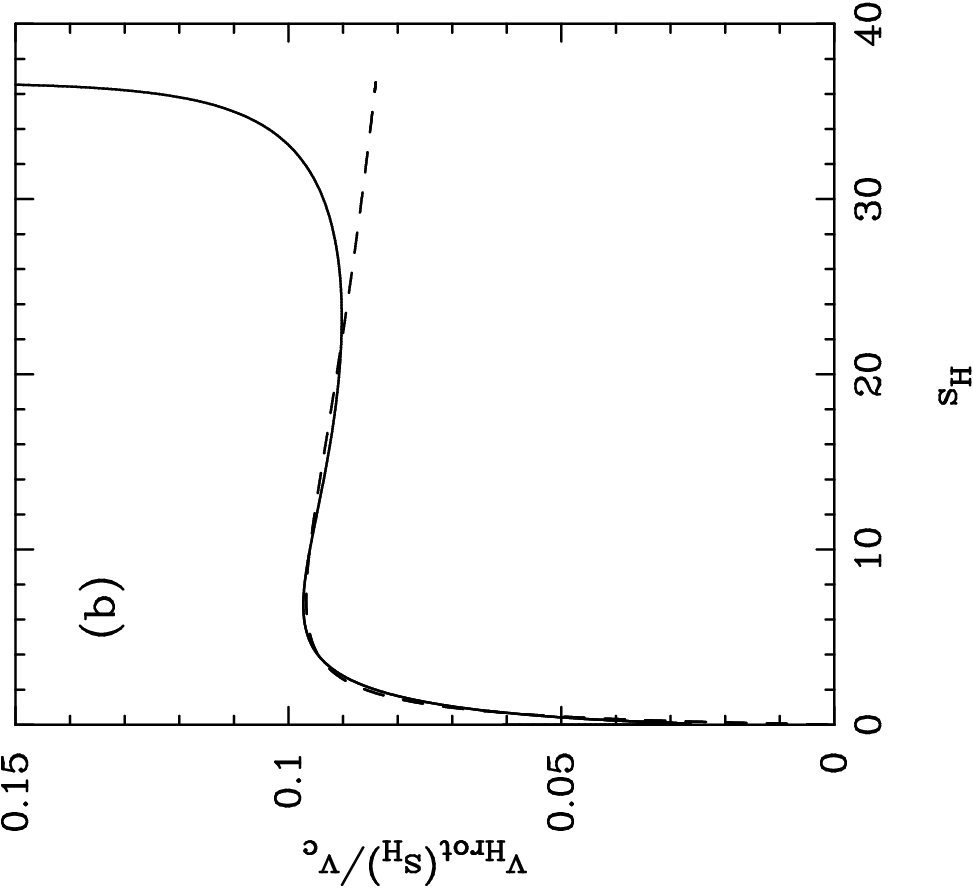}

\caption
{Figure 6a shows the solution of the mass conservation equation (30)
relating the initial halo radius $s_H = \varpi_H/r_D$ to the final
disc radius $s_D = \varpi_D/r_D$. The solid line in 
figure 6b shows the derived rotation
curve of the halo  in units of $v_c = (GM_D/r_D)^{1/2}$ assuming
conservation of specific angular momentum $h_H = h_D$. The dashed line
shows the fitting function of equation (39).}
\label{figure6}
\end{figure*}

Following Fall and Efstathiou (1980), the gas is assumed to follow the
spatial distribution of the halo component with the same distribution
of specific angular momentum prior to collapse. The halo is assumed to
rotate cylindrically with rotation speed $v^{rot}_H(\varpi_H)$, where
$\varpi_H$ is the radial coordinate in the cylindrical coordinate
system.  The gas is assumed to conserve its specific angular momentum
during its collapse, so that the final specific angular momentum of
the disc at radius $\varpi_D$, $h_D = \varpi_D v^{rot}_D$, is equal to
the specific angular momentum of the halo $h_D = \varpi_H v^{rot}_H$
at the radius $\varpi_H$ from which the gas originated. Mass
conservation relates the radii $\varpi_H$ and $\varpi_D$,
\begin{equation}
 {d \varpi_H \over d\varpi_D} = {\mu_D(\varpi_D)\over \mu_H(\varpi_H)}
{M_H \over M_D} {\varpi_D \over \varpi_H}, 
 \label{Sam1}
\end{equation}
where $M_H/M_D$ is the ratio of the halo to disc mass interior to the maximum
infall radius of the disc (see figure 6a below) and $\mu_H$ is the projected
surface mass density of the halo
\begin{equation}
 \mu_H(\varpi) = 2 \int_0^\infty \rho_H \left ( ( \varpi^2 + z^2)^{1/2}
\right ) dz.
 \label{Sam2}
\end{equation}
The solution of equation (\ref{Sam1}) yields $\varpi_D(\varpi_H)$
and the rotation speed of the halo follows from the conservation of
specific angular momentum, $v^{rot}_H = \varpi_D
v^{rot}_D(\varpi_D)/\varpi_H$. The results for the parameters of
models MW and DW are shown in figure
\ref{figure6}, where we have used the notation $s = \varpi/r_D$. When
expressed in the dimensionless units of Figure \ref{figure6}, the solutions
for models MW and DW are identical.

\begin{figure*}

\vskip 3.5 truein

\includegraphics{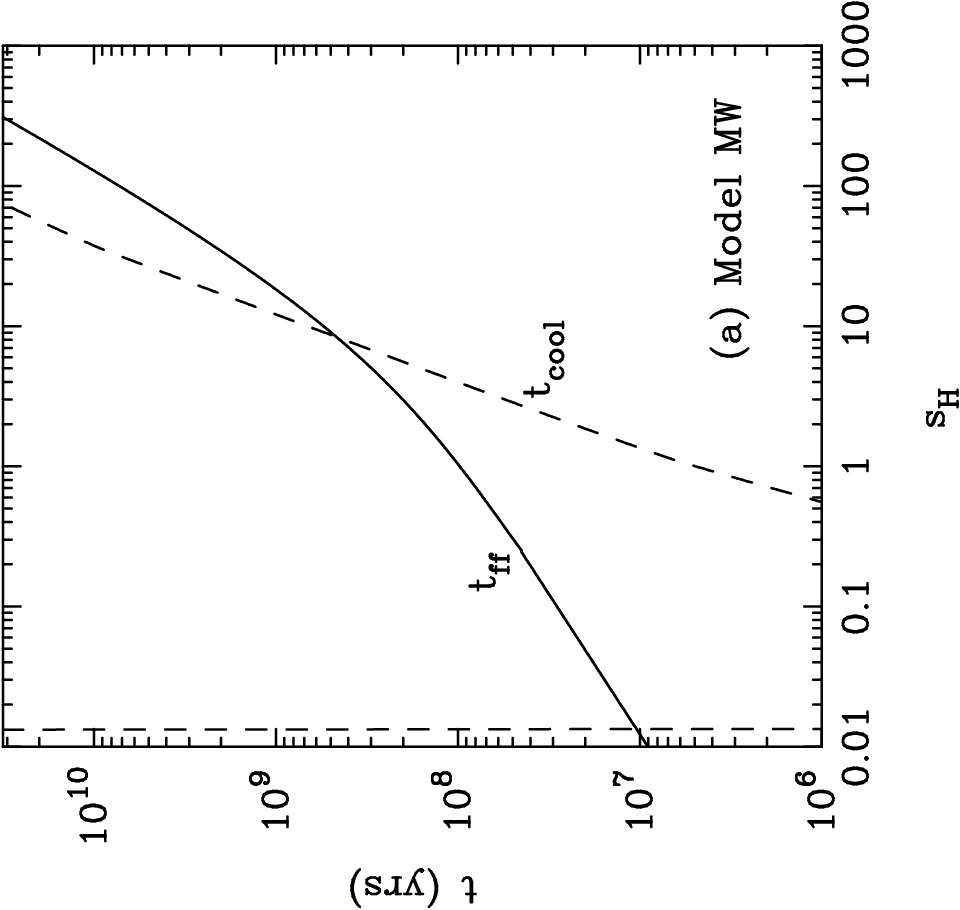}
\includegraphics{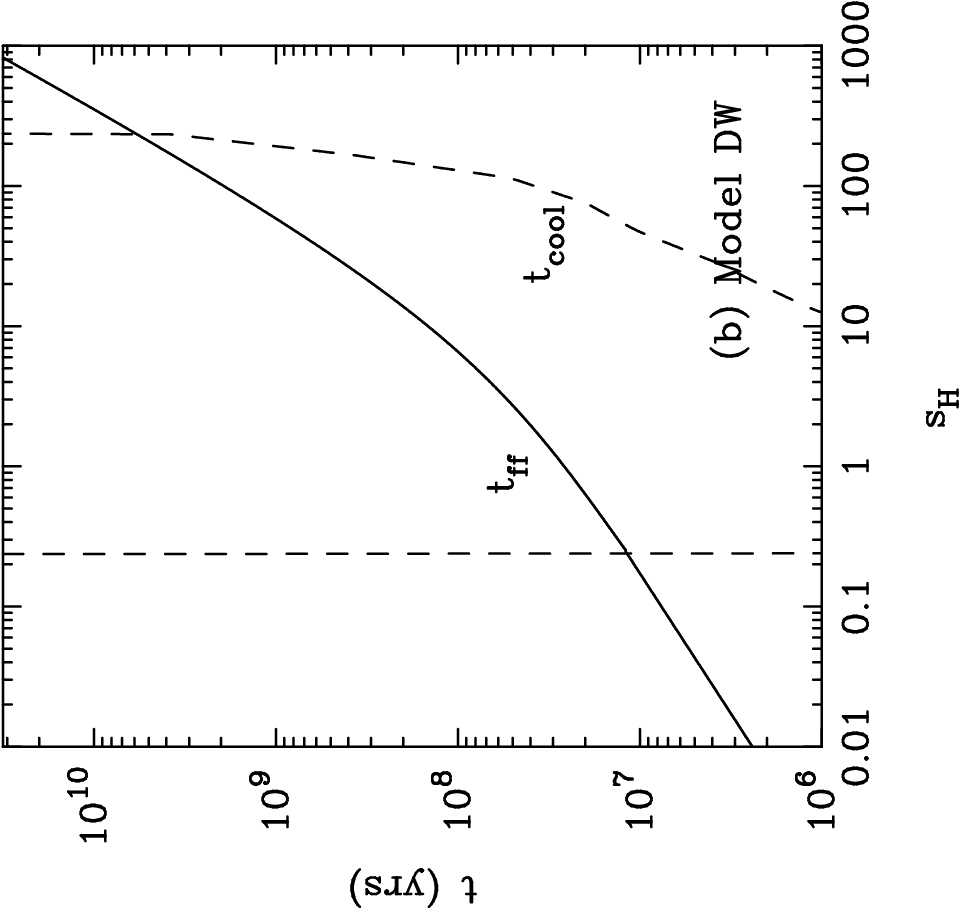}

\caption
{The free-fall  (solid lines) and cooling times (dashed lines) 
for the two model galaxies plotted as a function of halo radius $r_H/r_D$.
Note that for these models, the ratio of baryonic to dark mass within
the virial radius is $0.1$.}
\label{figure7}
\end{figure*}

This prescription is guaranteed to form an exponential disc with the
required parameters. The derived rotation velocity of the halo is
almost independent of radius in general agreement with what is found
in N-body simulations (Frenk \etal$\;$ 1988, Warren \etal$\;$
1992). The upturn in the halo rotation speed at $s_H(max)
\approx 30$ is caused by the rapid decline in the mass of the input
exponential disc at large radii and is of little consequence in the
discussion that follows.  The values of the spin parameter quoted in
Table 1 were derived from the mass and binding energy of the halo and
assuming that the the halo rotation velocity is constant at $0.095
v_c$ at large radii.

\subsection{Mass infall rate}

 To determine the gas infall rate we compute the free fall time
for a gas element at rest at radius $r_i$, 
\begin{equation}
 t_{\rm ff} =  \int_0^{r_i} {dr \over \sqrt 2 [\phi_H(r_i) - \phi_H(r)]^{1/2}},
 \label{Mir1}
\end{equation}
and the cooling time
\begin{equation}
 t_{\rm cool} =  {3 \over 2} {kT_v \times 1.92 \over \Lambda(T_v) n_e(r)},
 \label{Mir2}
\end{equation}
where $n_e(r)$ is the electron density. The temperature $T_v$ in equation 
(\ref{Mir2}) is set to the virial temperature derived from the equation
of hydrostatic equilibrium assuming that the temperature is slowly
varying with radius
\begin{equation}
 T_v \approx - v^2_H(r) {\mu_p \over k} {d {\rm ln} r \over d {\rm ln} \rho_b(r)},
 \label{Mir3}
\end{equation}
where we assume that the baryons follow the same spatial distribution
as the halo. The infall rate is given by
\begin{equation}
  \dot M_{inf} = 4 \pi \rho_b(r_H) r_H^2 \Bigg \{ \begin{array}{ll}
 dr_H(t_{\rm ff} = t)/dt & \quad t_{\rm ff} > t_{\rm cool} \\
 dr_H(t_{\rm cool} = t)/ dt & \quad t_{\rm cool} > t_{\rm ff} \end{array}  . 
 \label{Mir4}
\end{equation}
Finally, conservation of specific angular momentum specifies the final
radius in the disc for each gas element. Since the halo is assumed to
rotate on cylinders, the gas near to the poles in an infalling shell
has a lower specific angular momentum that the gas at the equator. The
infalling material is therefore distributed through the disc according
to 
\begin{equation} 2 \pi \varpi_D \dot \mu_D(\varpi_D) d\varpi_D =
\dot M_{inf} {\varpi_H d\varpi_H  \over r_H (r_H^2 - \varpi_H^2)^{1/2}},
 \label{Mir5}
\end{equation}
where $\varpi_D$ and $\varpi_H$ are related by the solution of equation
(\ref{Sam1}).

Equations (\ref{Mir1}) -- (\ref{Mir5}) specify the infall model. The
free-fall and cooling times of the two model galaxies are shown in
Figure \ref{figure7}.  In the larger galaxy, gas within $r_H/r_D
\approx 10$ infalls on the free-fall timescale and ends up within one
scale length of the final disc. The material in the outer parts of the
disc infalls on the cooling timescale. In contrast, apart from a small
amount of gas in the very central part of the halo
with virial temperature $< 10^4$ K, the gas in the dwarf
galaxy infalls on a free-fall timescale because the cooling time is so
short.

\subsection{Simple self-regulating model with inflow and outflow}

  The models described in this section are exactly the same as those
described in section 3.3, except that we grow the discs gradually
using the infall model of sections 4.1 and 4.2. In the models
described below, inflow and outflow are assumed to occur
simultaneously. This is often assumed in semi-analytic models
of galaxy formation ({\it e.g.} Cole \etal$\;$ 1994, Somerville and
Primack 1999) and may not be completely unrealistic if the
infalling gas is clumpy.  The dark matter haloes will contain
significant sub-structure ({\it e.g.} Moore \etal$\;$ 1999) which may
contain pockets of cooled gas. Furthermore, if the cooling time is
short  compared to
the dynamical time, the infalling gas will be thermally unstable (Fall
and Rees 1985) and will fragment into clouds. These will fall to the
centre on a free-fall timescale if they are sufficiently dense and
massive that gravity dominates over the ram pressure of the wind. This
requires clouds with masses
\begin{eqnarray}
m_{\rm cloud} \simgt 9.5 \times 10^5 M_\odot \Big ( { a_{\rm cloud} \over 1{\rm
kpc}} \Big ) \Big ( {r \over 10{\rm kpc}} \Big )^{-1} \times \nonumber \\
\qquad
\Big
({\dot M_w \over 1 M_\odot/{\rm
yr}} \Big ) \Big ( {v_w \over 100 {\rm km}/{\rm s}} \Big ) \Big (
{v_v \over 100 {\rm km}/{\rm s} } \Big )^{-2},  \label{SR1}
\end{eqnarray}
where $a_{\rm cloud}$ is the radius of the cloud. However, even if
(\ref{SR1}) is satisfied, the clouds may be sheared and disrupted into
smaller clouds by Kelvin-Helmholtz instabilities on a timescale of a
few sound crossing times as they flow through the wind ({\it e.g.}
Murray \etal $\;$ 1993). The wind energy will be partially
thermalized in shocks with the infalling clouds and dissipated in
evaporating small clouds. But for the typical mass outflow rates
expected from dwarf galaxies ($\dot M_w
\simlt 0.2 M_\odot/{\rm yr}$), the rate at which energy is supplied
by the wind $\dot E_w = {1/2}\dot M_w v^2_w$ is much smaller than the
energy lost in radiative cooling,
\begin{eqnarray}
{\dot E_{\rm cool} \over \dot E_w} \approx  50 \Lambda_{-23} 
\left ( {v_v \over 100 {\rm km}/{\rm s}} \right )^{4}
\left ( {r_{\rm cool} \over 10 {\rm kpc}} \right)^{-1} \times \nonumber \\
\qquad \left ( {v_w \over 100 {\rm km}/{\rm s}} \right )^{-2} 
\left ({\dot M_w \over 1 M_\odot/{\rm
yr}} \right )^{-1}, \label{SR2}
\end{eqnarray}
where $r_{\rm cool}$ is the radius at which the cooling time is equal to
the age of the system. 

The qualitative picture that we propose is as follows. In 
galaxies with a short cooling time, clouds formed by thermal
instabilities will infall ballistically if (\ref{SR1}) is
satisfied. If (\ref{SR1}) is not satisfied, the ram pressure of the
wind will drive out the infalling gas and infall will be
suppressed. With infall suppressed, the star formation rate in the
disc and the wind energy will decline until infall can begin again.
The wind will be partially thermalised before reaching $r_{\rm cool}$ and
completely thermalised at $\sim r_{\rm cool}$, but the energy supplied by
the wind will be small compared to the energy radiated by the gas at
$r \simgt r_{\rm cool}$ and so cannot prevent radiative cooling. If
(\ref{SR1}) is satisfied, some of the outflowing gas may fall back
down to the disc after shocking against infalling clouds. However, in
the models described here the efficiency of converting infalling gas
into stars is low in dwarf systems, so provided the gas does not cycle
around the halo many times, neglecting return of some of the
outflowing gas should not affect the qualitative features of the
models. The global geometry of the system, {\it e.g.} if the
wind is weakly collimated perpendicular to the disc, may also 
permit simultaneous inflow and outflow of gas.

\begin{figure*}
\vskip 6.7 truein

\includegraphics{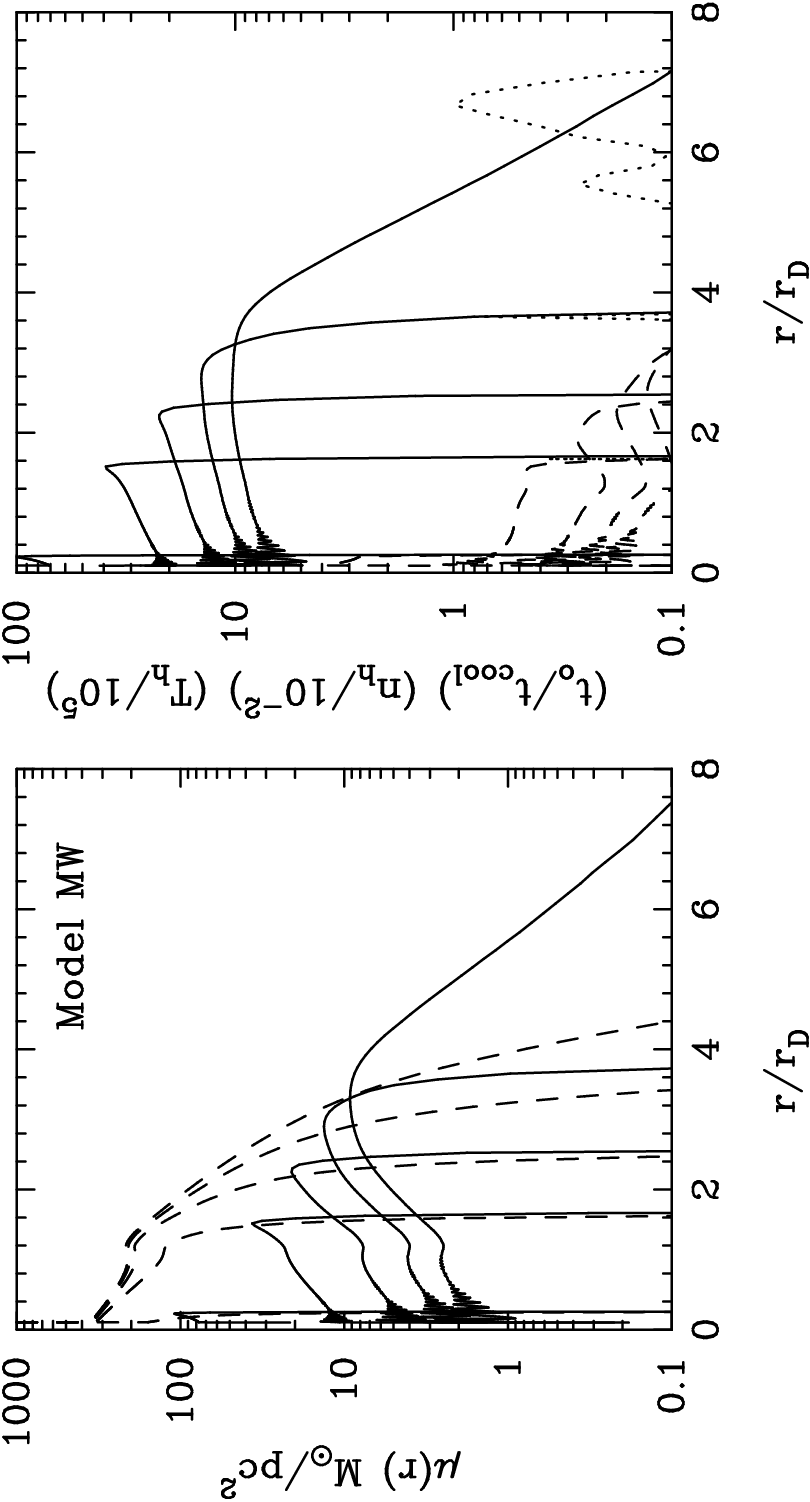}
\includegraphics{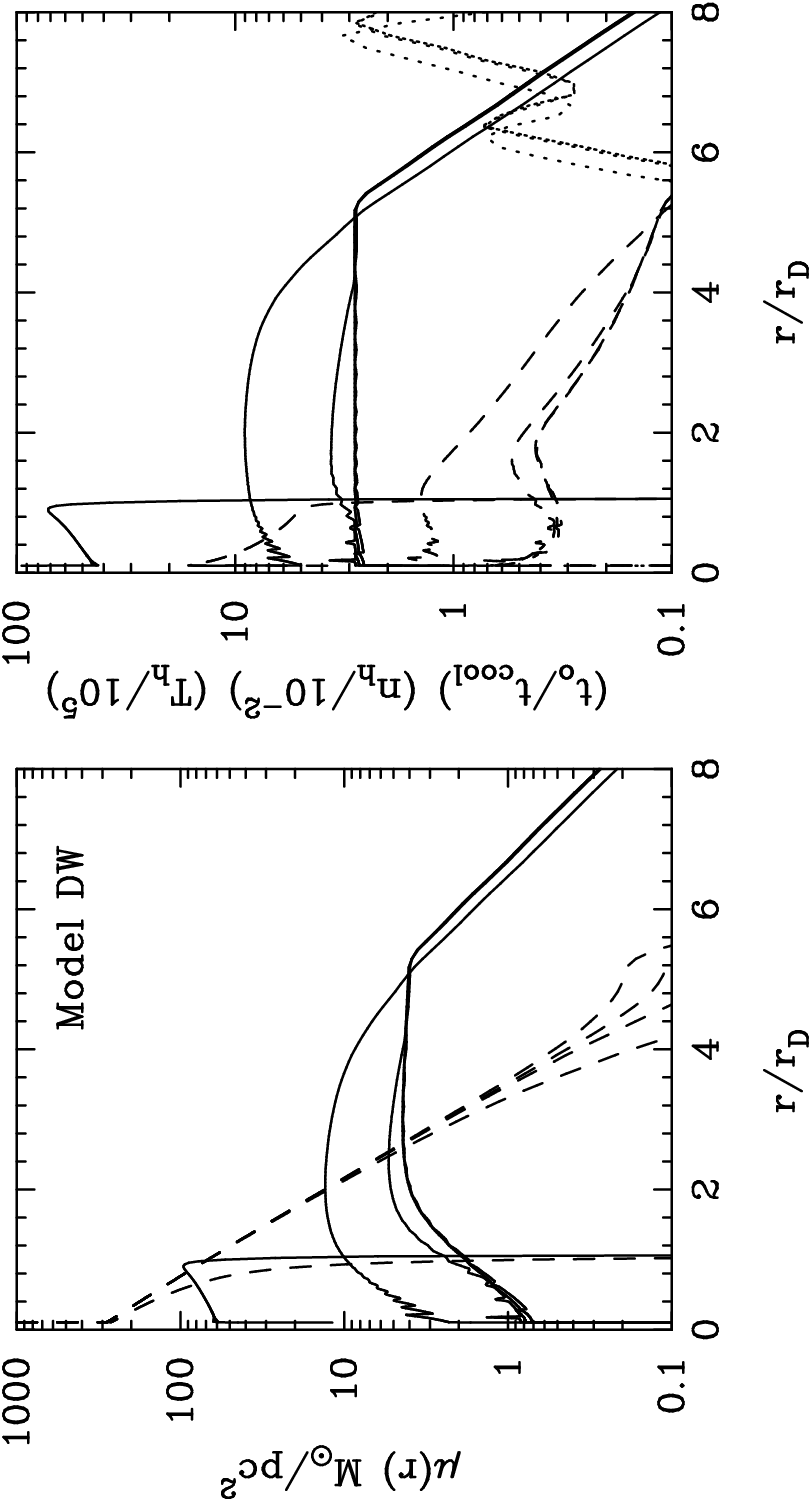}

\caption
{The left hand panels show the evolution of the gas (solid lines) and
stellar (dashed lines) surface mass density distributions for ages of
$0$, $0.1$, $1$, $3$, $6$ and $10$ Gyr as in figure 2. The panels to
the right show the radial distribution of density, temperature and
ratio of overlap to cooling timescales for the hot gas component.}
\label{figure8}
\end{figure*}

\begin{figure*}

\vskip 4.5 truein

\includegraphics{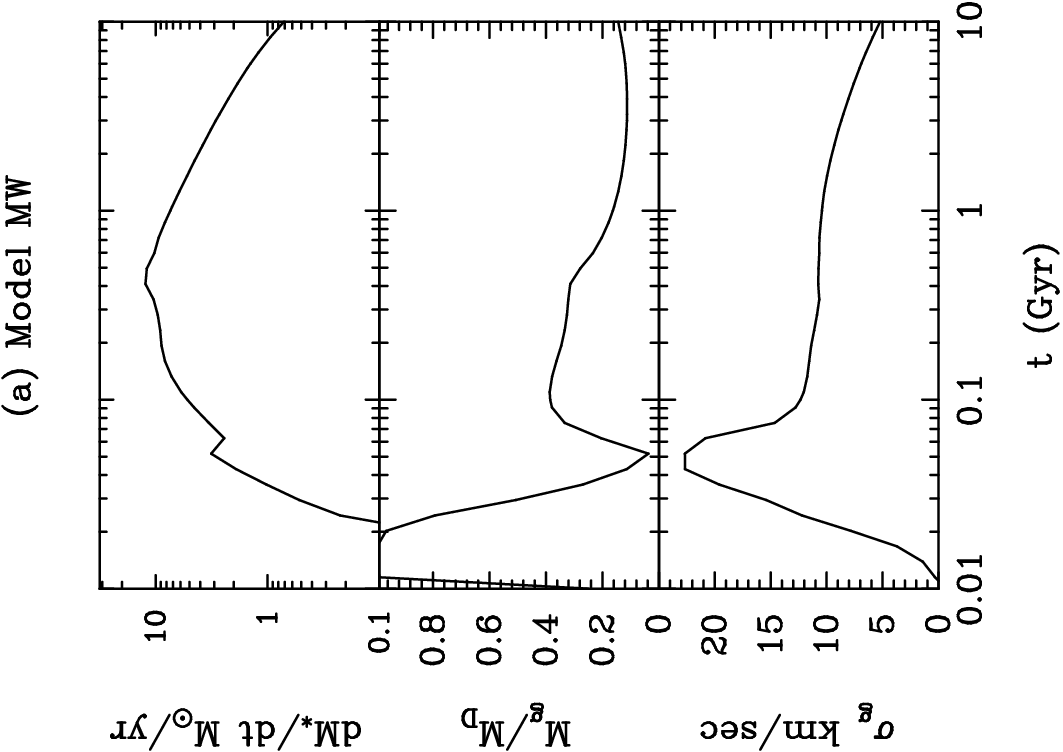}
\includegraphics{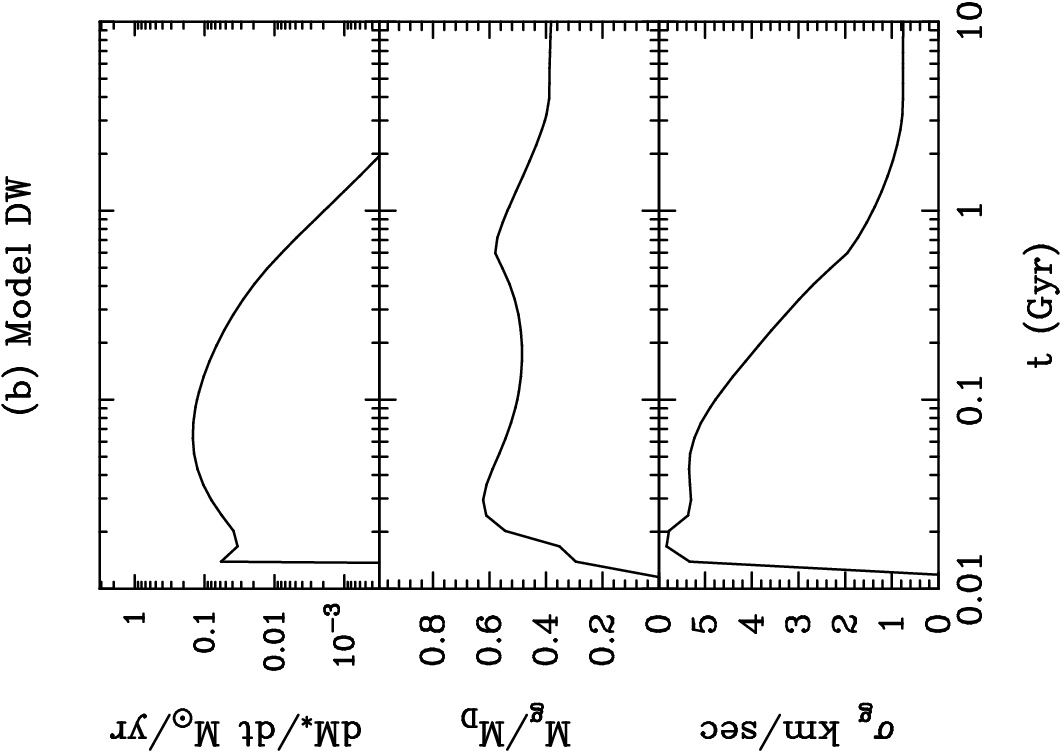}

\caption
{Evolution of the star formation rate, gas fraction and gas cloud
velocity dispersion for the models shown in figure 8.}
\label{figure9}
\end{figure*}	 
 
The interaction of an outflowing wind with an inhomogeneous infalling
gas clearly poses a complex physical problem. In reality, the process
may be far from steady, with outflow occurring in bursts accompanied
by infall from discrete sub-clumps containing cooled gas.  In the
models described below and in the rest of this paper, we will assume
that the infall and outflow occur simultaneously, steadily and without
any interaction between the inflowing and outflowing gas.  As the
discussion of the preceeding two paragraphs indicates, this is
obviously an over-simplification. It should be viewed as an
idealization, on a similar footing to some of the other
assumptions adopted in this paper ({\it e.g.}  spherical
symmetry, neglect of halo substructure and merging, steady star
formation rates {\it etc}) designed to give some insight into how a
quiescent mode of feedback might operate.

The analogues of figures \ref{figure4} and \ref{figure5} for the
models incorporating infall and outflow are shown in figures
\ref{figure8} and \ref{figure9}. The discs build up 
from the inside out, as in the models of disc formation described by
Fall and Efstathiou (1980) and Gunn (1982).  Most of the star
formation occurs in a propagating ring containing the most recently
accreted gas. The most significant differences from the models of
section 3.3, are the net rates of star formation (figure
\ref{figure9}) and the timescale of outflow.  The initial high rates
of star formation in the models of section 3.3 are suppressed in the
models with infall, and the timescale for outflow is now much longer
because it is closely linked to the gas infall timescale.  Apart from
these differences, the final states, gas fractions and mass-loss
fractions are similar to those in the models without infall. In model
MW some outflow occurs when $t \simlt 10^8$yrs and the temperature of
the hot gas is high enough that it can escape from the
system. Thereafter, the hot component cannot escape and the disc
builds up without further outflow. About 17\% of the total galaxy
mass is expelled in the early phases of evolution, but as we have
described above, this could be an overestimate since
some of  this gas may be returned to the galaxy if the
wind energy is thermalized before it reaches the virial radius.  In
contrast, model DW drives a wind until $t \sim 1$ Gyr and expels about
74\% of its mass. About half of the gas is lost within $3 \times
10^8$ yrs, {\it i.e.} on about the infall timescale for most of the gas
in the halo (see figure 7b).

\section{Refinements of the Model}

The models described in the previous sections contain a number of
simplifications, which we will attempt to refine in this section.  We
do not address the problem of the interaction of a wind with the
infalling gas, which is well beyond the scope of this paper.  Instead,
we introduce some simple improvements to the infall model (\S 5.1),
model for mass loss from the galactic disc (\S 5.2, \S 5.3) and the
pressure response of the cold ISM to the hot phase (\S 5.4, \S5.5).

\subsection{Infall Model}

In Section 4, we used a simplified model of infall that guarantees the
formation of an exponential disc if angular momentum is conserved
during collapse.  In this section, we assume a specific functional
form for the rotation velocity of the halo,
\begin{equation}
 v^{rot}_H(s) = c_1v_c {(s/c_2)  \over (1 + (s/c_2))(1 + (s/c_3)^{c_4})},
 \quad s \equiv r/r_D
 \label{Inf1}
\end{equation}
with $c_1 = 0.115$, $c_2= 0.6$, $c_3 = 16$, $c_4 = 0.25$. The functional form and
coefficients in equation (\ref{Inf1}) have been chosen to provide a good
fit to the halo rotation profile derived in section 4.1 from conservation
of specific angular momentum and is plotted as the dashed line in 
figure (6b). As in the previous section,
the gas is assumed to follow the same radial density
distribution and rotation velocity as the halo component, but its
final radius in the disc is computed by assuming conservation of
specific angular momentum and self-consistently solving for the
rotation speed of the disc component. The halo component
is assumed to be rigid and the contribution of the disc component to
$v^{rot}_D$  is computed using the
Fourier-Bessel theorem (see Binney and Tremaine 1987 \S 2.6)
\begin{eqnarray}
v^2(r) = - r \int_0^\infty S(k) J_1(kr)k\;dk, \quad \nonumber \\ 
S(k) = 
- 2\pi G \int_0^\infty J_0(kr) \mu_D(r) r \;dr. \label{Inf2}
\end{eqnarray}
Equation (\ref{Inf2}) is time consuming to evaluate accurately and in
our application $v^2(r)$ must be computed many times.  A fast
algorithm has therefore been developed as described in  Appendix A. The
epicyclic frequency $\kappa$ is required in equation (\ref{Stab3}) to
compute the instantaneous star formation rate  and is derived by
numerically differentiating the rotation speed.

With this formulation of the infall model, the infall rate is governed
by the dark matter profile and the ratio of dark to baryonic mass
within the virial radius, $M_v/M_D = (v_v/v_c)^2 f_{coll}$. For the
models described here, we adopt $M_v/M_D = 10$, consistent with the
parameters listed in Table 1. The final disc surface mass density will
be close to an exponential by construction, since the halo rotation
velocity (\ref{Inf1}) has been chosen to match the rotation profile
derived by assuming an exponential disc and conservation of specific
angular momentum.

%

\subsection{Galactic Fountain}

In previous sections, we have assumed that gas is lost from the disc
if the bulk velocity of the wind $v_w \approx \sqrt{2.5}c_i$, exceeds
the escape speed $v_{esc}$ from the halo, but is otherwise returned
instantaneously to the ISM. More realistically, gas with $v_w <
v_{esc}$ will circulate in the halo along a roughly ballistic
trajectory and will cool forming a galactic fountain (Shapiro and Field 1976,
Bregman 1980). In the models described in this section, hot gas with
$v_w < v_{esc}$ is returned to the disc at the radius from which it
was expelled after a time $t_{ret}$,
\begin{equation}
 t_{ret}  =  2 t_{ff}(r_{max}), \qquad v^2_w = 2[ \phi_H(r_{max}) -
\phi_H(0)],
 \label{Esc2}
\end{equation}
{\it i.e.} we ignore the gravity of the disc and the
angular momentum of the gas in computing the ballistic trajectory
of a gas element.

\subsection{Escape Velocity of the Wind}

The detailed dynamics of the hot corona itself is complicated and
beyond the scope of this paper. Type II supernovae at the upper and
lower edges of the gas disc will be able to inject their energy
directly into the hot gas, as will Type Ia supernovae forming in the
thicker stellar disc. In addition, the hot component will interact
with the primordial infalling gas in a complicated way as sketched in
\S 4.3.

In the absence of radiative cooling, the hot gas will extend
high above the galactic disc in an extended corona. For an
isothermal corona, the equation of hydrostatic equilibrium in the 
z-direction has the following approximate solution,
\begin{eqnarray}
 {\rho_H(\varpi, z) \over \rho_H(0)}  \approx {\rm sech}^{2p_g} \left (
{ z \over H_g} \right )\; {\rm sech}^{2p_s} \
\left (
{ z \over H_s} \right ) \times \nonumber  \qquad \\
 {\rm exp} \left ( -{1 \over c_i^2}
\int_0^z   {z v^2_H(r) \over r^2}\; dz \right ), \quad r^2 = \varpi^2 +
z^2,  \label{Win1}
\end{eqnarray}
\begin{eqnarray}
 p_g = {\mu_g \sigma^2_g \sigma_s \over c_i^2 (\mu_g \sigma_s + 
\mu_s \sigma_g)}, \;\; 
 p_s = {\mu_s \sigma^2_s \sigma_g \over c_i^2 (\mu_g \sigma_s + 
\mu_s \sigma_g)}, \;\;  \qquad \qquad   \nonumber
\end{eqnarray}
where we have assumed that both the stars and the gas follow ${\rm
sech^2}$ vertical distributions (equation \ref{Vert1}) and $c_i$ is
the isothermal sound speed of the hot gas. We define a characteristic
scale height for the hot component, $H_{hot}(\varpi)$, at which the
density drops by a factor $\sim e$ according to equation
(\ref{Win1}). If radiative cooling were negligible, we would expect a
sonic point in the flow at about $z \sim H_{hot}$.
It is interesting to compare some characteristic
numbers for the coronal gas:
\beglet
\begin{eqnarray}
\dot E_{in} \approx 1.1 \times 10^{40} T_{h6} \dot M_{ev}\; {\rm erg}\;{\rm
s}^{-1}  \qquad \qquad\qquad \qquad \quad  \\
\dot E_{\rm SNII} \approx 5.7 \times 10^{40} \epsilon_{\rm SNII} E_{51}\dot M_* \;{\rm
erg}\;{\rm s}^{-1} \qquad \qquad \qquad \;\; \\
\dot E_{\rm cool} \approx 2 \times 10^{39} n_{h-2}^{2} \Lambda_{-23} \left (
{H_{hot} \over 1\;{\rm kpc} } \right ) \left ( {R_{hot} \over 3\; {\rm
kpc}} \right )^2  {\rm erg}\;{\rm s}^{-1} 
 \label{Win2}
\end{eqnarray}
\endlet
where the rates $\dot M_{ev}$, $\dot M_*$ are in $M_\odot\;/{\rm yr}$.
Here $\dot E_{in}$ is the thermal energy injected into the hot corona
by evaporating cold gas at a rate $\dot M_{ev}$, $\dot E_{\rm SNII}$ is
the energy supplied to the corona by Type II supernovae forming above
and below one scale height of the cold gas layer and the parameter
$\epsilon_{\rm SNII}$ expresses the efficiency with which this energy
is coupled to the hot coronal gas.  $\dot E_{\rm cool}$ is the rate of
energy lost by a uniform density isothermal corona of scale height
$H_{hot}$ within a cylinder of radius $R_{hot}$.  For a large galaxy
such as the Milky Way that can sustain an evaporation rate of $\sim 10
M_\odot/{\rm yr}$, $\dot E_{\rm cool}$ is small compared to $\dot
E_{in}$ and it is a good approximation to neglect radiative cooling in
the early stages of the flow (see Appendix B). However, in a dwarf
galaxy $\dot E_{\rm cool}$ is typically larger than $\dot E_{in}$. In this
case, we expect that the hot component will develop a sonic point at a
characteristic cooling scale height $H_{\rm cool}$
\begin{equation}
  H_{\rm cool} \sim v_w t_{\rm cool} \sim 11 \left ( {v_w \over 100\; {\rm
 km}/{\rm s}} \right ) T_{h6} n_{h-2}^{-1} \Lambda_{-23}^{-1} \;\;
 {\rm kpc}.  \label{Win4}
\end{equation}
(see {\it e.g.} Kahn 1981 and Appendix B) and that most of the thermal
energy will be converted into kinetic energy by the time the gas flows
to $H_{\rm cool}$. We therefore ignore radiative cooling and estimate the 
bulk velocity of the wind at each radial shell in the disc from
\beglet
\begin{equation}
  {1 \over 2} \dot M_{out}^\Omega v^2_w = \dot E_{in}^\Omega + \dot
 E_{\rm SNII}^\Omega, \qquad\qquad 
 \label{Win3a}
\end{equation}
and to close the equations we assume that
\begin{equation}
\dot M_{out}^\Omega = \dot M_{in}^\Omega  = \dot M_{ev}^\Omega.
 \label{Win3b}
\end{equation}
\endlet
(Note that the numerical coefficient in equation (43a) has been
adjusted to give $v_w = \sqrt{2.5}c_i$ if $\dot E_{\rm SNII} =0$ so
that the criterion for the wind to escape, $v_w > v_{esc}$, is the same
as in  the preceeding Section).

Energy input from Type II supernovae exploding high above the cold gas
layer will make a small contribution to the thermal energy of the hot
coronal gas.  For values of $\epsilon_{SNII} \sim 1$, $\dot
E_{SNII}^\Omega$ will be about $20\%$ or so of $\dot E_{in}$ and
cannot be higher because $\dot M_{ev}^\Omega$ and the star formation
rate are nearly proportional to each other (equation
\ref{Temp3}). Type Ia supernovae will also supply energy to the
corona, with a time lag of perhaps $\simgt 1$ Gyr (Madau, Della Valle
and Panagia 1998). However, this effect will also make a small
perturbation to the energy budget of the corona and so it is ignored
here. Furthermore, in the models described here much of the gas is
expelled on a timescale of $\simlt 1$ Gyr, thus feedback is likely to
be more or less complete before energy injection from Type Ia
supernovae becomes significant.

\subsection{Pressure equilibrium and cold cloud radii}

In the models of Section 3 and 4, the cold cloud radii were kept constant
irrespective of the pressure of the confining hot phase. More
realistically, the cold cloud radii will adjust to maintain
approximate pressure equilibrium with the hot phase. Thus
\begin{equation}
 {a \over a_\odot}  \approx 0.53  \left ( {T_c \over 80K} \right )^{1/3}
(n_{h-2} T_{h6})^{-1/3},
 \label{Pe5}
\end{equation}
where $a_\odot$ denotes the cloud radii at the solar neighbourhood with
values as given in equation (\ref{Diss1}) and $T_c$ is the
internal temperature of the cold clouds.  In our own Galaxy,
photoelectric heating of dust grains is believed to be the main
heating mechanism of the cold clouds (see {\it e.g.} Wolfire \etal$\;$
1995) but other heating mechanisms may be important, for example,
cosmic-ray heating (Field, Goldsmith and Habing 1969). We therefore
expect that $T_c$ varies in a complex (and uncomputable) way as a
galaxy evolves. To assess the effects of the pressure
response of the cold clouds,  $T_c$ will be assumed to remain constant at
$80$K. The cloud radii are then determined solely by the pressure of
the hot component via equation (\ref{Pe5}).

The energy lost through cloud collisions (equation \ref{Diss2}) varies
as $a^2$ (for fixed cloud masses).  However, the cloud heating
efficiency factor $\epsilon_c$ will also change as the cloud radii
change in response to the pressure of the hot phase. In the model of
MO77, the energy acquired from momentum exchange with cooling
supernovae shells varies as $a^4$. The net effect of these variations
in the self-regulating star formation model is to introduce positive
feedback, since a higher rate of star formation is required to balance
energy lost through cloud collisions in regions where the ambient
pressure is higher. This is modelled by assuming $\dot E_{coll}
\propto (a/a_\odot)^2$ and $\epsilon_c = \epsilon_{c\odot}
(a/a_\odot)^4$, where $\epsilon_{c\odot}$ is a fiducial efficiency
factor.

\subsection{Induced star formation}

The maximum mass for an isothermal cloud in pressure equilibrium with
the confining medium of pressure $p_h$ is given by the Bonner-Ebert
criterion (Bonner 1956, Ebert 1955, Spitzer 1968),
\begin{eqnarray}
 m_{\rm BE}  =  1.18 \left ( {k  T_c \over \mu_p} \right )^2 G^{-3/2}
 p_h^{-1/2} \qquad \nonumber \\
 = 433 \left (T_c \over 80K \right )^2 (n_{h-2} T_{h6})^{-1/2} M_\odot.
 \label{Be1}
\end{eqnarray}
For our own Galaxy (MO77), $n_h \sim 1.5 \times 10^{-3} {\rm
cm}^{-3}$, $T_h \sim 4 \times 10^5$ and $p_h \sim 3 \times 10^{-12}
{\rm dyne}\;{\rm cm}^{-2}$, hence $m_{\rm BE} \approx 1700
M_\odot$. This is reasonably close to the upper mass limit, $m_u =
4300 M_\odot$, for the cold cloud mass spectrum adopted in this paper
($a_u = 10 {\rm pc}$ with $\rho_c = 7 \times 10^{-23}\; {\rm
g/cm^3}$).  Gravitational stability requires $m_u \approx m_{\rm BE}$
and we will henceforth impose this condition in determining the upper
mass limit of the cold cloud spectrum.  An increase in the pressure of
the hot phase will lead to a decrease in $m_u$ and hence to some
pressure induced star formation. If the over-pressured clouds fragment
into stars with an efficiency $\epsilon_{\rm BE}$, the induced star
formation rate is given by
\begin{equation}
{dM_s^\Omega \over dt} = \epsilon_{\rm BE} {M_g^\Omega \over 2 p_h 
{\rm ln } (m_u/m_L)} \Bigg \{ \begin{array}{ll}
 dp_h/dt & \quad dp_h/dt > 0 \\
  0  & \quad  dp_h/dt \le 0  \end{array} ,
 \label{Be2}
\end{equation}
where $m_L$ is the lower limit to the cloud mass spectrum, $m_L
\approx 0.5 M_\odot$. This provides an additional source of positive
feedback, since as the pressure of the hot component rises the star
formation rate of the self-regulating model is enhanced by pressure
induced star formation.

\subsection{Chemical evolution}

It is straightforward to include chemical evolution in the models using the
instantaneous recycling approximation. We distinguish between `primordial'
infalling gas accreting at a rate $d\mu_I/dt$ with metallicity $Z_I$,  and
processed gas from the galactic fountain of metallicity $Z_F$ accreted at
a rate $d\mu_F/dt$. The equation of chemical evolution is then
\begin{equation}
 \mu_g dZ  =  p d\mu_s + (Z_I - Z) d\mu_I + (Z_F - Z) d\mu_F,
 \label{CE1}
\end{equation}
(see {\it e.g.} Pagel 1997), where $p$ is the yield. 
We adopt a yield of $p = 0.02$ and assume that the primordial gas has
zero metallicity ($Z_I =0$). Gas ejected in a galactic fountain is
assumed to have the same metallicity as the ISM at the time that 
the gas was ejected.  Within the disc, the ISM gas is assumed to be perfectly
mixed at all times. We normalize the metallicities to the solar value,
for which we adopt $Z_\odot = 0.02$.

\section{Results and Discussion}

\subsection{Variation of input parameters}

In addition to the many simplifying assumptions introduced in previous
sections, the model described here has $4$ key parameters: (i)
$\phi_\kappa$, determining the efficiency of heat conduction (equation
\ref{Eva1}); (ii) $\epsilon_{c\odot}$, controlling the star formation
rate (equation \ref{Diss4}); (iii) $\epsilon_{\rm SNII}$, determining the
efficiency with which energy from Type II supernovae couples directly
to the gas (equation \ref{Win3a}) ; (iv) $\epsilon_{\rm BE}$, setting the
efficiency with which over-pressured ISM clouds collapse to make stars
(equation \ref{Be2}). In addition, the ISM cloud radii can be allowed
to vary in response to the pressure of the ISM as described in Section
5.4.

\begin{table*}
\bigskip
\centerline{\bf Table 2: Feedback Efficiency: Variation of Input Parameters}
\begin{center}
\begin{tabular}{|c|cc|cc|cc|} \hline \hline
Model & MW1 & DW1 & MW2 & DW2 & MW3 & DW3   \\
$\phi_\kappa$ &  \multicolumn{2}{c|}{$0.1$}  & \multicolumn{2}{c|}{$0.01$} & \multicolumn{2}{c|}{$0.1$}  \\
$\epsilon_{c\odot}$  &  \multicolumn{2}{c|} {$0.01$} & \multicolumn{2}{c|} {$0.01$} & \multicolumn{2}{c|} {$0.03$}  \\
$\epsilon_{\rm SNII}$ & \multicolumn{2}{c|} {$0.0$} & \multicolumn{2}{c|} {$0.0$}  &\multicolumn{2}{c|} {$1.0$}    \\
$\epsilon_{\rm BE}$  & \multicolumn{2}{c|} {$0.0$} & \multicolumn{2}{c|} {$0.0$}  &\multicolumn{2}{c|} {$0.0$}   \\
\S5.4 & \multicolumn{2}{c|} {no} & \multicolumn{2}{c|} {no} & \multicolumn{2}{c|} {no}  \\
$M_s$ ($M_\odot$) & $2.8\times 10^{10}$ & $4.2 \times 10^7$ & $2.3
\times10^{10}$& $7.1 \times 10^7$ & $2.3 \times 10^{10}$& $3.6 \times
10^7$ \\
$M_g$ ($M_\odot$) & $5.4\times 10^{9}$  & $1.3\times 10^8$ & $5.4
\times10^9$& $9.9 \times 10^7$& $6.9 \times 10^9$& $1.3 \times 10^8$\\
$M_{ej}$ ($M_\odot$) &  $4.6 \times 10^9$& $2.6 \times 10^8$ & $7.9
\times 10^9$ & $3.1 \times 10^8$ & $6.5 \times 10^9$& $3.0 \times
10^8$ \\
$f_{ej}$ & $0.12$& $0.59$ & $0.22$ & $0.64$ & $0.18$& $0.64$ \\
$\tau_{ej}$ (Gyr) & $0.25$& $0.82$ & $0.40$& $1.8$ & $0.30$ &$1.24$  \\
$f_{*}$ & $0.74$& $0.10$ & $0.63$ & $0.15$ & $0.63$ & $0.08$  \\
$\langle Z_g/Z_\odot \rangle$ & $0.65$& $0.03$ & $0.57$ & $0.04$ & $0.56$ & $0.02$  \\
$\langle Z_s/Z_\odot \rangle$ & $0.55$& $0.20$ & $0.43$ & $0.30$ & $0.42$ & $0.19$  \\
$\langle Z_{ej}/Z_\odot \rangle$ & $0.29$& $0.11$ & $0.38$ & $0.14$ & $0.26$ & $0.09$  \\
\hline
\noalign{\medskip}
\end{tabular}

\begin{tabular}{|c|cc|cc|cc|} \hline \hline
Model & MW4 & DW4 & MW5 & DW5 & MW6 & DW6   \\
$\phi_\kappa$ &  \multicolumn{2}{c|}{$0.1$}  & \multicolumn{2}{c|}{$0.1$} & \multicolumn{2}{c|}{$0.1$}  \\
$\epsilon_{c\odot}$  &  \multicolumn{2}{c|} {$0.01$} & \multicolumn{2}{c|} {$0.01$} & \multicolumn{2}{c|} {$0.01$}  \\
$\epsilon_{\rm SNII}$ & \multicolumn{2}{c|} {$1.0$} & \multicolumn{2}{c|} {$1.0$}  &\multicolumn{2}{c|} {$1.0$}    \\
$\epsilon_{\rm BE}$  & \multicolumn{2}{c|} {$0.0$} & \multicolumn{2}{c|} {$0.05$}  &\multicolumn{2}{c|} {$0.05$}   \\
\S5.4 & \multicolumn{2}{c|} {yes} & \multicolumn{2}{c|} {no} & \multicolumn{2}{c|} {yes}  \\
$M_s$ ($M_\odot$) & $2.5\times 10^{10}$ & $4.3 \times 10^7$ & $2.4
\times10^{10}$ & $4.3 \times 10^7$ & $2.5 \times 10^{10}$& $4.5 \times
10^7$ \\
$M_g$ ($M_\odot$) & $2.1\times 10^{9}$  & $3.5\times 10^7$ & $4.9
\times10^9$& $9.5 \times 10^7$& $1.9 \times 10^9$& $2.9 \times 10^7$\\
$M_{ej}$ ($M_\odot$) &  $7.1 \times 10^9$& $3.4 \times 10^8$ & $6.8
\times 10^9$ & $3.1 \times 10^8$ & $7.2 \times 10^9$& $3.4 \times
10^8$ \\
$f_{ej}$ & $0.21$& $0.82$ & $0.19$ & $0.69$ & $0.21$& $0.82$ \\
$\tau_{ej}$ (Gyr) & $0.30$& $1.19$ & $0.30$& $1.19$ & $0.21$ &$1.12$  \\
$f_{*}$ & $0.73$& $0.10$ & $0.67$ & $0.09$ & $0.74$ & $0.11$  \\
$\langle Z_g/Z_\odot \rangle$ & $0.62$& $0.03$ & $0.59$ & $0.02$ & $0.64$ & $0.02$  \\
$\langle Z_s/Z_\odot \rangle$ & $0.47$& $0.17$ & $0.44$ & $0.21$ & $0.48$ & $0.19$  \\
$\langle Z_{ej}/Z_\odot \rangle$ & $0.26$& $0.09$ & $0.27$ & $0.10$ & $0.27$ & $0.10$  \\
\hline
\noalign{\medskip}
\end{tabular}
\end{center}
\end{table*}

\noindent

\begin{figure*}

\vskip 6.7 truein

\includegraphics{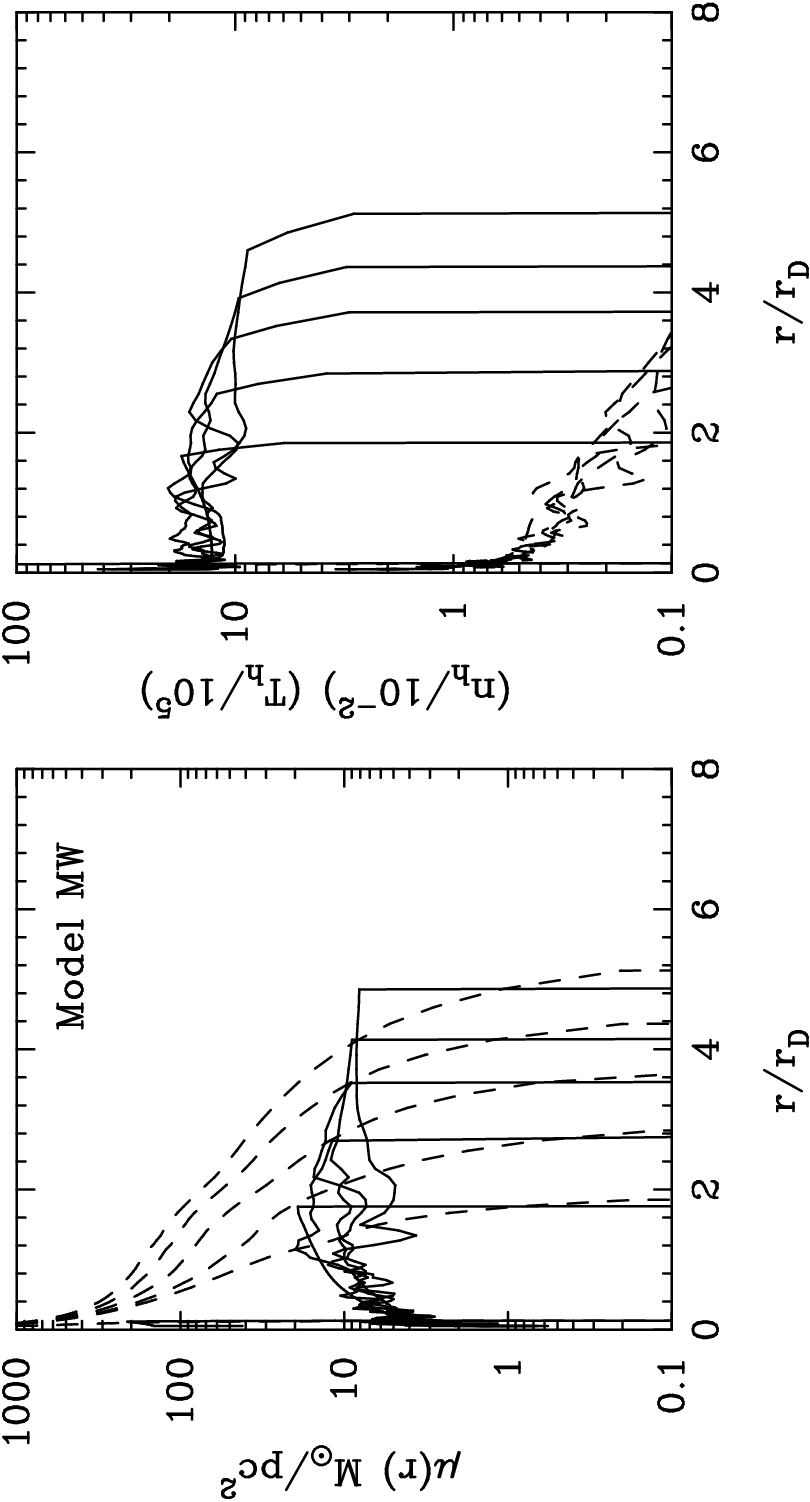}
\includegraphics{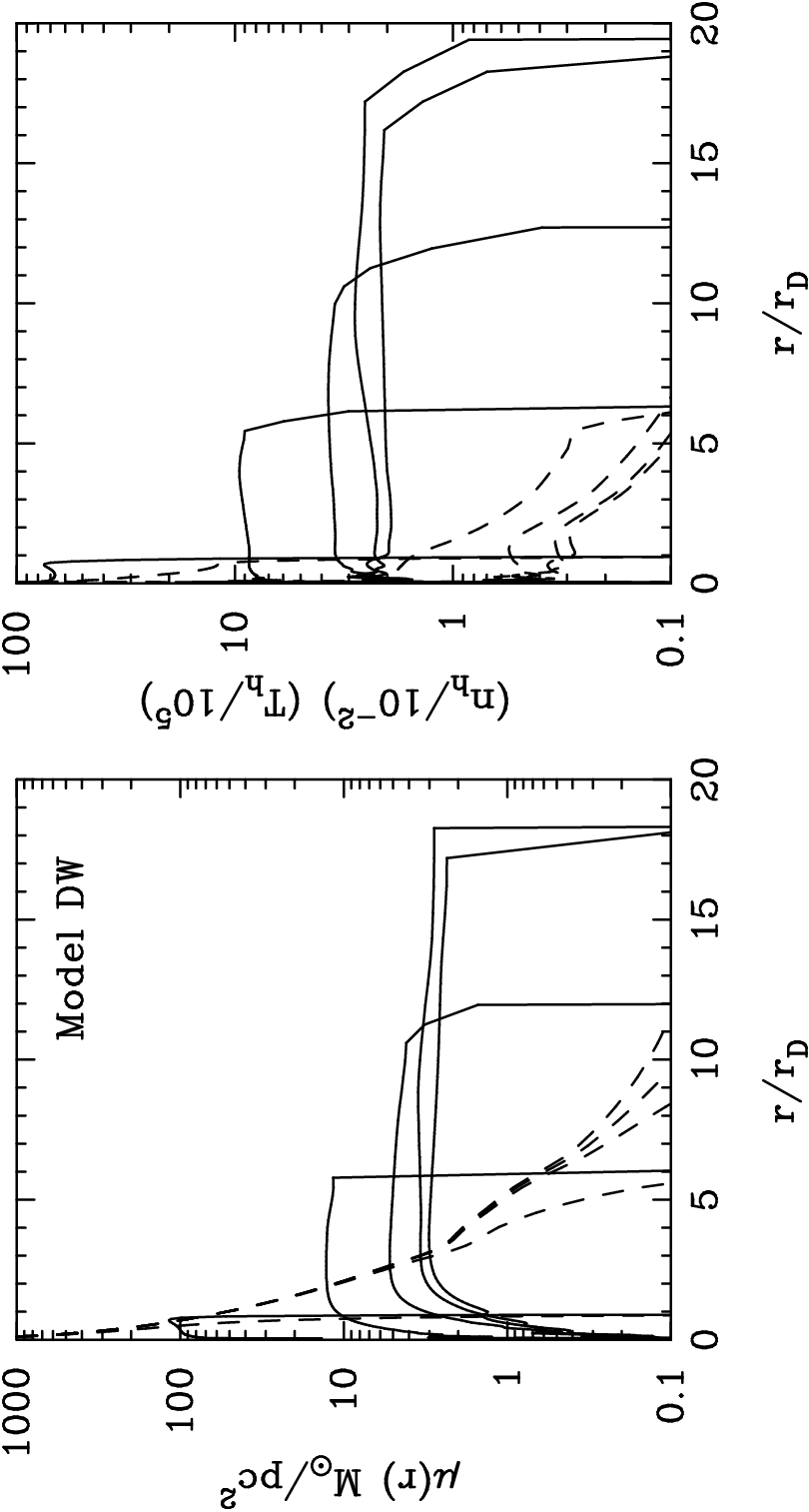}

\caption
{The left hand panels show the evolution of the stellar (dashed lines)
and cold gas (solid lines) profiles
 in models MW1 and DW1. The right hand panels
show the temperature (solid lines) and density (dashed lines) of the
hot component. The radius $r_D$ is the fiducial radius listed in
Table 1. The results are plotted at ages of $0.1$, $1$, $3$, $6$,
$10$ Gyr and (for MW1 only) at $15\;$ Gyr.}
\label{figure10}
\end{figure*}

The effects of varying these parameters are summarised in Table
2. Here we have run six models of galaxies MW and DW varying the input
parameters. We list the final stellar mass $M_s$, gaseous disc mass
$M_g$, and ejected mass $M_{ej}$ after $10$ Gyr for model DW (there is
very little evolution after this time) and after $15$ Gyr for model
MW. The parameters $f_{ej}$ and $f_*$ are the final ejected and
stellar masses divided by the total baryonic mass ($M_{ej} + M_* +
M_g$).  $\tau_{ej}$ is the time when half the final ejected mass is
lost. The last three numbers list the final mean  metallicities of
the cold ISM, the stars and the ejected gas.

The most important result from this table is that the final parameters
of the models are remarkably insensitive to variations of the input
parameters. For models DW, the final stellar disc mass varies between
$\sim 4 \times 10^7$ and $7 \times 10^7 M_\odot$ and the gas ejection
fraction varies from $0.59$ to $0.82$.  For models MW, the final
stellar disc mass varies between $\sim 2.3 \times 10^{10}$ and $2.8
\times 10^{10} M_\odot$ and the gas ejection fraction varies from
$0.12$ to $0.22$. Figure \ref{figure10} shows the evolution of the
radial density profiles of models MW1 and DW1 and figure
\ref{figure11} shows the time evolution of the star formation rates,
gas fractions and gas velocity dispersions. The models of Table 2 
behave in similar ways, and so these two figures are representative of
the behaviour of all of the models. These figures are qualitatively similar
to those of the simple model of Section 4 (figure \ref{figure8} and
\ref{figure9}).  The main differences are:

\begin{figure*}

\vskip 4.5 truein

\includegraphics{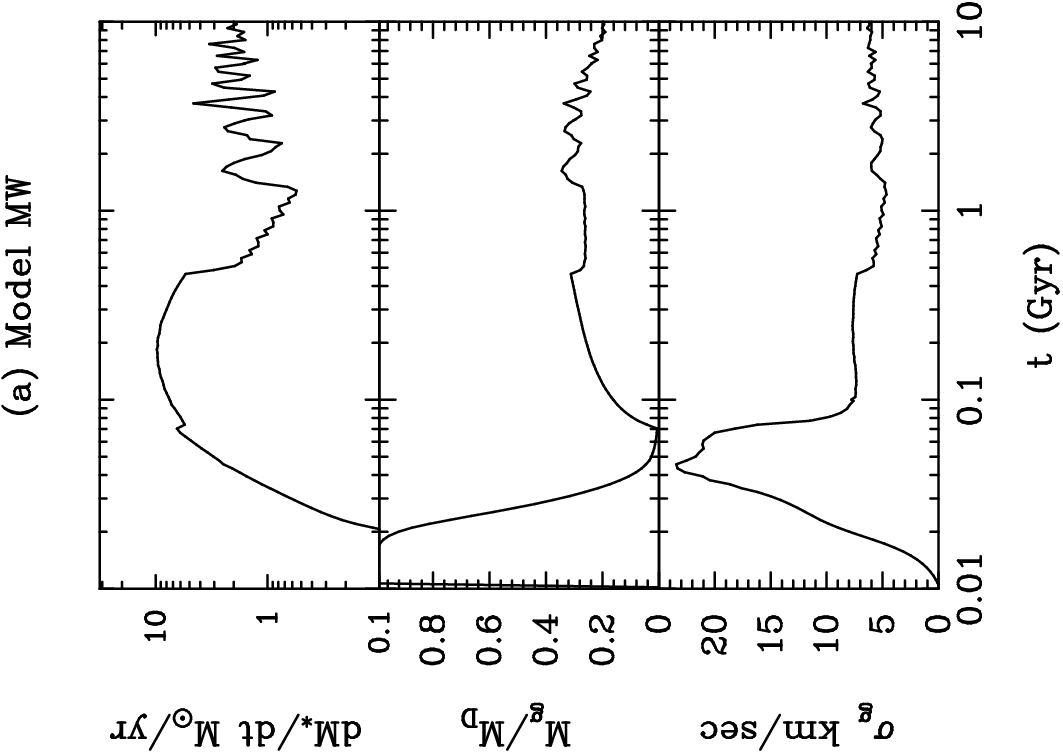}
\includegraphics{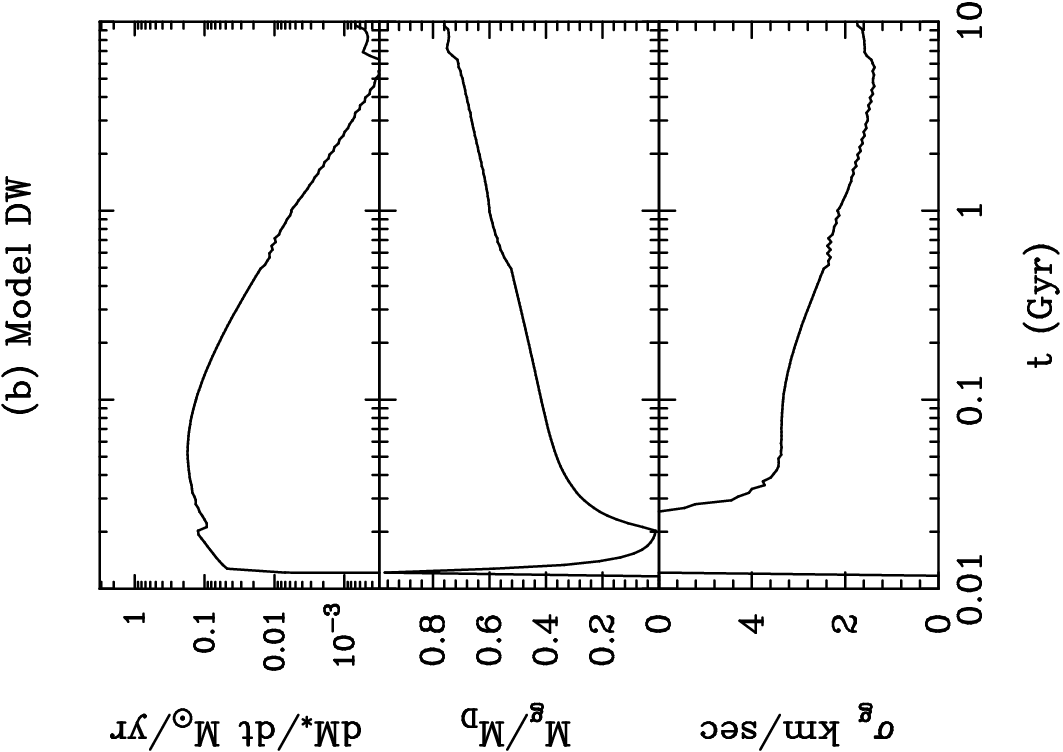}

\caption
{Evolution of the star formation rate, gas fraction and gas cloud
velocity dispersion for the models shown in figure 10.}
\label{figure11}
\end{figure*}

\smallskip

\noindent
(a) The gas discs have a sharper outer edge. This is a consequence of the
infall model; the outer edge is determined by the final time of the model
which sets the maximum cooling radius within the halo ({\it cf.} figures
\ref{figure7}). 

\smallskip

\noindent
(b) The radial profiles of models MW show oscillatory behaviour near
their centres, and the star formation rates and gas fractions show
oscillatory behaviour as a function of time. Both of these effects are
a consequence of the galactic fountain.

In these models, the star formation rate begins to rise as the gas
disc builds up from infalling gas. As the star formation rate rises,
the cold ISM is converted efficiently into a hot phase and this is
either driven out of the halo or becomes part of the galactic
fountain. In models MW, most of the gas that escapes from the system
is lost within this early ($\simlt 0.2$ Gyr) period of star formation
when the net star formation rate is close to its peak of $\sim 10
M_\odot$/yr.  After about $0.2$ Gyr, the temperature of the hot phase 
in models MW settles to $\sim 10^6$ K very nearly
independent of  radius ({\it cf} figure \ref{figure10}), and
so the galactic fountain cycles on a characteristic time-scale of
$\sim 4\times 10^8$ yr. Models DW behave in much the same way as the
simpler models of Section 4.3, except that infall, by construction,
extends over a longer period of time. In these models, the escape
criterion for the wind is satistfied over most of the lifetime of the
disc and hence the model of a  galactic fountain is unimportant.

We discuss briefly the effects of varying the input parameters:

\smallskip

\noindent
$\phi_k$: The evaporation rate $\dot M_{ev}$ has a weak dependence on
the evaporation parameter  $\phi_k$ ($\propto \phi_k^{0.29}$, equation
\ref{Temp4}) and obviously decreases as $\phi_k$ is reduced. However,
the temperature of the hot component is proportional to
$\phi_k^{-0.29}$ and hence rises as $\phi_k$ is reduced. The net
effect is that the mass of gas ejected is relatively insensitive to
$\phi_k$, but the mass of the final stellar disc increases as 
$\phi_k$ is reduced.

\smallskip

\noindent
$\epsilon_{c\odot}$: Increasing  this parameter reduces the
star formation rate in the self-regulating model for a fixed gas
surface density and velocity dispersion (equations \ref{SN2} and
\ref{Diss3}). However, a lower past star formation rate leads to
a higher gas surface density which increases the star formation rate.
These effects tend to cancel and so the models are 
insensitive to variations in $\epsilon_{c\odot}$.

\smallskip

\noindent
$\epsilon_{\rm SNII}$: Setting this parameter to unity increases the
temperature of the hot component slightly and hence increases the efficiency
of feedback.  As explained in Section 5.3,  energy injection by Type
II supernovae at large vertical scale heights will always be small
compared to the internal energy of the hot phase.

\smallskip

\noindent
$\epsilon_{BE}$: Values of $\epsilon_{\rm BE} \sim 0.05$ have little
effect on the evolution. Provided that $\epsilon_{\rm BE}$ is not too
large (so that it does not dominate the net star formation rate),
pressure enhanced star formation is self-limiting because it increases
the velocity dispersion of the cold clouds (reducing the star
formation rate in the self-regulating model) and converts cold gas to
hot gas. Both of these effects tend to reduce the net star formation
rate.

\smallskip

\noindent
{\it Pressure response of cold cloud radii:} Allowing the cold gas
radii to respond to the pressure of the hot phase provides strong
positive feedback in the very early stages of galaxy formation when
the pressure of the hot phase is high. However, most of the cold ISM
is ejected on a much longer timescale ({\it cf.} values of $\tau_{ej}$
in Table 2) when the typical pressure of the ISM is similar to that in
our own Galaxy. The pressure response of the cold cloud radii
therefore has little effect on the final feedback efficiency.

The models described here involve a complex set of coupled equations
and a number of  parameters. However, one of the most
interesting aspects of this study is that the equations interact in
such a way that the evolution of the models is insensitive to the
parameters. Most importantly, the efficiency of feedback is
insensitive to the thermal conduction parameter $\phi_\kappa$. The
possible severe suppression of thermal conduction by tangled magnetic
fields in astrophysical environments is a long standing theoretical
problem. However, our results show that even a reduction of $\kappa$
by a factor of $100$ or more will not significantly alter the
efficiency of feedback.

\begin{figure*}
\vskip 4.05 truein

\includegraphics{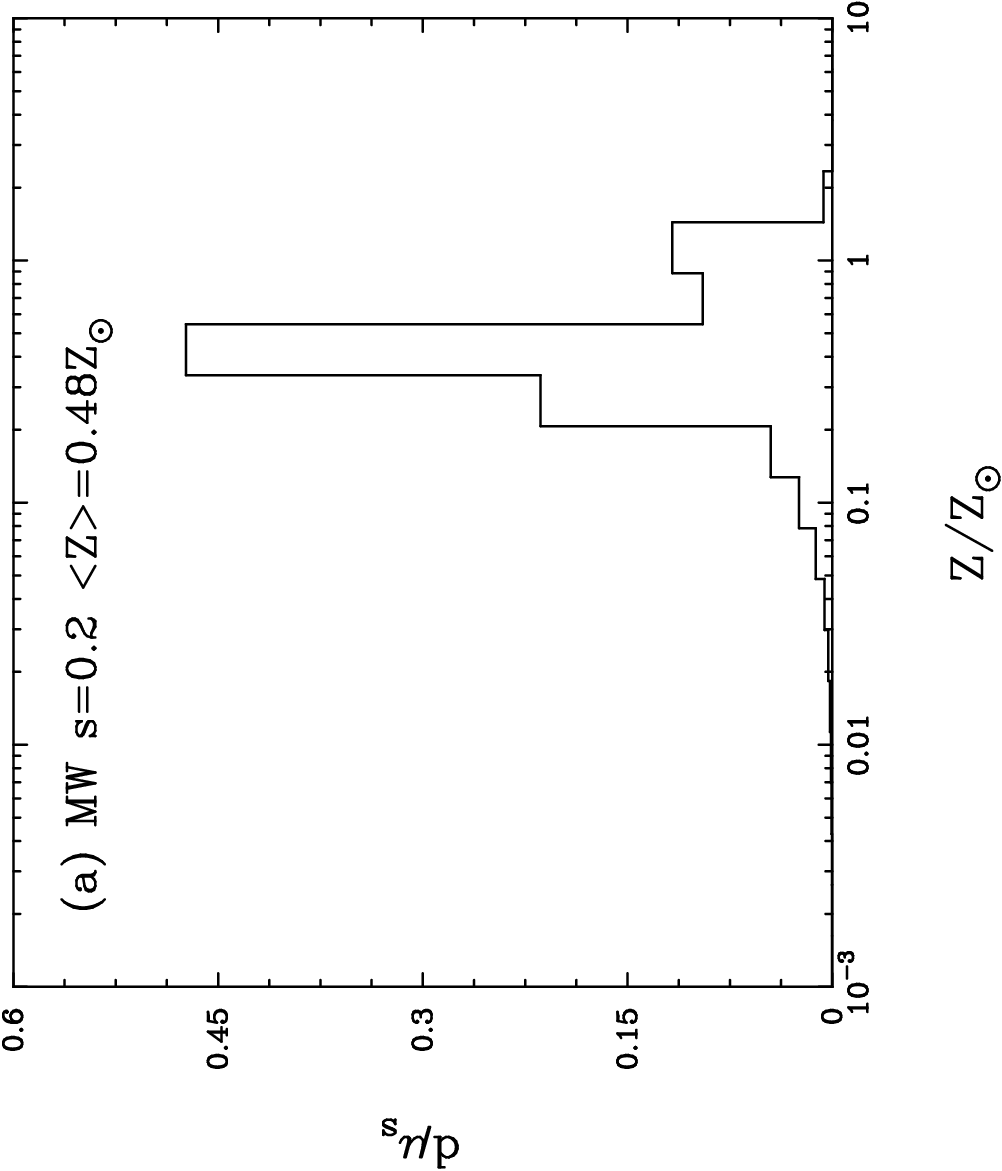}
\includegraphics{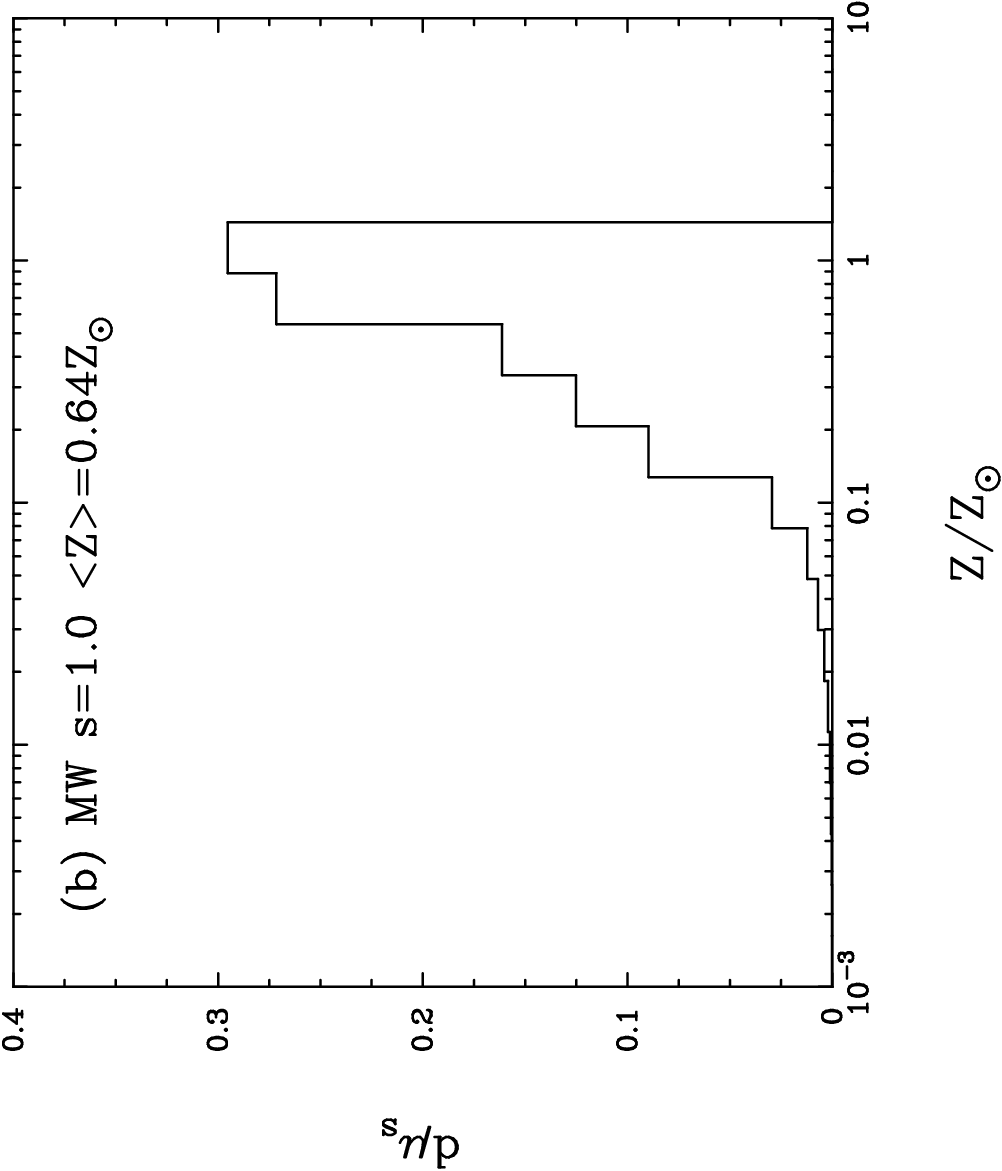}

\includegraphics{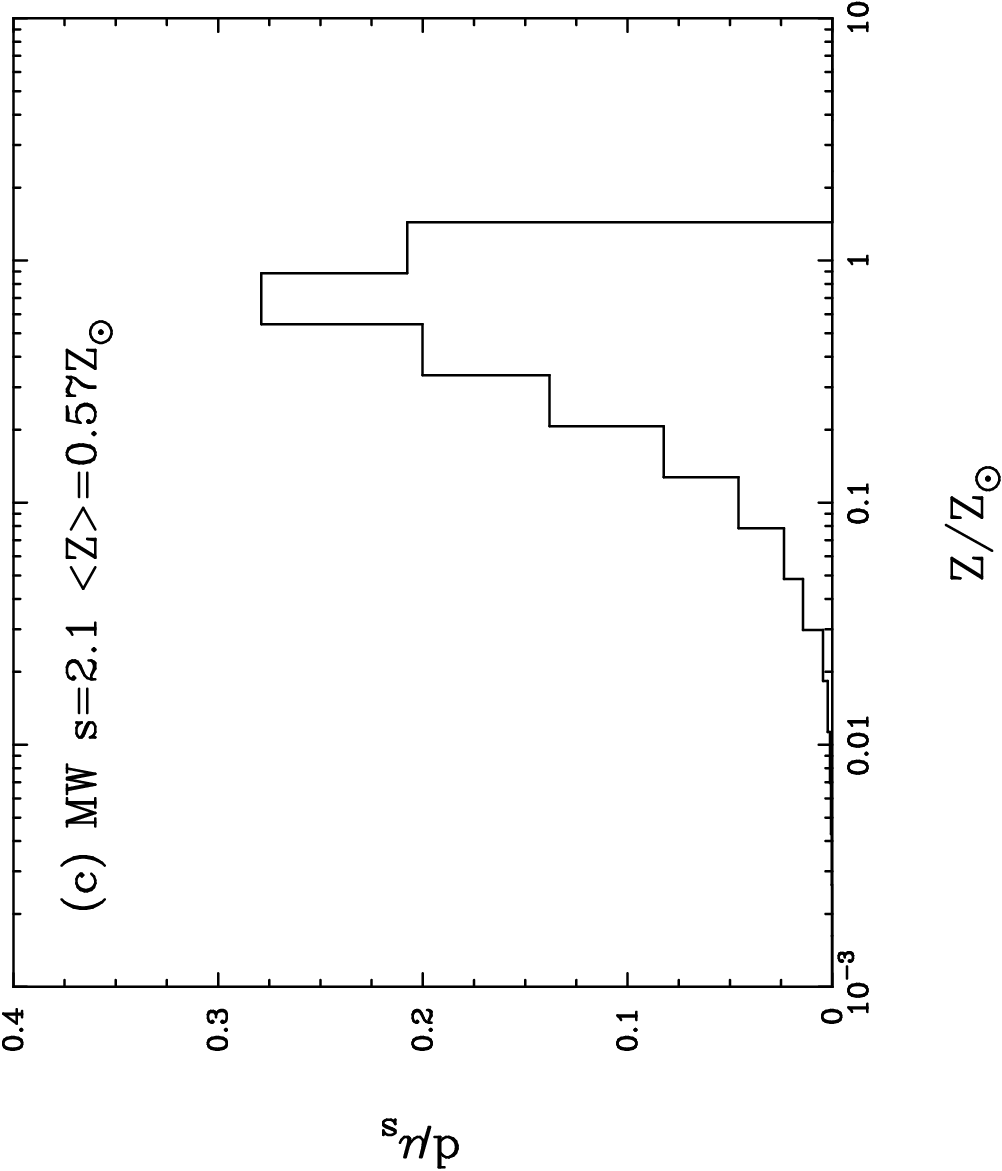}
\includegraphics{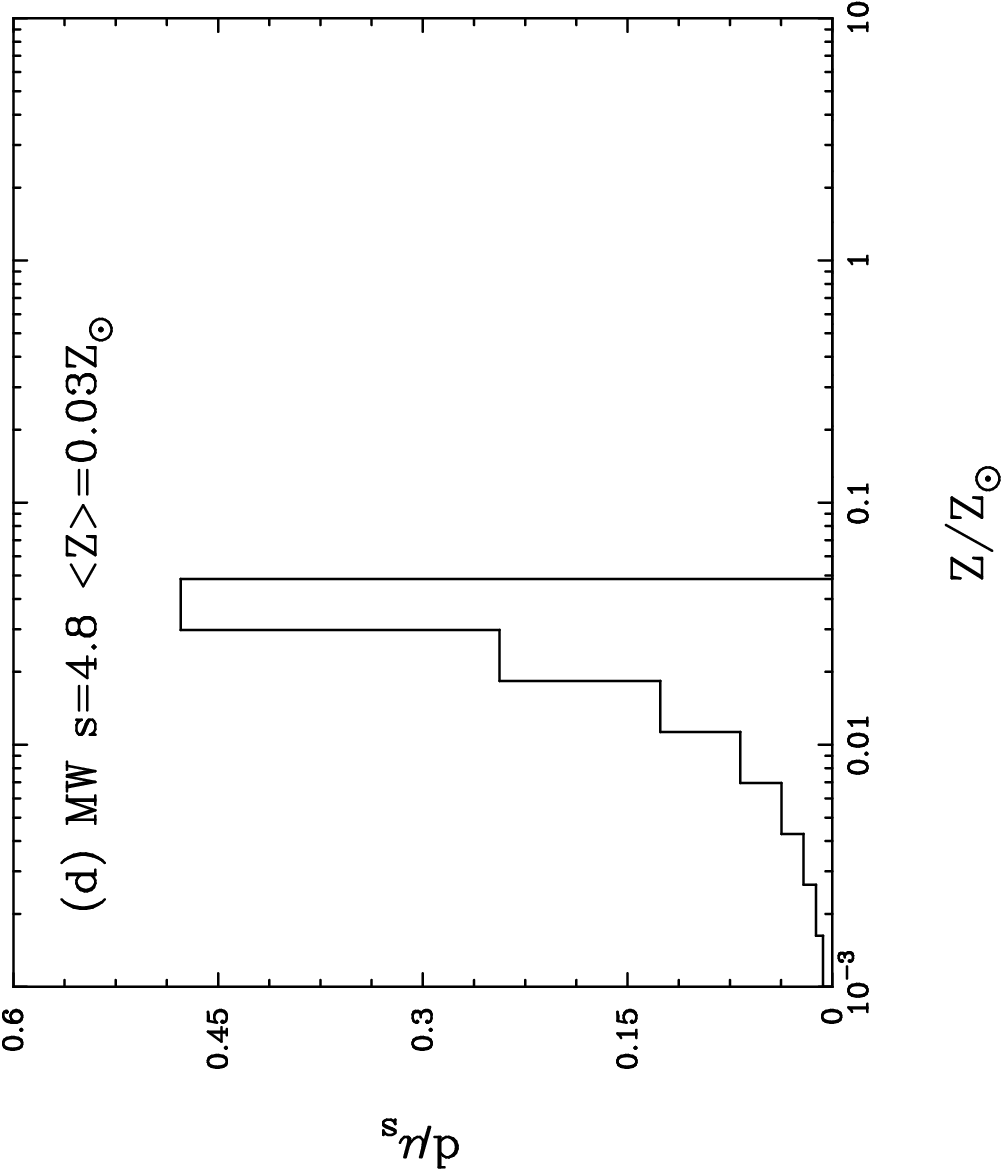}

\caption
{The distribution of stellar metallicities at four radii in model MW1
at an age of $15$ Gyr.}
\label{figure12}
\end{figure*}

\begin{figure*}
\vskip 4.0 truein

\includegraphics{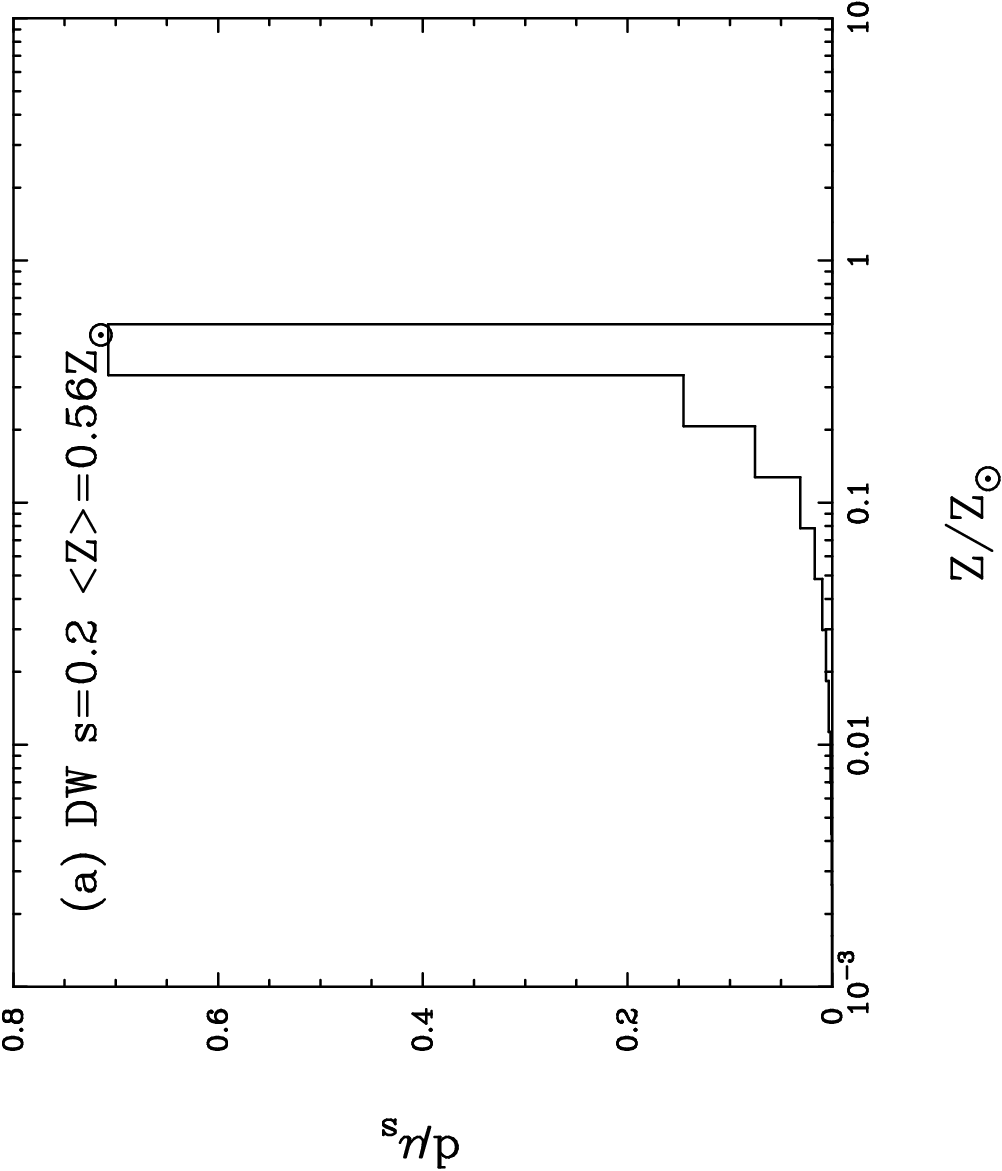}
\includegraphics{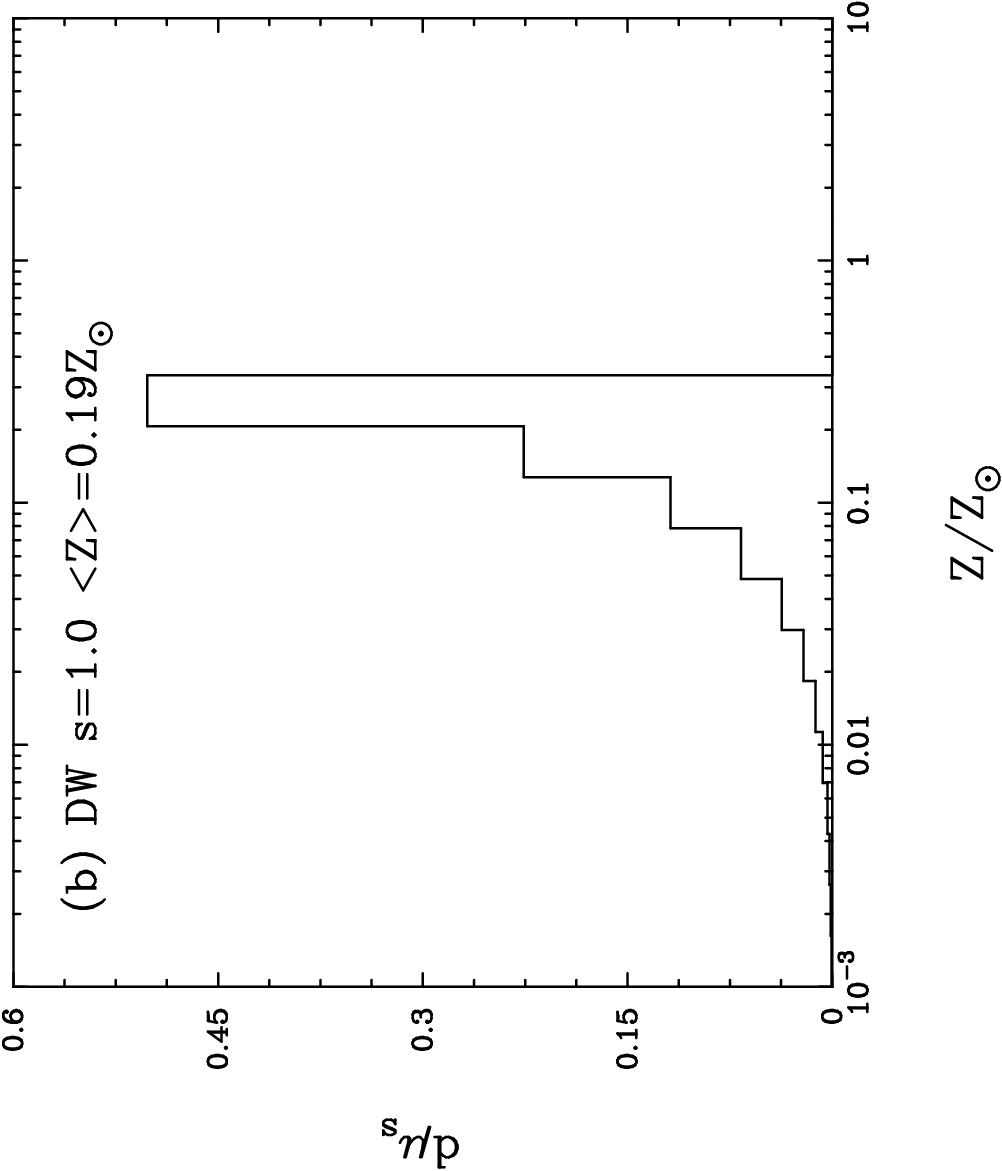}

\includegraphics{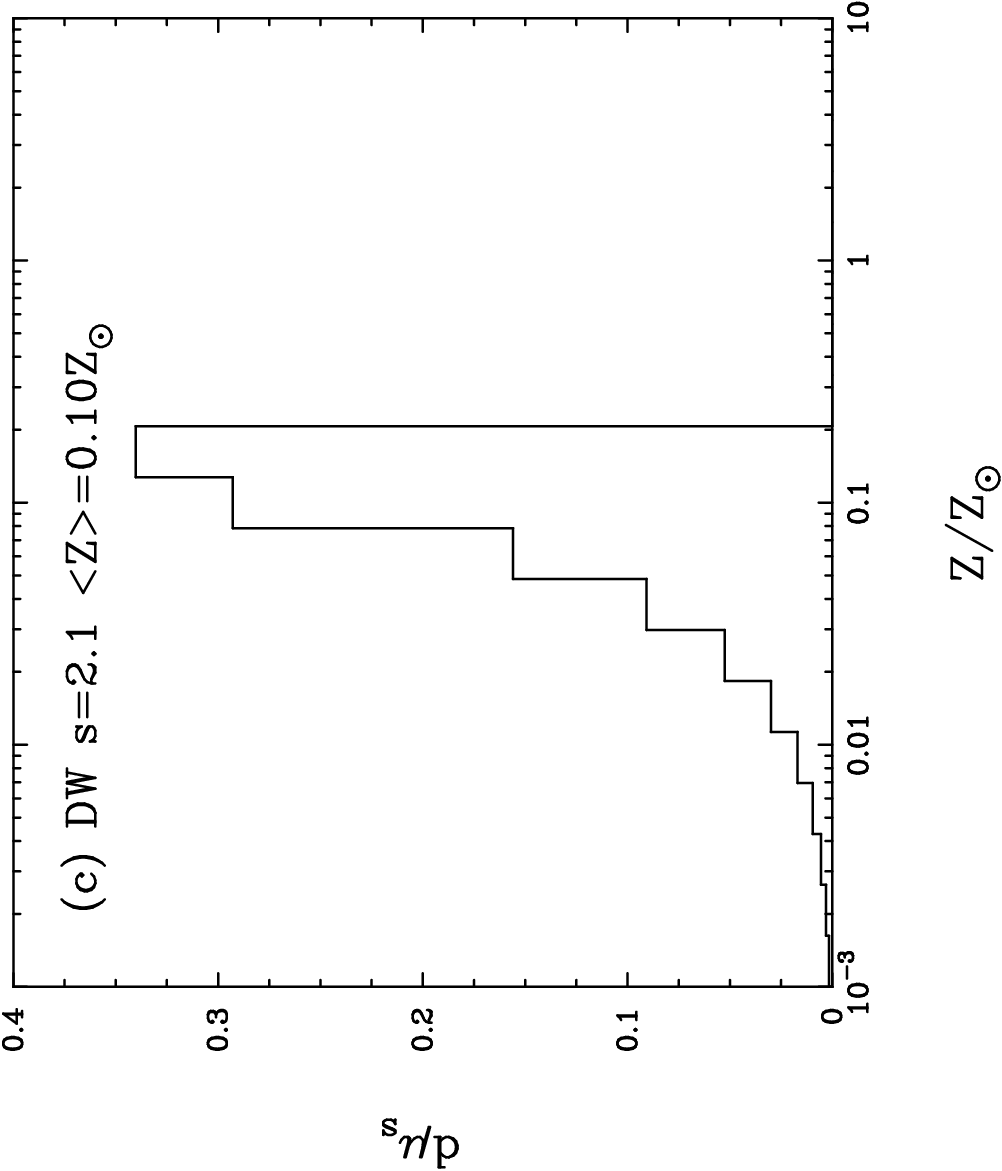}
\includegraphics{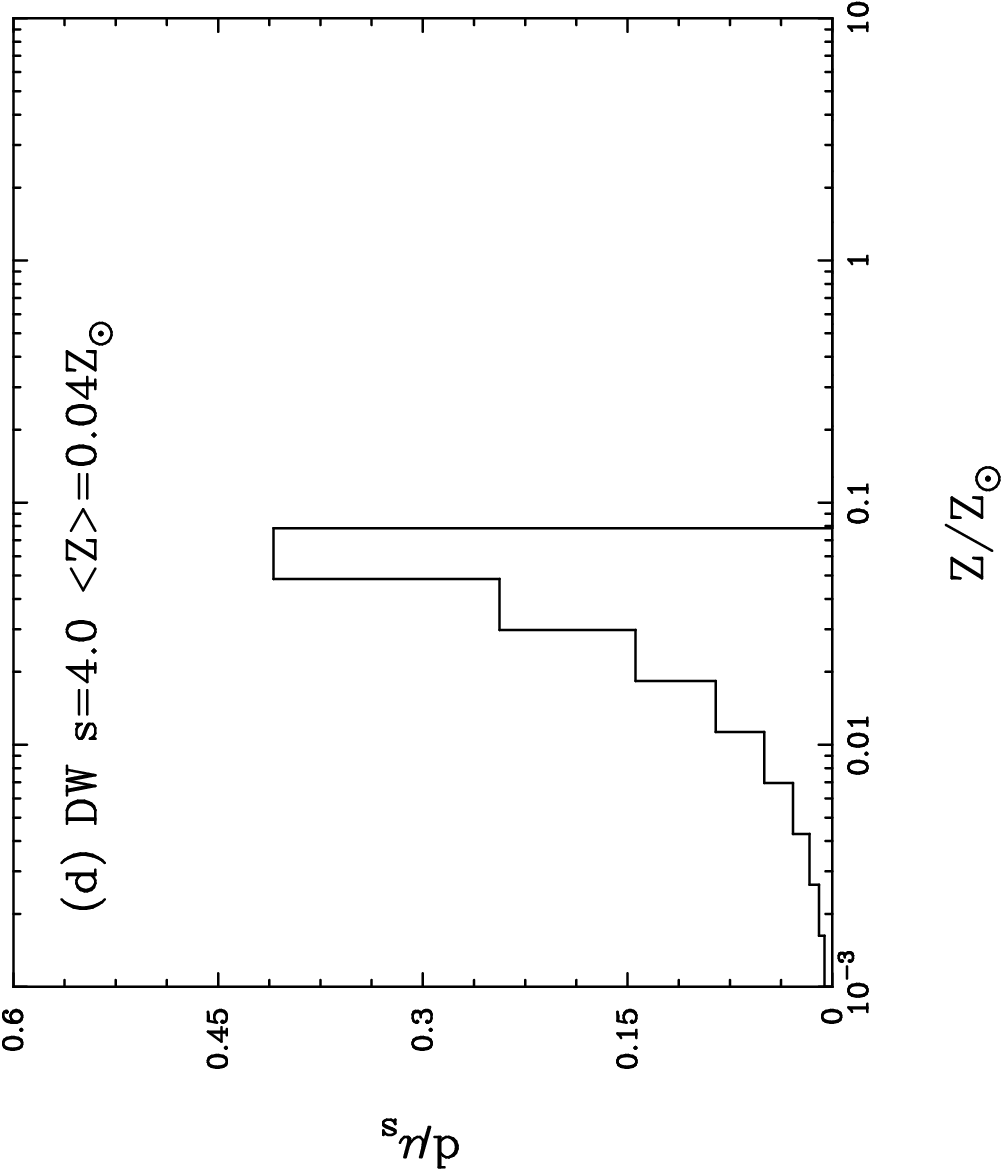}

\caption
{The distribution of stellar metallicities at four radii in model DW1
at an age of $10$ Gyr.}
\label{figure13}
\end{figure*}

\begin{figure*}
\vskip 3.0 truein

\includegraphics{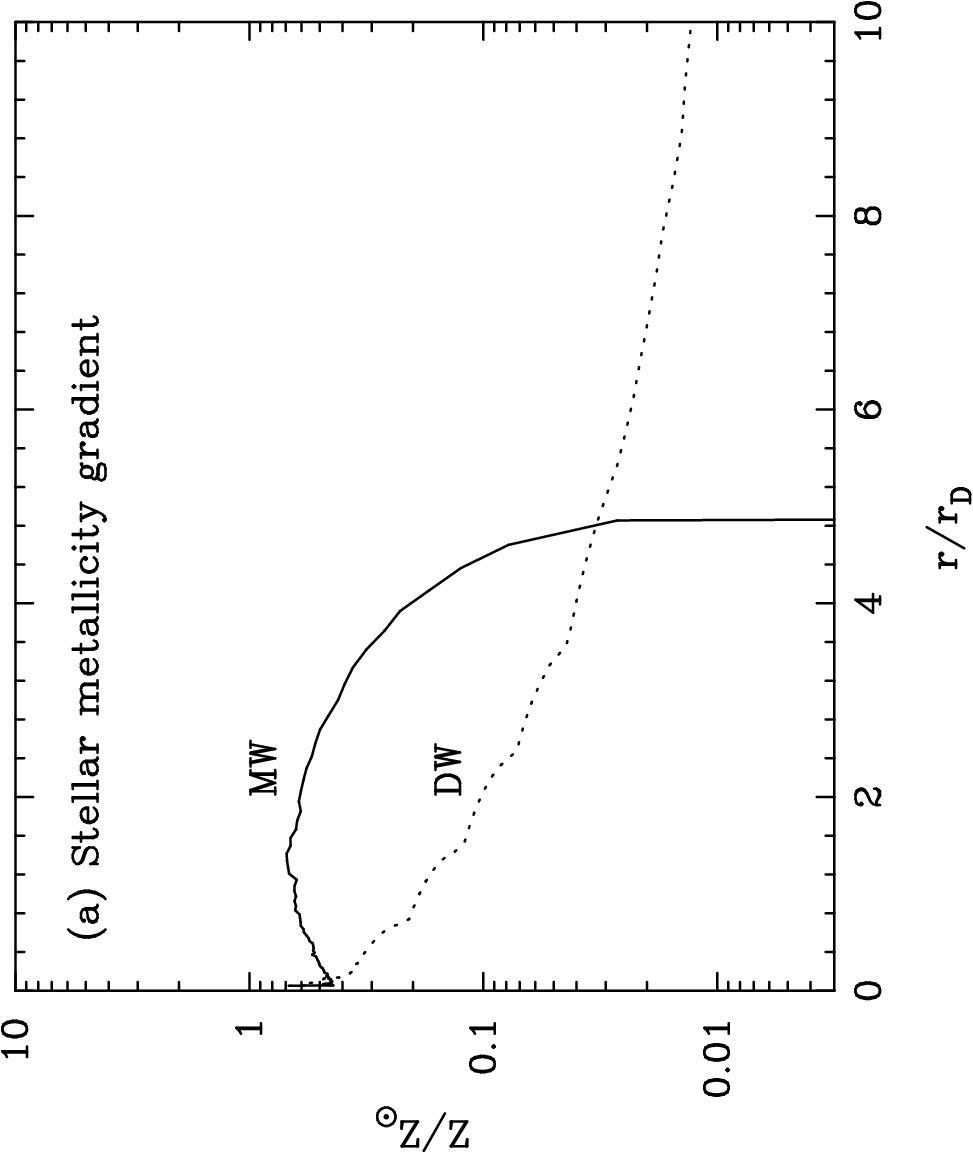}

\includegraphics{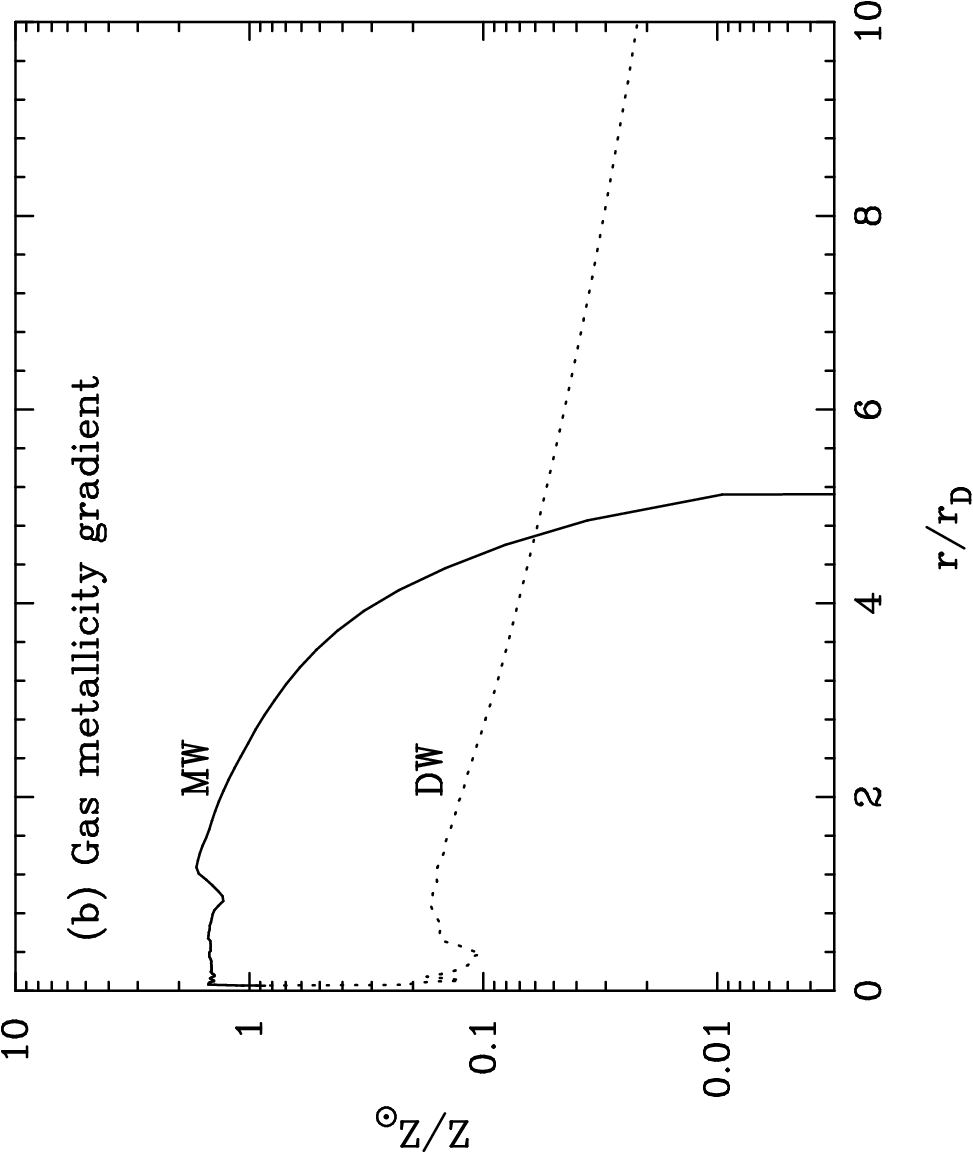}

\caption
{Metallicity gradients in the stars and gas at the final times in
models MW1 and DW1}
\label{figure14}
\end{figure*}

\subsection{Chemical evolution}

In this section, we summarize some of the results relating to chemical
evolution in these models. Our intention is not to present a detailed
model of chemical evolution in disc systems along the lines of, for
example, Lacey and Fall (1983, 1985) but to investigate some of the
general features of chemical evolution with physically motivated
models of inflow and outflow. The chemical evolution model is based on
the instantaneous recycling approximation as described in Section
5.6. This is probably a reasonable approximation since the timescales
of star formation and outflow are $\sim 1\;$Gyr, but will overestimate
the gas metallicities where the gas density is low. As in the previous
Section, results from models MW1 and DW1 are used to illustrate the
general features of the models. The other models listed in Table 2
behave in very similar ways.

\subsubsection{Stellar metallicity distribution}

The final  mean stellar metallicities are typically $Z_s/Z_\odot
\approx 0.5$ in models MW and $\approx 0.2$ in models DW. Models DW
have a lower stellar metallicity because a larger fraction of the ISM
is expelled in a wind. The stellar metallicity distributions are shown
in Figures (\ref{figure12}) and (\ref{figure13}). Figure
(\ref{figure12}c) is particularly interesting because this radius is
close to the solar radius. This metallicity distribution is quite
similar to that of G-dwarfs in the solar cylinder (see {\it e.g.}
figure 8.19 of Pagel 1997), showing that the infall model solves the
`G-dwarf' problem of closed box models of chemical evolution. The
metallicity distributions of model DW1 plotted in figure
(\ref{figure13}) also show a lack of stars with low metallicities.

\subsubsection{Metallicity gradients}

Over most of the stellar disc, model MW1 has a fairly weak stellar
metallicity gradient (figure \ref{figure14}a) except at the very outer
edge where the stellar density and metallicity fall abrubtly. This
differs from the metallicity gradients seen in large disc systems
which show linear gradients (see Vila-Costas 1998 for a recent review).  It
is possible that this problem might be resolved by including radial
gas flows in the models (Lacey and Fall 1985, Pitts and Tayler 1985).
The stellar metallicity gradients in model DW1 are steeper, in
qualitative with observations which indicate that the abundance
gradients in Scd and Irr galaxies are steeper than those in earlier
type galaxies.

The radial gas metallicity profiles are shown in figure
(\ref{figure14}b). Model DW contains a gasesous disc extending well
beyond the edge of the stellar disc. This gas disc has a low
metallicity in the outer parts, with $Z/Z_\odot \simlt 10^{-2}$ at $r
\simgt 2\;$kpc. At these large radii, the star formation rate is
always low and the gas disc can survive for much longer than a Hubble
time without converting into stars. This is unlikely to happen in all
galaxies for at least two reasons: (i) the energy injection from
supernovae into this gas will not be uniform as assumed in this paper;
(ii) the extended gas disc is susceptible to external disturbances and
so could be tidally stripped or transported towards the centre of the
system in a tidal interaction. Nevertheless, it is possible that dwarf
galaxies at high redshift possess extended gaseous discs, some of
which survive to the present day.

\begin{figure}

\vskip 3.0 truein

\includegraphics{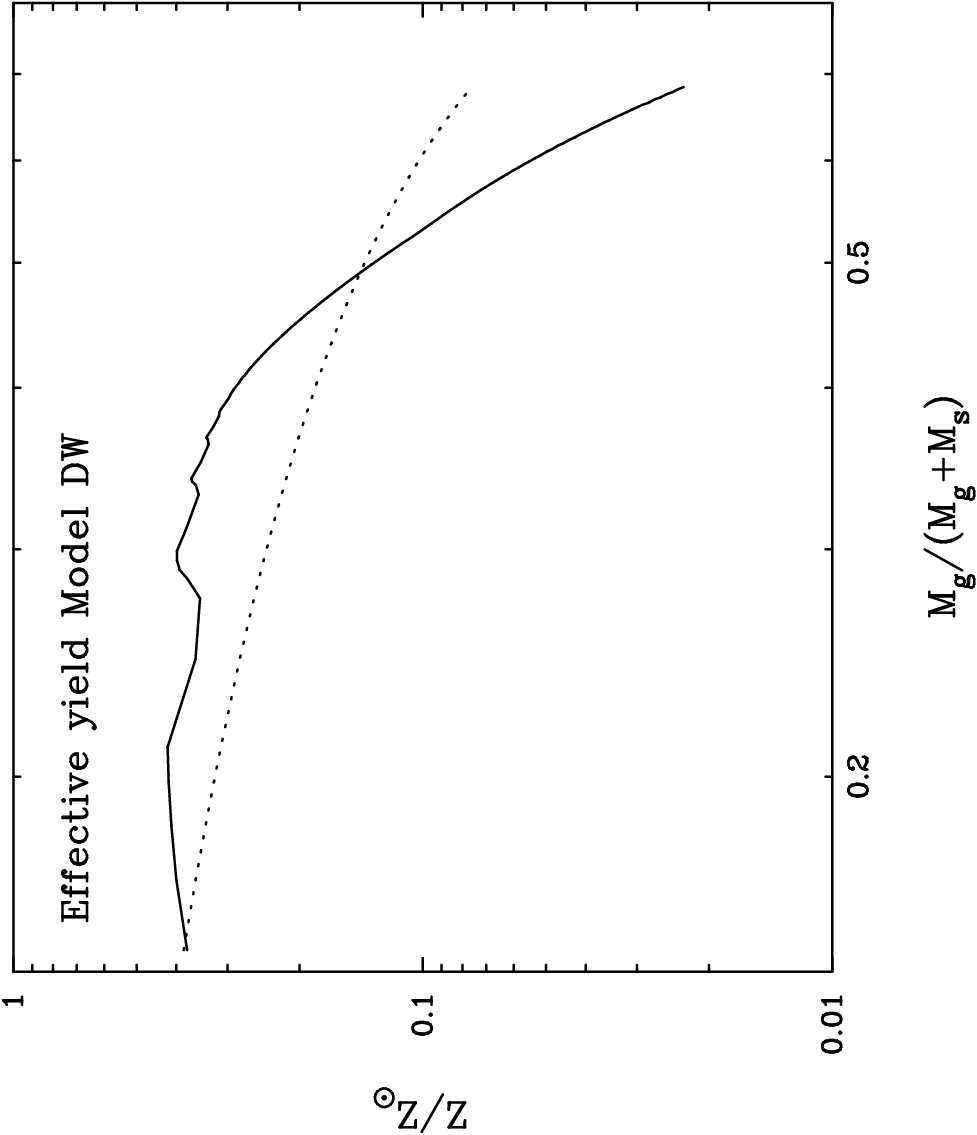}

\caption
{The effective yield for model DW1. For each radial ring in the galaxy
we plot the gas metallicty $Z_g$ against the gas fraction in that
ring. According to the simple closed box model of chemical evolution,
$Z_g = -p{\rm ln}(M_g/(M_g + M_s))$, where $p$ is the yield. The
dashed line shows this relation, but using an effective yield $p_{\rm
eff} = p/5$.}
\label{figure15}
\end{figure}

\subsubsection{Effective yields}

According to the simple closed box model of chemical evolution, the
metallicity of the ISM is related to the gas fraction according to
\begin{equation}
Z_g = -p\;{\rm ln}(M_g/(M_g + M_s)).
 \label{Met1}
\end{equation}
It is well known that the yields
derived from applying this relation to gas rich galaxies (usually
dwarf systems) result in ``effective yields'', $p_{\rm eff}$, that are
much lower than the yield expected from a standard Salpeter-like IMF.
For example, Vila-Costas and Edmunds (1992) find effective yields in
the range  $p_{\rm eff} \sim 0.004$--$0.02$ and that the effective yield
decreases with increasing radius.

The solid line in figure (\ref{figure15}) shows the final gas
metallicity in radial rings in model DW1 plotted against the gas
fraction within each ring. The dashed line shows equation (\ref{Met1})
with an effective yield of $0.004$ ({\it i.e.} one-fifth of the true
yield). The strong outflows in this model suppress the effective yield
well below the true yield and produce a strong radial variation of the
effective yield,  in qualitative agreement with observations.

\subsubsection{Metallicity of ejected gas}

The last line in Table 2 lists the mean metallicity of the gas that
escapes from the galaxy. The mean metallicity of the ejected gas is
about $0.3 Z_\odot$ for model MW1 and about $0.1 Z_\odot$ for model
DW1. In model DW1 this value is about $3$ to $5$ times higher than the
mean metallicity of the final gas disc. The ejected gas in this model
is therefore `metal enhanced' relative to the gaseous disc. The
mechanism for this metal enhancement is physically different to that
in the models  of Vader (1986, 1987) and Mac Low and Ferrara
(1999). In the models of these authors, metal enhancement arises from
incomplete local mixing between the supernovae ejecta and the ISM. In
our models, the gas is assumed to be well mixed locally, but metal
enhancement arises because the gas is lost preferentially from the
central part of the galaxy, which has a higher metallicity than the gas
in the outer parts of the system.

\subsection{Connection with damped Lyman alpha systems}

The column density threshold for the identification of damped
Ly$\alpha$ systems is $N({\rm HI}) \simgt 2 \times 10^{20}\;{\rm cm}^{-2}$
(Wolfe 1995) corresponding to a neutral gas surface mass density of
$\sim 1.6\; M_\odot/{\rm pc}^2$. Comparison with Figure 10 shows that the
extended cold gasesous discs around dwarf galaxies would be detectable
as damped Ly$\alpha$ systems. Furthermore, in CDM-like models, such
extended discs around dwarf galaxies would dominate the cross-section
for the identification of damped Ly$\alpha$ systems at high redshift
because the space density of haloes with low circular speeds is high
(Kauffman and Charlot 1994, Mo and Miralda-Escude 1994).  If this is
the case, then the metallicities of damped Ly$\alpha$ systems would be
expected to be low at high redshift, $Z/Z_\odot \sim {\rm few} \times
10^{-2}$,  with occasional lines-of-sight intersecting the central
regions of galaxies where the metallicity rises to $Z/Z_\odot \simgt
0.1$. At lower redshifts, the metallicities of damped systems would 
be expected to show a similarly large scatter, but with
perhaps a trend for the mean metallicity to increase as disc
systems with higher circular speeds form and the extended gaseous
discs around dwarfs are disrupted by tidal encounters.

This is qualitatively in accord with what is observed (Pettini
\etal$\;$ 1997, Pettini \etal$\;$ 1999, Pettini 1999). These authors
find that the typical metallicity of a damped Ly$\alpha$ system 
at $z \sim 2$--$3$ is about
$0.08Z_\odot$ with a spread of about two orders of
magnitude. Comparing the metallicities of  high redshift systems
with those of $10$ damped Ly$\alpha$ systems with redshifts $z =
0.4$--$1.5$, Pettini \etal$\;$ (1999) find no evidence for evolution
of the column density weighted metallicity. Whether these and other
properties of the damped Ly$\alpha$ systems can be reproduced with the
feedback model described here requires more detailed `semi-analytic'
calculations along the lines described by Kauffmann (1996). However,
the key point that we wish to emphasise is that according to the
models described here, most of the cross-section at any given redshift
will be dominated by largely unprocessed gas in the outer parts of
galaxies that does not participate in the star formation
process. The metallicity distributions and the evolution of
$\Omega_{\rm HI}$ as a function of redshift are therefore more likely to
tell us about feedback processes and the outer parts of dwarf
galaxies than about the history of star formation. Attempts to use 
the properties of damped Ly$\alpha$ systems to constrain the cosmic
star formation history ({\it e.g.} Pei, Fall and Hauser 1999) should
therefore be viewed with caution.

\begin{figure}

\vskip 3.0 truein

\includegraphics{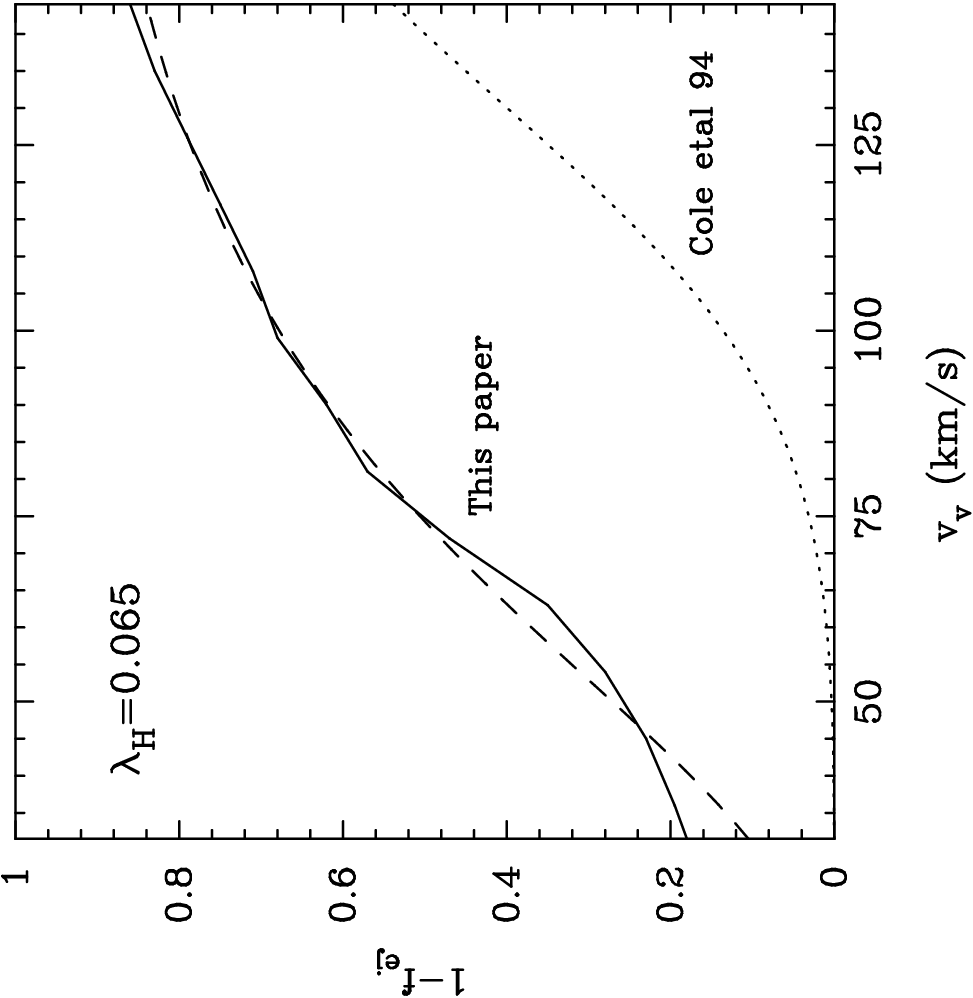}

\caption
{The solid line shows the retained baryonic fraction $1 - f_{ej}$ as a
function of the halo circular speed $v_v$.  The parameters
$\phi_\kappa$, $\epsilon_{c\odot}$ {\it etc.} adopted are the same as
those of models MW4 and DW4 listed in Table 2. The dotted line shows
the relation adopted by Cole \etal$\;$ (1994), equation (51) with
$v_{hot} = 140\; \kms$ and $\alpha_{hot} = 5.5$. The dashed line shows
equation (51) with $v_{hot}=75\; \kms$ and $\alpha_{hot}=2.5$.}
\label{figure16}
\end{figure}

\subsection{Feedback efficiency as a 
function of halo circular speed and semi-analytic
models of galaxy formation}

In this section we investigate the efficiency of feedback as a
function of halo circular speed.  We have adopted the parameters of
models 4 in Table 1 and run a series of models varying the halo
circular speed $v_v$. The virial radius of the halo is set to $r_v =
150 (v_v/126\; \kms)^2\;{\rm kpc}$, the concentration parameter $c=10$
and the ratio of gas to halo mass within the virial radius is set to
$1/10$. The halo rotation speed is set by equation (\ref{Inf1}) with
the fiducial disc scale length equal to $r_v/50$.  With these
parameters, the family of models has a constant value for the halo
spin parameter of $\lambda_H = 0.065$.

The retained baryonic fraction, $1-f_{ej}$, is plotted as a function
of halo circular speed in Figure 16. The dotted line shows the
relation used by Cole {\it et al.}  (1994, hereafter C94) in their
semi-analytic models,
\begin{equation}
1 - f_{hot}  = {1 \over 1 + \beta(v_v)}, \qquad \beta(v_v) =
\left( {v_v \over v_{hot} } \right )^{-\alpha_{hot}},
 \label{Cole1}
\end{equation}
where $f_{hot}$ is the fraction of the cooled gas that is reheated and
$v_{hot}$ and $\alpha_{hot}$ are parameters.  C94
adopt a severe feedback prescription with $\alpha_{hot} = 5.5$
and $v_{hot} = 140\;\kms$ to reproduce the flat faint end slope of the
B-band galaxy luminosity function in a critical density CDM model. The
C94 feedback  model does not agree at all well with the models
described here. There is a slight ambiguity in the appropriate value
of $v_v$ to use in equation (\ref{Cole1}) because C94 adopt
an isothermal rather than an NFW halo profile; the halo circular speed
at $\sim 0.1r_v$ may be $20 \%$ higher than the circular speed at the
virial radius, but this is far too small a difference to reconcile
the C94 feedback prescription with the models of this paper.

In fact, the dashed line in Figure 16 shows that our models are
reasonably well described by equation (\ref{Cole1}) with $v_{hot} =
75\; \kms$ and $\alpha_{hot} = 2.5$. Our results therefore suggest a
much  gentler feedback prescription than assumed in C94.
Note that with the C94 parameters, a Milky Way
type galaxy with $v_v \approx 130 \; \kms$ would have lost about $60
\%$ of its baryonic mass in a wind. This is well outside the range
found from our models for plausible choices of the input parameters
({\it cf.} Table 2).

Recently Baugh \etal$\;$ (1999) and Cole \etal$\;$ (1999) describe
semi-analytic models applied to $\Lambda$-dominated CDM cosmologies
that employ a gentler feedback model. The prescription for their
reference model is similar to that of equation (\ref{Cole1}) with
$v_{hot} = 150 \; \kms$ and $\alpha_{hot} = 2.0$, but with $v_v$
replaced by the disc circular speed $v_{\rm disc}$. This model is
closer to the results found here. Assuming angular momentum
conservation, a halo with $\lambda_H \approx 0.06$ will produce a disc
with a circular speed $v_{\rm disc} \approx 1.7 v_v$ ({\it cf.} Table
1) and so their model can be approximated by equation (\ref{Cole1})
with $v_{hot} \approx 90 \; \kms$ and $\alpha_{hot} = 2.0$. With these
parameters, their model gives somewhat stronger feedback than found in
our models, but is well within the range of physical
uncertainties. Kauffmann \etal$\;$ (1993) and Kauffmann, Guiderdoni
and White (1994) also adopt a much less severe feedback prescription
than C94 in their semi-analytic models. For a detailed analysis of the
effects of varying the feedback prescription (and other parameters) in
semi-analytic models see Somerville and Primack (1999).

The change from an Einstein-de Sitter CDM cosmology in C94
to a $\Lambda$-dominated CDM model in Cole \etal$\;$ (1999) partly
explains why the revised models provide a reasonable match to
observations using less efficient feedback. However, the revised
models predict a faint end slope for the B-band luminosity function
that is consistent with the observations of Zucca \etal$\;$ (1997) but not
with those of other authors ({\it e.g.} Loveday \etal$\;$ 1992, Maddox
\etal$\;$ 1998).  (The earlier paper of Cole \etal$\;$ 1994 attempted to
reproduce the flat faint end slope of the Loveday \etal$\;$ luminosity
function).  The observational differences in estimates of the faint
end slope of the optical luminosity function are not properly
understood and so it remains unclear whether a gentle feedback model,
of the type proposed here and used in Cole \etal$\;$ (1999), can account
for galaxy formation in CDM-type models.

\bigskip
\centerline{\bf Table 3: Dependence of Feedback Efficiency}
\centerline{\bf  of Model DW on Halo Angular Momentum}
\begin{center}
\begin{tabular}{cccc} \hline \hline
\noalign{\medskip}
 $f_{coll}$  & $v_c$ (km/s) & $\lambda_H$  & $f_{ej}$  \\
 $\;\;25$   &  $\;\;50$ & $0.12$  & $0.64$  \\
 $\;\;50$   &  $\;\;70$ & $0.065$ & $0.59$  \\
 $150$      &  $120$    & $0.031$ & $0.82 $  \\ \hline
\noalign{\medskip}
\end{tabular}
\end{center}

With the Cole \etal$\;$ (1999) parameterization the efficiency of
 feedback depends, by construction, on the surface density of the
 galaxy and hence on the angular momentum of the parent halo.  In
 their model, higher angular momentum haloes lead to more efficient
 feedback because they form low surface density discs with low disc
 circular speeds.  This is not what is found in our models. Table 3
 lists the ejected gas fraction as a function of the halo spin
 parameter $\lambda_H$.  Here, the halo circular speed and virial
 radius, $v_v$ and $r_v$, are the same as for model DW in Table 1, but
 the amplitude of the halo rotation speed (or equivalently the
 parameter $f_{coll}$) is adjusted to change the spin parameter of the
 halo. The parameters of the feedback model are the same as those for
 model DW1 in Table 2.  The feedback efficiency depends weakly (and
 non-monatonically) on $\lambda_H$, and is greater in systems with low
 values of $\lambda_H$.  This is because higher surface densities in
 low $\lambda$ galaxies result in higher star formation rates and a
 higher temperature hot component that can escape more easily from the
 halo.

The timescale for feedback in C94 and Cole \etal$\;$ 1999 is closely
linked to the star formation timescale which is assumed to be
shorter in galaxies with high circular speeds. This is not what is
found in the models of Table 2. The timescale for star formation is 
of order several Gyr in models MW (which have a roughly constant star
formation rate at late times, see Figure 11), yet the ejection of hot
gas occurs only in the initial stages of formation with a
characteristic timescale of $\sim 0.3$ Gyr. In models DW, the
situation is reversed with star formation occuring on a somewhat shorter
timescale than that for outflow.

\section{Conclusions}

The main aim of this paper has been to show that supernovae driven
feedback can operate in a quiescent mode and that high rates of star
formation are not required to drive efficient feedback. In dwarf
galaxies feedback occurs on an infalling timescale and so can extend
over a period of $\sim 1$ Gyr. In the feedback model developed here,
cold gas clouds are steadily evaporated in expanding supernovae
remnants and converted into a hot component.  Critically, the rate at
which cold gas is evaporated can exceed the rate at which mass is
converted into stars. If the temperature of the hot component is high
enough, a wind will form and the hot gas can escape from the halo
(provided the interaction with infalling gas can be ignored). If the
temperature of the hot component is not high enough for it to escape
from the halo, it will cool and fall back down to the  disc in
a galactic fountain. Some characteristic features of the models are as
follows:

\smallskip

\noindent
(i) In a Milky Way type system, feedback from supernovae may drive out
some of the gas from the halo in the early phases of evolution ($t
\simlt 0.3$ Gyr) when the star formation rate is high and the
temperature of the hot phase exceeds $\sim 5 \times 10^6$ K.  For
plausible sets of parameters, perhaps $20$ -- $30 \%$ of the final
stellar mass might escape from the galaxy. At later times, the
temperature of the hot phase drops to $T \sim 10^6$ K and the
evaporated gas cycles within the halo in a galactic fountain.

\smallskip

\noindent
(ii) In a dwarf galaxy with a circular speed $\sim 50 \; \kms$, expanding
supernovae remnants can convert the cold interstellar medium
efficiently into a hot component with a chacteristic temperature of a
few times $10^5$ K. This evaporated gas can escape from the halo in a
cool wind. Typically, only about $10\%$ of the baryonic material forms
stars. Gas  accreted from the halo at  $\simgt 1$ Gyr forms
an extended gaseous disc which, according to the self-regulating star
formation model used here, can  survive for longer than a Hubble time
without converting into stars.

\smallskip

\noindent
(iii) The feedback model developed here is meant to provide a sketch
of how feedback might operate in a multi-phase interstellar medium.
The model contains a number of obvious over-simplifications. For
example, we have neglected any interaction of the outflowing gas with
the infalling medium, we have not addressed the origin of the cold
cloud spectrum, ignored the dense molecular cloud component of the ISM
and neglected any local dissipation of supernovae energy in star
forming regions.  These effects, and other processes, are undoubtedly
important in determining the efficiency of feedback.  Nevertheless,
the simplified model presented here contains some interesting
features. Firstly, the model shows how positive feedback (via pressure
induced star formation) and negative feedback (via outflowing gas) can
occur {\it simultaneously}. Secondly, the models are remarkably
insensitive to uncertain physical parameters, in particular, thermal
conduction would need to be suppressed relative to its ideal value by
factors of much more than $100$ to qualitatively change the model.  If
thermal conduction is highly suppressed, it may be possible to
construct a qualitatively similar model to the one presented here in
which cold gas is converted into hot gas in shocks.

\smallskip

\noindent
(iv) The self-regulated star formation and feedback models described
here provide physically based models for the star formation timescale
and feedback efficiency as a function of the parameters of the
halo. The star formation timescale and feedback efficiency (or
timescale) are taken as free parameters in semi-analytic models of
galaxy formation ({\it e.g.} Cole \etal$\;$ 1999, Kauffmann \etal$\;$
1994) and are critically important in determining some of the key
predicted properties of these models, for example, the faint end slope
of the galaxy luminosity function and the star formation history at
high redshifts (see {\it e.g.}  Somerville and Primack 1999). It is
therefore important that we develop a theoretical understanding
of these parameters (as attempted crudely here) and also that ways
are found to constrain these parameters observationally. The results
presented here show that supernovae feedback is much less effective
than assumed in some earlier semi-analytic models (Cole \etal$\;$
1994, Baugh \etal$\;$ 1996) but is closer to the more gentler feedback
prescriptions used in more recent models (Cole \etal$\;$ 1999,
Somerville and Primack 1999).

The feedback model described in this paper has a number of consequences and
raises some problems which are summarized below.

\smallskip

\noindent
(i) {\it Evidence for outflows:} According to the models described
here, outflows with speeds of $\sim 200\; T_{h6}^{1/2}\kms$ should be
common in high redshift galaxies. There is evidence for an outflow of
$\sim 200 \; \kms$, a mass loss rate of $\sim 60 M_\odot/{\rm yr}$ and a star
formation rate of $\sim 40 M_\odot/{\rm yr}$ in the gravitationally lensed
Lyman break galaxy MS1512-cB58 (Pettini \etal$\;$ 2000). The outflow
velocity in this galaxy is consistent with our models, but the star
formation and mass loss rates (which are highly uncertain) are high.
The most likely explanations are either that MS1512-cB58 is a massive galaxy
driving an outflow that will remain bound to the system, or that it is
a less massive system undergoing a burst of star formation. In
addition to direct detection of outflowing gas, winds may have other
observational consequences.  The winds from dwarf galaxies will cool
rapidly (see Appendix B). Wang (1995b) has suggested that photoionized
gas clouds formed in the cooling wind might contribute to the 
Ly$\alpha$ forest. Nulsen, Barcon and Fabian (1998) suggest that
outflows caused by bursts of star formation in dwarf galaxies might
even produce damped Ly$\alpha$ systems.

\smallskip

\noindent
(ii) {\it Damped Ly$\alpha$ systems:} The extended gaseous discs that
form around dwarf galaxies in our models have low metallicities
because they have low rates of star formation. If this is correct,
then this largely unprocessed gas would dominate the cross section for
the formation of damped Ly$\alpha$ absorbers. The metallicities
of most of these systems would be low, but would
show a large scatter because some lines of sight will pass close to
the central regions of galaxies containing  gas of high
metallicity. This is broadly in agreement with what is
observed. Extended gaseous discs would be vulnerable in tidal
interactions. Some of the gas might be stripped and some might be
transported into the central regions to be  converted into stars
and hot gas. The evolution of $\Omega_{\rm HI}$ determined from damped
Ly$\alpha$ systems ({\it e.g.}  Storrie-Lombardi, McMahon and Irwin
1996) might have more to do with infall, feedback and tidal
disruption than with the cosmic star formation history.

\smallskip

\noindent
(iii) {\it Angular momentum conservation:} In hydrodynamic
simulations, 
gas is found to cool effectively in sub-units during the formation of
a protogalaxy. These sub-units lose their orbital angular momentum to
the halo as they spiral towards the centre and merge. Hence the gas
does not conserve angular momentum during the formation of a massive
galaxy.  (Navarro and Benz 1991, Navarro and Steinmetz 1997, Weil Eke
and Efstathiou 1998, Navarro and Steinmetz 2000). In fact, in the
absence of feedback it has proved impossible to form discs with
angular momenta similar to those of real disc galaxies starting from
CDM initial conditions.  In the models described here, it has been
assumed for simplicity that the specific angular momentum of the gas
is conserved during collapse. This assumption could easily be
relaxed. However, the feedback model decribed here suggests that the
numerical simulations miss some important physics. Firstly, it is
preferentially the low angular momentum gas, infalling in the early
stages of evolution, that is most likely to be ejected from a
developing protogalaxy. Secondly, supernovae driven feedback may help
to solve the angular momentum momentum problem by ejecting gas
efficiently from sub-units. The ejected gas may then infall at later
times when the halo is less sub-structured, approximately conserving
its angular momentum (Weil
\etal 1998, Eke, Efstathiou and Wright 2000).

\smallskip

\noindent
(iv) {\it Implementing feedback in numerical simulations:} There have
been a number of attempts to implement supernovae feedback in gas
dynamical numerical simulations ({\it e.g.} Katz 1992, Navarro and
White 1993, Navarro and Steinmetz 2000). These involve either heating
the gas around star forming regions (which is ineffective because the
energy is quickly radiated away) or reversing the flow of infalling
gas. An implementation of the feedback model described here is well
beyond the capabilities of present numerical codes. It would require
modelling several gas phases, a cold interstellar medium, a hot
outflowing medium and an infalling component, including 
mass transfer between each phase. It might be worth
attempting simpler simulations in which cold high density gas is added
to the halo beyond the virial radius at a rate that is determined by
the local star formation rate.

\smallskip

\noindent
(vi) {\it Starbursts vs quiescent feedback:} It is likely that
starburts are more common at high redshift because of the increased
frequency of galaxy interactions. Starbursts could contribute to
supernovae driven feedback in addition to the quiescent mode described
here. However, at any one time, our models suggest that the cold gas
component will have a mass of only $20$ -- $50\%$ of the the mass of the
stellar disk. Even if a substantial fraction of this gas is
transported towards the centre of a galaxy in a tidal encounter (see
{\it e.g.}  Barnes and Hernquist 1996) and is subsequently ejected in
a superwind, this mode of feedback will be inefficient because the
mass of gas involved is a small fraction of the total gas mass ejected
in the quiescent feedback mode over the lifetime of the galaxy.

\smallskip

\noindent
(vii) {\it Metallicity ejection:} The mean metallicity of the gas
ejected from a dwarf galaxy is typically about $Z_\odot/10$ in our
models, and comparable to the mean metallicity of the stars in the
final galaxy. Yet typically a dwarf galaxy is predicted to expel $5$
to $10$ times its residual mass in stars. Dwarf galaxies can therefore
pollute the IGM with metals to a much higher level than might be
inferred from their stellar content. The high metallicity of gas in
the central regions of clusters $\sim Z_\odot/3$ ({\it e.g.}
Mushotsky and Loewenstein 1997) may require a top-heavy IMF and gas
ejection from massive galaxies.

\vskip 0.2 truein

\noindent
{\bf Aknowledgments:} The author aknowledges the award of a  PPARC Senior
Fellowship.

\begin{appendix}

\section{Fast Computation of the Rotation Curve of a Thin Disc}

We begin with the expression for the potential of a thin disc
at $z=0$
\begin{equation}
 \phi(r, z=0)  =  -2 \pi G \int_0^\infty \int_0^\infty J_0(kr) J_0(kr^\prime)
r^\prime \mu(r^\prime) dr^\prime dk.
 \label{App1}
\end{equation}
The integral over $k$ is a well-known discontinuous integral ({\it e.g.}
Watson 1944) 
\begin{equation}
  \int_0^\infty J_0(kr) J_0(kr^\prime) dk = {2 \over \pi}
 \left\{ \begin{array}{ll}
 (1/r) K(r^\prime/r)        & r^\prime < r \\
 (1/r^\prime) K(r/r^\prime) & r^\prime > r \end{array} \right . 
 \label{App2}
\end{equation}
where $K$ is the complete elliptic integral of the first kind.
Differentiating  equation (\ref{App2}), we find
\begletA
\begin{eqnarray}
 v^2(r)  =  
4  G r \Bigg  \{  \int_0^{r - \epsilon} I_{<}(r, r^\prime)
r^\prime \mu(r^\prime) dr^\prime   \qquad \qquad \nonumber  \\
 +  \int_{r+ \epsilon}
^\infty I_{>}(r, r^\prime)
r^\prime \mu(r^\prime) dr^\prime \Bigg \}, \qquad 
\\
 I_<(r, r^\prime) = {E(r^\prime /r ) \over (r^2 - {r^\prime}^2)},
\qquad \qquad \qquad \qquad \qquad \qquad \qquad\\
 I_>(r, r^\prime) = {K(r/r^\prime) \over r r^\prime} - 
{ r^\prime E(r/r^\prime ) \over r ({r^\prime}^2 - r^2)},  \qquad \qquad \qquad \qquad \quad
\end{eqnarray}
\endletA
where $E$ is the complete elliptic integral of the second kind. This integral
is convergent in the limit $\epsilon \rightarrow 0$.
The functions $I_<$ and $I_>$ can be evaluated once and stored, 
reducing the computation of $v^2(r)$ to a simple
integral over the surface density of the disc multiplied by the tabulated
functions. We evaluate the epicyclic frequency by differentiating 
equation (A3) numerically.

\section{Steady Spherical Winds}

\begin{figure*}

\vskip 3.7 truein

\includegraphics{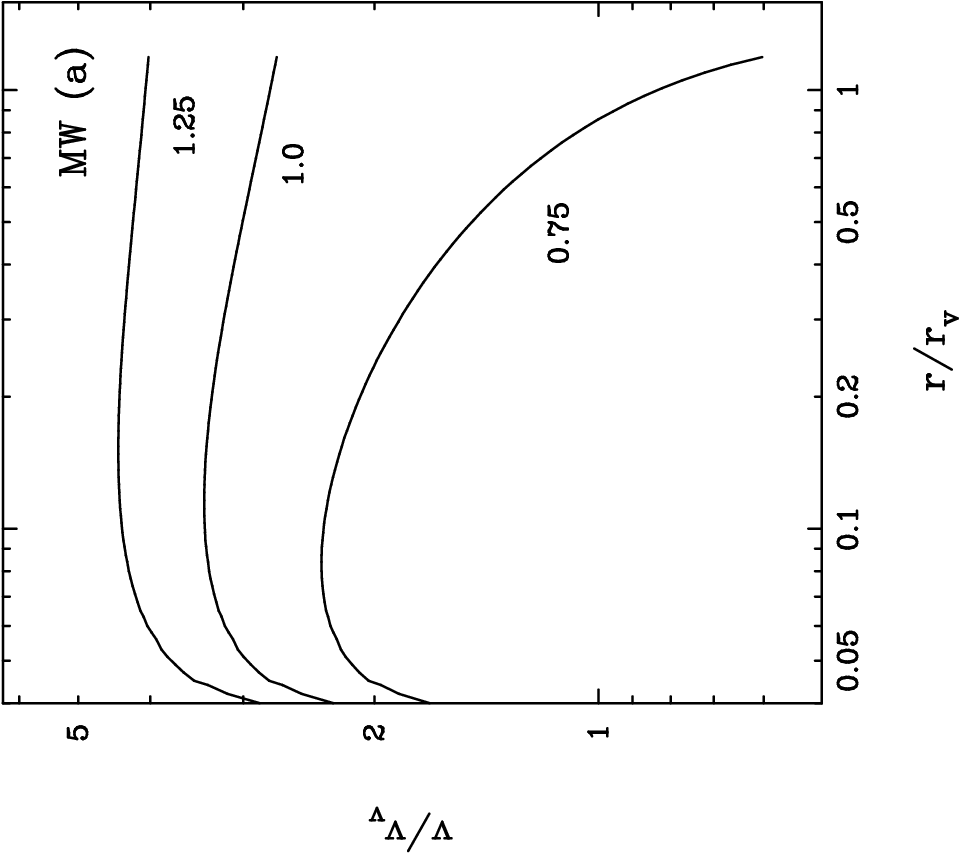}
\includegraphics{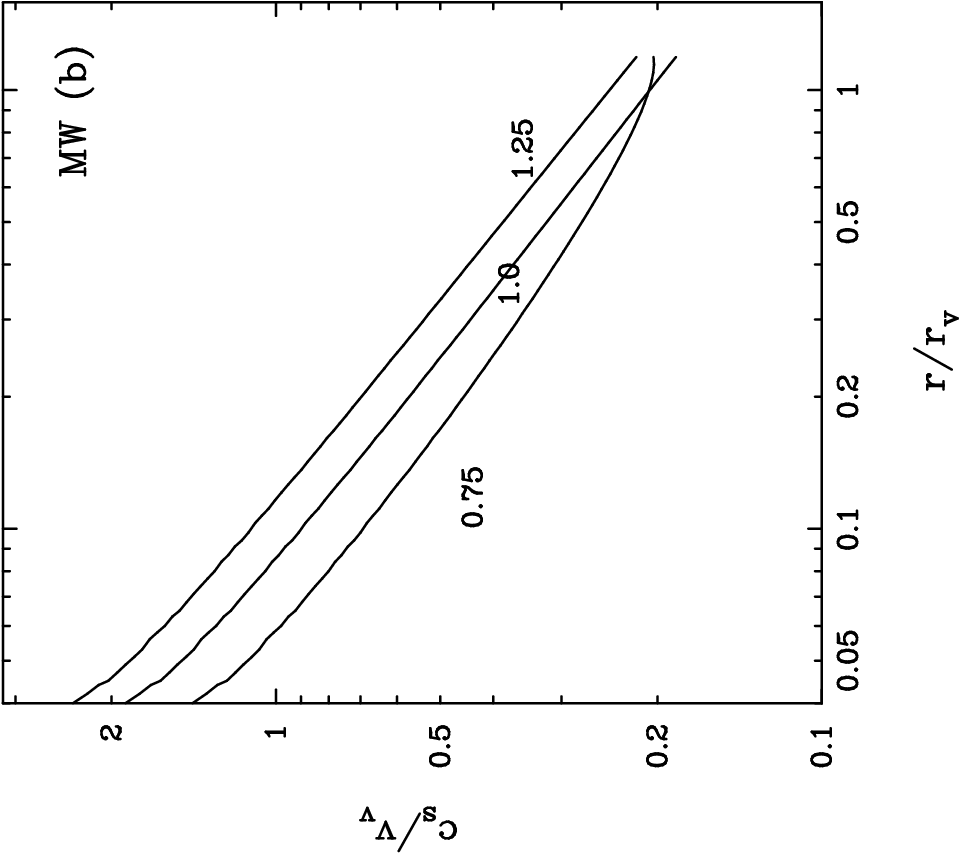}

\vskip 3.7 truein

\includegraphics{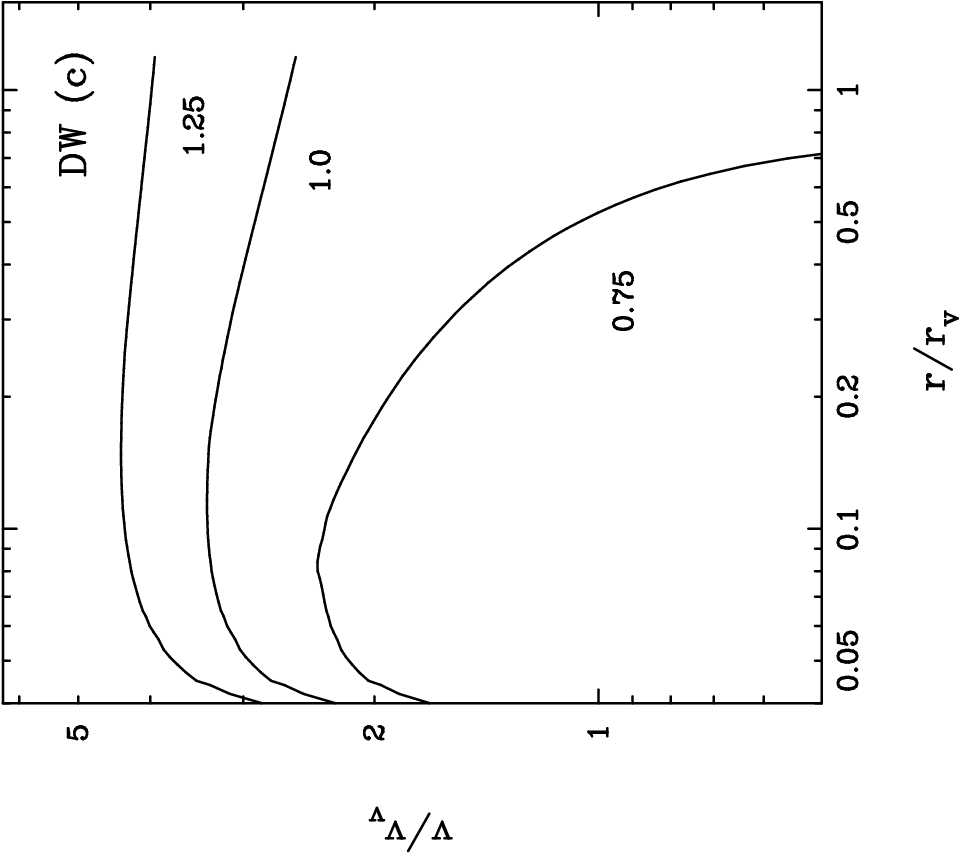}
\includegraphics{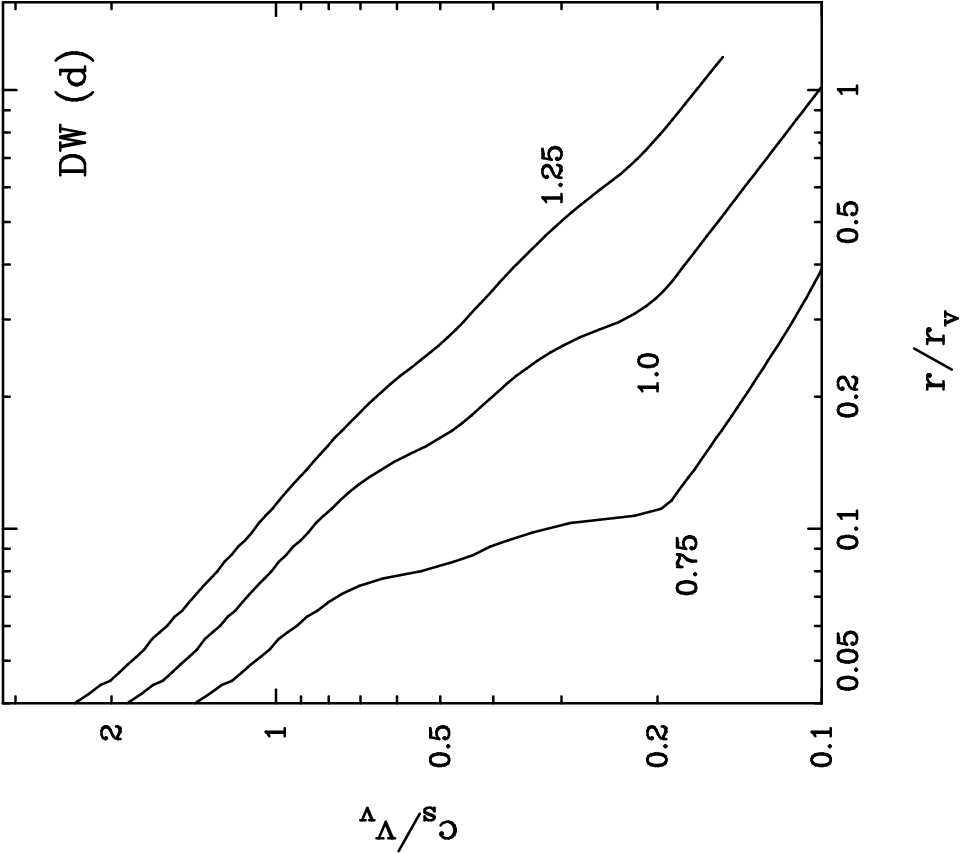}

\caption
{Steady wind solutions for models MW and DW including radiative
cooling. The curves show the wind velocity (figs a,c) and adiabatic
sound speed (figs b, d) assuming that the flow begins at $r_i =
0.04r_v$ with a Mach number of unity. The numbers give the initial
isothermal sound speed in units of the escape speed $v_{esc}$ from the
centre of the halo. These curves are for a mass injection rate of $10
M_\odot/{\rm yr}$ for model MW and $0.2 M_\odot/{\rm yr}$ for model
DW.}
\label{figure17}
\end{figure*}

The equations governing a steady spherically symmetric wind are
\begletB
\begin{eqnarray}
 {1 \over r^2} {d \over dr}( \rho v r^2) = q(r),  \\
 \rho v {dv \over dr} = -{dp \over dr} - \rho {d \Phi \over dr} - q(r)v, \\ 
 {1 \over r^2} {d \over dr} \left [ \rho v r^2 \left ({1 \over 2} v^2 + 
{5 \over 2} {p \over \rho} \right )\right] + \rho v {d \Phi \over dr}
 = {\cal H} - {\cal C},
\end{eqnarray}
\endletB
where $q(r)$ is the mass density injected per unit time and ${\cal H}$
and ${\cal C}$ are the heating and cooling rates per unit volume
({\it e.g.} Burke 1968, Holzer and Axford 1970). We assume that the
gravitational force is given by the NFW halo potential (equation 3),
$d\Phi/dr = v^2_H(r)/r$,  and rewrite these equations as two
dimensionless first order equations
\begletB
\begin{eqnarray}
  {d v^2\over dx} = {1 \over 2 \pi x^2 (c^2 - v^2)}
\Big [ - 8 \pi x c^2 v^2 + 4 \pi x v^2 v^2_H  \nonumber \\
+ {4 \over 3} \gamma v^2
+  \gamma c^2_i - {2 \over 3} \kappa \Big ]
\\
  {d c^2\over dx} = {-1 \over 6 \pi x^2 (c^2 - v^2) v^2}
\Big [  -8 \pi x c^2 v^4 + 4 \pi x c^2 v^2 v^2_H \nonumber \\
-  \gamma (v^2c^2
- {3 \over 2} c^4 - {5 \over 6} v^4) \cr
- {3 \over 2} \gamma c^2_i (c^2 - {5 \over 3} v^2 )
+ \kappa (c^2 - {5 \over 3} v^2) \Big ]
\end{eqnarray}
where $c$ is the adiabatic sound speed, $x = r/r_v$, and all
velocities are expressed in units of $v_v$. The quanities $\gamma$ and
$\kappa$ in these equations are related to the mass injection and
cooling rates according to
\begin{eqnarray}
\gamma (x, v) = {q(r) \dot M(r) \over \rho^2 r_v v_v^2}, \quad
\kappa(x, c) =  {\dot M(r) \Lambda(T) n_e^2 \over \rho^2  v_v^4 r_v}, \cr
 \dot M(r) = 4 \pi \int_0^r q(r) r^2 \;dr.
\end{eqnarray}
\endletB
and  the injected gas is assumed to have a uniform initial isothermal
sound speed of $c_i = (k T_i/(0.61 m_p))^{1/2}$.

We illustrate the behaviour of the wind solutions by studying two
regimes. Firstly, we assume that $q=0$ beyond an initial radius $r_i =
0.04r_v$ defining the base of the flow
({\it i.e.} two disc scale lengths for $f_{coll} = 50$).
 Equations (B2) do not have a transonic
point when $q=0$ (Wang 1995a, see also the discussion below) and so we
begin the integrations at a Mach number slightly greater than unity
with $c^2 = 5c_i^2/3$.  We adopt the parameters of models MW and DW
given in Table 1 and integrate the equations (B1) adopting $\dot M =
10 M_\odot/{\rm yr}$ for model MW and $\dot M = 0.2 M_\odot/{\rm yr}$
for model DW.  These mass injection rates are close to the maximum
rates at times $t
\sim \tau_{ej}$ for the models described in Section 6. The curves in
Figure \ref{figure17} show solutions for initial isothermal sound speeds of
$0.75$, $1.0$ and $1.25$ times the escape velocity from the centre of
the halo ($v_{esc}=430\;\kms$ for model MW and $107\;\kms$ for model DW).

The figure shows that the criterion $v_w \approx \sqrt{2.5} c_i
\simgt v_{esc}$ is about
right if the wind is to reach beyond the virial radius. For $c_i
\approx v_{esc}$ the wind in model MW begins at a high temperature of
$T_i \approx 1.4 \times 10^7\;{\rm K}$ and cools almost adiabatically
initially, reaching a temperature of $\sim 1.5 \times 10^5\;{\rm K}$
at the virial radius. The timescale for the flow to reach the virial
radius, $\sim 2 \times 10^8 {\rm yrs}$, is slightly longer than the
cooling time at $r_v$.  The behaviour of models DW is quite
different. For $c_i \approx v_{esc}$ the initial temperature of the
gas is $T_i \approx 8 \times 10^5{\rm K}$ and cools to $\simlt 10^4
{\rm K}$ by $r = 0.3r_v$. As expected from the discussion in Section
6, cooling is important in outflows from dwarf galaxies (see {e.g.}
Kahn 1981, Wang 1995a, b).

\begin{figure}

\vskip 3.2 truein

\includegraphics{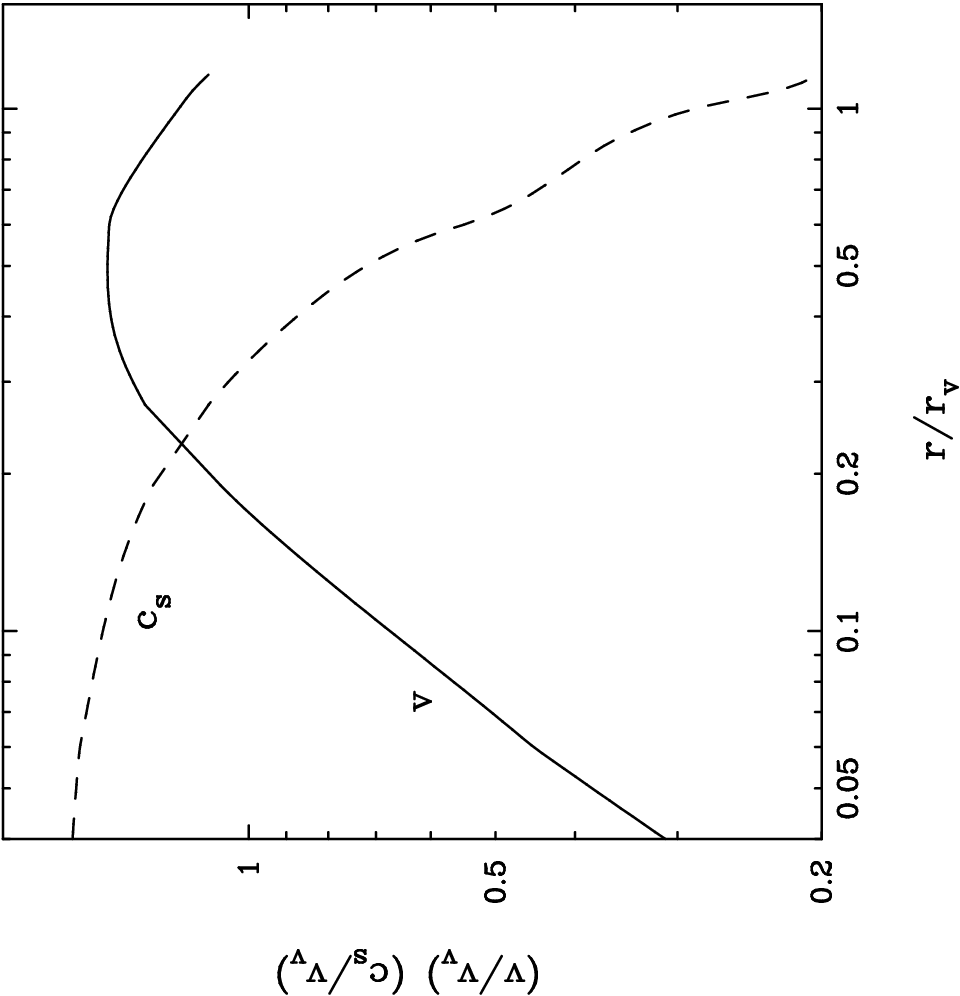}

\caption
{Critical solution for a wind in model DW with $c_i=v_{esc}$, $\dot M(\infty)
= 0.2 M_\odot/{\rm yr}$, and $q(r)$ given by equation (B3).}
\label{figure18}
\end{figure}

An investigation of transonic solutions of equations (B2) require a
model for $q(r)$. An example is illustrated in Figure \ref{figure18}
for model DW,
using
\begletC
\begin{equation}
q(r) = {\dot M(\infty) \over 8 \pi r_w^3} {\rm exp}(- r/r_w),
 \qquad r_w= 0.04r_v.
\end{equation}
\endletB
In this solution, $\dot M(\infty) = 0.2 M_\odot/{\rm yr}$ and the
central gas density was adjusted to obtain a critical solution for the
case $c_i = v_{esc}$. If the gas is to escape from a dwarf galaxy the
transonic point must occur before cooling sets in. For such systems,
the wind parameters would adjust so that a sonic point exists at a
characteristic cooling scale height as shown in Figure \ref{figure18}. The wind
will then cool radiatively just beyond the sonic point forming a cold
wind as discussed above. It is also likely that the wind will be
heated to a temperature of $T \sim 10^4$K by photoionizing radiation
from the galaxy and the general UV background. These sources of
heating have not been included in the models of Figures \ref{figure17}
and \ref{figure18}.

The wind will be thermally unstable when cooling sets in, and may form
clouds. However, in the absence of a confining medium, the clouds
would have a filling factor of order unity so the wind is likely to
maintain its integrity until it meets the surrounding IGM.  The
external pressure required to balance the ram pressure of the wind is
\begletC
\begin{eqnarray}
{ p_{ext} \over k} \approx 80 
\left ( {\dot M \over 0.2 M_\odot/{\rm yr}} \right )  
\left ( { r \over 10\;{\rm kpc}} \right ) ^{-2} \times \qquad \nonumber \\
\left ( {v_w \over 100 {\rm km}/{\rm s} } \right )   
\; {\rm cm}^{-3}\; {\rm K}, 
\end{eqnarray}
\endletB
which is about equal to the pressure of the IGM with a temperature of $10^4$K
and an overdensity of
\begletC
\begin{eqnarray}
\Delta  \approx 4500 \left ( {2 \over 1+z} \right )^3 T_4^{-1} 
\left ( {\dot M \over 0.2 M_\odot/{\rm yr}} \right )  \times \qquad \nonumber
\\
\left ( { r \over 10\;{\rm kpc}} \right ) ^{-2} 
\left ( {v_w \over 100 {\rm km}/{\rm s} } \right )  .
\end{eqnarray}
\endletB
Provided that the halo is devoid of high pressure gas, the cool wind
will propagate beyond the virial radius and will be halted either by
the ram pressure of infalling gas or after sweeping up a few times its
own mass. As pointed out by Babul and Rees (1992), if a dwarf galaxy is 
embedded in a group or cluster of galaxies with a pressure exceeding
$\sim 100 \; {\rm cm}^{-3} {\rm K}$, the bulk motion of the outflowing gas
would be thermalized in a shock and the cooled shocked gas could fall
back onto the galaxy generating a new burst of star formation. The
efficiency of feedback is therefore likely to be a function of
local environment.

\end{appendix}

\end{document}